\documentclass[]{aa}

\input psfig.tex
\usepackage{graphicx}
\usepackage{natbib}
\usepackage{array}
\usepackage{graphics}
\usepackage{latexsym}
\usepackage{amssymb}
\usepackage{amsmath}
\usepackage{fancyhdr}
\usepackage{morefloats}
\bibpunct{(}{)}{;}{a}{}{,}
\include{hyphe}

\begin{document}

\title{Formation and Evolution of the Dust in Galaxies. III. \\
The Disk of the Milky Way }

\author{L. Piovan \inst {1,2}, C. Chiosi \inst{1}, E. Merlin \inst{1}, T. Grassi \inst{1},
R. Tantalo \inst{1}, U. Buonomo \inst{1} and L. P. Cassar\`{a}
\inst{1}}

\institute{    $^1$ Department of Astronomy, Padova University,
Vicolo dell'Osservatorio 3, I-35122, Padova, Italy\\
$^2$Max-Planck-Institut f\"ur Astrophysik, Karl-Schwarzschild-Str.
1, Garching bei M\"unchen, Germany\\
\email{  {lorenzo.piovan\char64unipd.it} }   }

\date{Received: July 2011; Revised: *** ****;  Accepted: *** ***}

%=================== BEGIN ABSTRACT =====================%
\abstract   {Models of chemical evolution of galaxies including the
 dust are nowadays required to decipher the high-z universe. In a
series of three papers we have tackled the problem and set a modern
chemical evolution model in which the interstellar medium (ISM) is
made of gas and dust. In the first paper (Piovan et al., 2011a) we
revised the condensation coefficients for the elements that
typically are present in the dust. In the second paper (Piovan et
al., 2011b) we have implemented the dust into  the Padova model with
infall and radial flows  and tested it against the observational
data for the Solar Neighbourhood (SoNe). In this paper (the third of
the series) we extend it to the whole Disk of the Milky Way (MW).}
{The Disk  is used as a laboratory to analyze the spatial and
temporal behaviour of (i) several dust grain families with the aid
of which we can describe the ISM dust in a simple way, (ii) the
abundances in the gas, dust, and total ISM of the elements present
in the dust emitted by stars and the dust grains grown by accretion
in the ISM, and finally (iii) the depletion of the same elements.} {
The temporal evolution of the dust and gas across the Galactic Disk
is calculated under the effect of radial flows  and a Bar in the
central regions. The gradients of the abundances of C, N, O, Mg, Si,
S, Ca and Fe in gas and dust across the Disk are derived  as a
function of time.} {The theoretical gradients nicely reproduce those
derived from Cepheids, OB stars, Red Giants and HII regions. This
provides the backbone for the companion processes of dust formation
and evolution across the Galactic Disk. We examine in detail the
contributions to dust by AGB stars, SN{\ae} and grain accretion in
the ISM at different galacto-centric distances across the Disk.
Furthermore, we examine the variation of the ratio between silicates
and carbonaceous grains with time and position in the Disk. Finally,
some hints about the depletion of the elements in regions of high
and low SFR (inner and outer Disk) are presented.} {The results
obtained for the Disk of the MW make it possible  to extend the
model we have developed to other astrophysical situations such as
galaxies of different morphological types, like ellipticals or
star-busters, or even to a different theoretical modelling, like the
chemo-dynamical N-Body simulations.} 

\titlerunning{Formation and evolution of dust in galaxies}

\authorrunning{L. Piovan  et al.}
\maketitle

\section{Introduction}
\label{Introduction}

New generation, powerful telescopes and space instrumentation have
unveiled a high-z universe surprisingly obscured by large amounts of
dust, and consequently spurred a new generation of chemical and
spectral galaxy models in which the formation, presence  and
evolution of dust is taken into account \citep[See for instance][to
mention the most recent
ones]{Zhukovska08,Calura08,Valiante09,Pipino11,Gall11a,Gall11b,Mattsson11,Valiante11,Acharyya11,
Kemper11,Dwek11}. Given  the interest in understanding the
properties of the early universe, most of these models are tuned
toward QSOs or star-burst galaxies. They try to get clues about the
role of dust  in the early stages and to estimate how much dust can
be produced both by stars and other processes in the interstellar
medium (ISM) on a given timescale in order to match the amount of
dust observed in high-z objects
\citep{Gall11a,Gall11b,Mattsson11,Dwek11}. However, because of the
huge distances, the information on  dust  to our disposal for these
extremely far objects comes from (i) the extinction curve, that
suggests clues about dust composition by the comparison with the one
for galaxies in the  local such as the SMC, LMC, and M; (ii)
estimates of the dust mass from the fluxes observed in the
pass-bands to our disposal. Other pieces of information are
typically related to the upper limit to the age of the target, its
metallicity and star formation rate (SFR) and, finally, the total
masses in form of  stars, molecular gas (if CO observations are
available), and  Dark Matter \citep{Valiante11}. Since we are
observing the integrated properties of these galaxies, the history
of chemical enrichment for individual elements and different
distances from the centre of the galaxy are hardly available. A
detailed information about the way in which the various elements
behave is however required if we want to cross compare data with
theoretical modelling of dust formation/evolution, because it would
correlate the processes affecting the dust with the gas composition
existing in that part of the galaxy we intend to model
\citep{Zhukovska08,Piovan11b}. This indeed would be the typical case
in which  the relative proportion of silicates/carbonaceous grains
and iron dust develop with time \citep{Dwek05,Zhukovska09} in
different regions of a galaxy, thus getting hints on the dust
mixture. For its proximity and hence richness of data, the best
possible laboratory is the SN. Another useful laboratory in which we
can trace the evolution of the element abundances by means of the
observations of different types of objects is the Disk of the MW.
Nevertheless, till now, only an handful of models have considered
the presence of dust in the Galactic Disk and its evolution in space
and time \citep{Dwek98,Zhukovska09}.\\

\indent Basing on the observational data we have referred to above,
in this study we plan to examine  the formation/evolution of dust
and  depletion of the elements along the MW Disk, trying to satisfy
the constraints derived from the radial gradients in  matter and
abundances  of  various chemical elements as measured in different
types of sources. This paper is the last one of a series of three in
which we analyzed the formation and evolution of dust in the MW. It
is worth briefly summarizing the aims and results  of the first two
papers, and also briefly comment on the gradual development of the
chemical model. In the first paper \citep{Piovan11a} we examined and
presented a compilation  of condensation efficiencies to be adopted
to study the dust content of the gas. In the second paper
\citep{Piovan11b} we apply our condensation coefficients to a
detailed chemical model of the Solar Neighborhood and reproduced the
local depletions of the chemical elements together with other
observational data getting clues on the effect induced by important
parameters such as the IMF, the star formation law, and the
accretion model adopted to simulate the grain growth in the ISM. In
addition to this, the same analysis has been made for the inner and
outer regions of the MW Disk, trying to understand how they
would behave in a high/low star forming environment.\\

\indent The plan of the present paper is as follows. In Sect.
\ref{chemical} we summarize the main features of the chemical model,
more details can be also found in \citep{Piovan11b}. In particular,
we cast the equations governing the model in presence of radial flow
of matter and dust. In Sect. \ref{discrete} we present the equations
for the radial flows in presence of dust, gas, and ISM (gas+dust).
In Sect. \ref{boundaryCond} we discuss the boundary conditions and
the final equation for the innermost shell (Sect. \ref{shell1}), the
generic shell (Sect. \ref{boundary}), and the outermost shell (Sect.
\ref{shellN}). Some computational details are discussed in Sect.
\ref{numericalradf}. In Sect. \ref{RadialMCs} we present a simple
algorithm  based on the Artificial Neural Networks (ANNs) to predict
the amount of molecular gas (cold regions where we assume that dust
accretion takes place) in the different rings of the Disk taking
into account the information about SFR and  surface mass density. In
Sect. \ref{RadialAbundances} we compare the results of the chemical
models, whose properties for the Solar Vicinity have been already
discussed in \citet{Piovan11b}, extended now the whole Disk. The
theoretical predictions are compared with the observational data for
the gradients in gas content, chemical abundances, and abundance
ratios. Furthermore, we present and discuss the radial and time
evolution of many dust-related quantities. In particular, we analyze
the radial behaviour of (i) some typical grain families; (ii) the
elements involved in the dust formation, and (iii) the depletion in
the innermost and outermost regions. Finally, in Sect.
\ref{Discus_Concl} we discuss the results and draw some general
conclusions.

\section{The chemical model} \label{chemical}

The model we are going to develop stems from the original infall
model proposed long ago by \citep{Chiosi80}, in the latest
multi-ring formulation by \citet{Portinari00} and
\citet{Portinari04a} for disk galaxies. This latter includes the
radial flows of matter and the galactic bar that were introduced to
reproduce the radial gradients and the density profile. The adopted
stellar yields are the original ones by \citet{Portinari98} in its
latest version (Portinari 2010 - private communication). The model
is able to describe the evolution of the abundances of the different
refractory elements trapped into dust, properly simulating the
process of injection of the star-dust \footnote{Thereinafter, by
star-dust we refer to the dust directly produced and emitted into
the ISM, by AGB stars and supernov{\ae} (SN{\ae}) both of type I and
type II. } into the ISM \citep{Piovan11a} and dust
accretion/destruction processes whose net balance determines the
amount of dust in the ISM. The model traces the evolution of the
abundance of  some typical grain families supposed to represent the
dust  in the ISM and  single of elements trapped into dust
\citep{Piovan11b}.\\

\indent The Disk of the MW is subdivided in N concentric circular
rings where $r$ is the galacto-centric distance. Each ring or shell
is identified by the mid radius $r_k$ with $k=1,.....,N$. In most
cases, radial flows of interstellar gas and dust are neglected, so
that each ring / shell evolves independently from the others, but in
our model the exchange of matter between contiguous shell is
allowed. The Disk is described by the surface mass density as a
function of the radial coordinate $r$ and time $t$: $\sigma
\left(r_{k},t \right)$. Depending on the case, $\sigma$ can refer to
the ISM ($\sigma^{\mathcal{M}}$), in turn split into  dust or gas
($\sigma^{D}$ or $\sigma^{G}$ respectively), to the stars
($\sigma^{*}$) or to the total mass (simply $\sigma$). At every
radius $r_{k}$,  the surface mass density is supposed to slowly grow
by infall of either primordial or already enriched gas and to fetch
at the present age $t_G$ the mass density profile across the
Galactic Disk for which an exponential profile is best suited to
represent  the surface mass density distribution: $\sigma
\left(r_{k}, t_{G}\right) \varpropto
\exp{\left(-r_{k}/r_{d}\right)}$, where $r_{d}$ is the scale radius
of the Galactic Disk, that is typically estimated of the order of
$4-5$ Kpc. Since the final density profile is a priori known, one
may normalize to it the current total surface mass density of the
ISM. The evolution of the generic elemental species $i$-th in the
dust at the radial distance $r_{k}$ from the centre of the galaxy
and at the time $t$, is described as:

\vspace{-7pt}
\begin{small}
\begin{flalign} \label{DUST}
\frac{d}{dt}&D_{i}\left(r_{k},t \right)  = -\chi_{i}^{D}\psi+ \nonumber \\
&+ \int_{0}^{t-\tau_{M_{B,l}}}\psi\left[\phi
\delta^{w}_{c,i}R_{i}\cdot\left(-\frac{dM}{d\tau_{M}}\right)\right]
_{M\left(\tau\right)}dt^{\prime} +  \nonumber  \\
&+ \left(1-A
\right)\int_{t-\tau_{M_{B,l}}}^{t-\tau_{M_{SNe}}}\psi\left[\phi
\delta^{w}_{c,i}R_{i}\cdot\left(-\frac{dM}{d\tau_{M}}\right)\right]
_{M\left(\tau\right)}dt^{\prime} + \nonumber \\
&+ \left(1-A
\right)\int_{t-\tau_{M_{SNe}}}^{t-\tau_{M_{B,u}}}\psi\left[\phi
\delta^{II}_{c,i}R_{i}\cdot\left(-\frac{dM}{d\tau_{M}}\right)\right]
_{M\left(\tau\right)}dt^{\prime} +  \nonumber \\
&+ \int_{t-\tau_{M_{B,u}}}^{t-\tau_{M_{u}}}\psi\left[\phi
\delta^{II}_{c,i}R_{i}\cdot\left(-\frac{dM}{d\tau_{M}}\right)\right]
_{M\left(\tau\right)}dt^{\prime} +  \nonumber  \\
&+ A\int_{t-\tau_{M_{SNe}}}^{t-\tau_{M_{1},max}}\psi\left[\textit{f}
\left(M_{1}\right)\delta^{II}_{c,i}R_{i,1}\cdot\left(-\frac{dM_{1}}{d\tau_{M_{1}}}\right)\right]
_{M\left(\tau\right)}dt^{\prime} +  \nonumber   \\
&+ A\int_{t-\tau_{M_{1},min}}^{t-\tau_{M_{SNe}}}\psi\left[\textit{f}
\left(M_{1}\right)\delta^{w}_{c,i}R_{i,1}\cdot\left(-\frac{dM_{1}}{d\tau_{M_{1}}}\right)\right]
_{M\left(\tau\right)}dt^{\prime} +  \nonumber   \\
&+ R_{SNI}E_{SNI,i}\delta^{I}_{c,i}+ \nonumber  \\
&-\left[\frac{d}{dt}D_{i}\left(r_{k},t \right) \right]_{out}
+\left[\frac{d}{dt}D_{i}\left(r_{k},t \right) \right]_{rf}+
\nonumber \\
&+\left[\frac{d}{dt}D_{i}\left(r_{k},t \right) \right]_{accr}
-\left[\frac{d}{dt}D_{i}\left(r_{k},t \right) \right]_{SN}
\end{flalign}
\end{small}

\noindent where $\chi_{i}^{D}=\chi_{i}^{D}\left(r_{k},t\right)$. The
first term at the r.h.s. of eqn. (\ref{DUST}) is the depletion of
dust because of the star formation that consumes both gas and dust
(assumed uniformly mixed in the ISM). The second term is the
contribution by stellar winds from low mass stars to the enrichment
of the $i$-th component of the dust. With respect to standard
equations for the sole gas component, the condensation coefficients
$\delta^{w}_{c,i}$ determine the fraction of material in stellar
winds that goes into dust with respect to that in gas (local
condensation of dust in the stellar outflow of low-intermediate mass
stars). The third term is the contribution by stars not belonging to
binary systems and not going into type II SN{\ae} (the same
coefficients $\delta^{w}_{c,i}$ are used). The fourth term is the
contribution by stars not belonging to binary systems, but going
into type II SN{\ae}. The condensation efficiency in the ejecta of
type II SN{\ae} are named as $\delta^{II}_{c,i}$. The fifth term is
the contribution of massive stars going into type II SN{\ae}. The
sixth and seventh term represent the contribution by the primary
star of a binary system, distinguishing between those becoming  type
II SN{\ae} from those failing this stage and using in each situation
the correct coefficients. The eighth term is the contribution of
type Ia SN{\ae}, where the condensation coefficients are named as
$\delta^{I}_{c,i}$ to describe the mass fraction of the ejecta going
into dust. For all the $\delta_{c,i}$ coefficients we  adopt the
prescriptions presented in \citet{Piovan11a}. The last four terms
describe: (1) the outflow of dust due to galactic winds (in the case
of disk galaxies this term can be set to zero); (2) the radial flows
of matter between contiguous shells; (3) the accretion term
describing the accretion of grain onto bigger particles in cold
clouds; (4) the destruction term taking into account the effect of
the shocks of SN{\ae} on grains, obviously giving a negative
contribution. The infall term in the case of dust can be neglected
because we can assume that the primordial material entering the
galaxy is made by gas only without a solid dust component mixed to
it. Equations similar to eqns. (\ref{DUST}) must be written for the
ISM and gas. They are identified by the suffix $\mathcal{M}$ (ISM)
and $G$ (gas). These equations are given in detail in
\citet{Piovan11b} and are shortly summarized below.\\

\noindent Indicating the contribution to the yields by stellar winds
and type Ia and II SN{\ae}  with the symbols
$W_{i,D}\left(r_{k},t\right)$, (and similar expressions for gas and
ISM as a whole) and neglecting the outflow term, eqns. (\ref{DUST}),
the equations for the the gas  and ISM  are

\vspace{-7pt}
\begin{flalign} \label{GISM_B}
\frac{d}{dt}&\mathcal{M}_{i}\left(r_{k},t \right)  =
-\chi_{i}^{\mathcal{M}}\left(r_{k},t \right)\psi\left(r_{k},t
\right)+ \nonumber+W_{i,\mathcal{M}}\left(r_{k},t\right) \nonumber \\
&+\left[\frac{d}{dt}\mathcal{M}_{i}\left(r_{k},t \right)
\right]_{rf}
\end{flalign}
\vspace{-7pt}

\vspace{-7pt}
\begin{flalign} \label{DUST_B}
\frac{d}{dt}&D_{i}\left(r_{k},t \right)  =
-\chi_{i}^{D}\left(r_{k},t \right)\psi\left(r_{k},t \right)+
\nonumber+W_{i,G}\left(r_{k},t\right) \nonumber \\
&+\left[\frac{d}{dt}D_{i}\left(r_{k},t \right) \right]_{accr}
-\left[\frac{d}{dt}D_{i}\left(r_{k},t \right) \right]_{SN} \nonumber
\\ &+\left[\frac{d}{dt}D_{i}\left(r_{k},t \right) \right]_{rf}
\end{flalign}
\vspace{-7pt}

\vspace{-7pt}
\begin{flalign} \label{GAS_B}
\frac{\displaystyle d}{\displaystyle dt}&G_{i}\left(r_{k},t\right)
= -\chi_{G,i}\left(r_{k},t \right)\psi\left(r_{k},t \right)+W_{i,G}\left(r_{k},t\right) \nonumber \\
&-\left[\frac{d}{dt}D_{i}\left(r_{k},t \right) \right]_{accr}
+\left[\frac{d}{dt}D_{i}\left(r_{k},t \right) \right]_{SN} \nonumber
\\ &+\left[\frac{d}{dt}G_{i}\left(r_{k},t \right) \right]_{inf}
+\left[\frac{d}{dt}G_{i}\left(r_{k},t \right) \right]_{rf}
\end{flalign}
\vspace{-2pt}

\noindent It is soon evident that the dust creation/destruction and
the radial flows  make the system of differential equations more
complicated than the original one by \citet{Talbot75} for a one-zone
closed-box model. As the ISM is given by the sum of gas and dust,
only two of these equations are required, furthermore Eqn. in the
case of radial flows eqn. (\ref{GISM_B}) can be used only if the gas
and dust flow with the same velocity.

%%%%%%%%%%%%%%%%%%%%%%Table 1
\renewcommand{\arraystretch}{1.3}
\setlength{\tabcolsep}{2.8pt}
\begin{table*}
\scriptsize
\begin{center}
\caption[]{\footnotesize Parameters of the models. Column (1) is the
parameter number, column (2) the associated physical quantity, and
column (3)  the source and the italic symbols are the identification
code we have adopted. See the text for some further information and
\citet{Piovan11b} for a detailed description of each choice.}
%\begin{tabular}{llllll}
\begin{tabular}{ccl}
\hline \noalign{\smallskip}
n$^{o}$         & \small Parameter     &\small Source and identification label  \\
\hline \noalign{\smallskip} \small 1 & \small IMF           &\small
Salpeter $(\mathcal{A})$,  Larson $(\mathcal{B}$), Kennicutt
$(\mathcal{C})$
Kroupa orig. $(\mathcal{D})$, \\
&&\small  Chabrier $(\mathcal{E})$, Arimoto $(\mathcal{F})$, Kroupa
2007 $(\mathcal{G})$,
Scalo $(\mathcal{H})$, Larson SoNe $(\mathcal{I})$ \\

\small 2 &\small SFR law              &\small Constant SFR
$(\mathcal{A})$, Schmidt $(\mathcal{B})$,
Talbot \& Arnett $(\mathcal{C})$, Dopita \& Ryder $(\mathcal{D})$, Wyse \& Silk $(\mathcal{E})$ \\
\small 3 & \small $\chi_{MC}$ model    &\small Artificial Neural
Networks model $(\mathcal{A})$, Constant $\chi_{MC}$
as in the Solar Neigh. $(\mathcal{B})$               \\

\small 4 &\small Accr. model          &\small Modified
\citet{Dwek98} and \citet{Calura08} $(\mathcal{A})$;
adapted \citet{Zhukovska08} model $(\mathcal{B})$         \\

\small 5 &\small SN{\ae} Ia model &\small Dust injection adapted
from: \citet{Dwek98}, \citet{Calura08} $(\mathcal{A})$,
\citet{Zhukovska08}  $(\mathcal{B})$                   \\

\small 6 &\small SN{\ae} II model     &\small Dust injection adapted
from: \citet{Dwek98} $(\mathcal{A})$, \citet{Zhukovska08}
$(\mathcal{B})$, \\
&& \small \citet{Nozawa03,Nozawa06,Nozawa07}
 $(\mathcal{C})$\\

\small 7 &\small AGB model            &\small Dust injection adapted
from: \citet{Dwek98}
$(\mathcal{A})$, \citet{Ferrarotti06}  $(\mathcal{B})$  \\

\small 8 &\small Galactic Bar         &\small No onset
$(\mathcal{A})$, onset at $t_{G}-4$ Gyr
$(\mathcal{B})$, onset at $t_{G}-1$ Gyr $(\mathcal{C})$ \\

\small 9 &\small Efficiency SFR       &\small Low efficiency
$(\mathcal{A})$, medium efficiency $(\mathcal{B})$
, high efficiency $(\mathcal{C})$  \\
\hline \label{Parameters}
\end{tabular}
\end{center}
\renewcommand{\arraystretch}{1}
%\begin{flushleft}\footnotesize
%Abundances $A(X)$ of the element $X$ are in units of
%$\log_{10}\left(\textrm{N(X)}/\textrm{N(H)}\right)+12$.
\footnotesize
%$^{1}${This is the same low photosphere value as in
%\citet{Lodders03}, selected from \citet{AllendePrieto02}, and
%confirmed in \citet{Asplund05b} and \citet{Scott06}.}\,
%\end{flushleft}
\end{table*}
\renewcommand{\arraystretch}{1}

In the model by \citet{Piovan11b}, many prescriptions/values for key
quantities are explored. In brief we consider (i) nine cases for the
IMF and five cases for the SFR, and for each SFR different specific
efficiencies can be used; (ii) two possible models of accretion of
dust grains in the cold regions of the ISM; (ii) different cases of
star-dust formation in AGB stars, type II and type Ia SN{\ae}; (4)
the age at which the effects induced by the Galactic Bar are taken
into account. The presence of the Bar is mimicked by a change in the
pattern of the velocities for the matter flowing from one shell to
another \citep{Portinari00}. The main motivation for the Bar  is  to
reproduce the molecular ring at  4-5 Kpc in the gas distribution
across the Galactic Disk. The velocity pattern is taken from
\citet{Portinari00} and is suitably chosen for every SFR and age
$t_{Bar}$ at which the effect of the Bar is supposed to begin (the
reader should  refer \citet{Portinari00}  for all the details). In
this paper, we keep constant the prescription for the velocity
patter (Bar) assigned by \citet{Portinari00} to
each case of SFR and age $t_{Bar}$.\\
\indent Table \ref{Parameters}  summarizes the many combinations of
the parameters and associated models that are possible. For a
thorough discussion of the effects induced by changing one at a time
the various parameters on the formation, destruction and evolution
of (the Solar Neighborhood, in particular) the reader should refer
to  \citet{Piovan11b}. High number of parameters implies a high
number of possible models. So a code must be found to establish a
easy to read correspondence between the particular combination of
parameters and the associated model.  Each model  is identified  by
a string of nine letters (the number of parameters) in italic face
whose position in the string and  alphabetic order  corresponds to a
particular parameter and choice for it. The code is presented in
Table \ref{Parameters}.  The sequence should be read from top to
bottom: for instance, the string $\mathcal{DBAABABAB}$ corresponds
to Kroupa 1998 IMF, Schmidt SFR, ANN model for the fraction of
molecular clouds in the ISM $\chi_{MC}$, \citet{Dwek98} accretion
model, \citet{Zhukovska08} recipe for the yields of dust from type
Ia SN{\ae}, \citet{Dwek98} condensation efficiencies for type II
SN{\ae}, \citet{Ferrarotti06} condensation efficiencies for AGB
stars, no effect of the Bar on the inner regions, and high
efficiency $\nu$ of the SFR. Radial flows are always be included by
default. We do not explore here the effects of different IMFs
and/SFR that have been already examined by \citet{Piovan11b}.\\
\indent We focus here  on the radial behaviour of dust formation and
evolution and seek to answer the following questions. What are the
dominant types of dust at different ages of the Galaxy and radial
distances? Are the predicted radial gradient of element abundances
in agreement with the observations in such a way to allow a correct
calculation of dust formation on consistent initial conditions? What
are the differences between the inner and outer regions of the
Galaxy? To highlight these and other related issues we set up the
so-called "reference model" in  which we vary the key parameters.
This reference model is identified by the string
$\mathcal{GDABBCBBB}$ and it includes the most refined recipes for
the production of dust by stars  and the accretion of dust in the
ISM.  It is worth noticing here the combination of the parameters we
have chosen  does  not necessarily yield the model best fitting the
observations across the Galactic Disk, but it only meant to provide
the gross features on the physical processes for dust formation
across the Disk and clues on the general behaviour of the system.
Clearly, the model must be in satisfactory agreement with the
observational data, even though it does not perfectly match it.

\section{Modelling radial flows}
\label{discrete}

A key term in the description of the properties of gas and dust
along the Disk is the one of radial flow  in eqn. \label{DUST}
driving  the exchange of gas/dust between contiguous shells. The
radial flows  are described according to the model developed by
\citet{Portinari00} however extended to  a two components fluid made
of gas and dust each of which in principle moving with its own
velocity.  Recalling that each circular ring /shell is identified by
the mid-point radius $r_k$ and the inner and outer borders by the
radii $r_{k-\frac{1}{2}}$ and $r_{k+\frac{1}{2}}$, the material can
flow across the inner and outer borders with velocity
$v_{k-\frac{1}{2}}$ and $v_{k+\frac{1}{2}}$, respectively. Flow
velocities are taken positive outward and negative inward. Let
denote with $F_{D}\left(r\right)$ the radial flow of  dust  and
$F_{G}\left(r\right)$ that of gas, so that the total flux of ISM is
$F_{\mathcal{M}}\left(r\right)=F_{D}\left(r\right)+F_{G}\left(r\right)$.
The motion of this two-components fluid  alter the gas surface
density in the $k$-th shell. The most convenient description of it
is as follows:  If there is no velocity drift between gas and dust,
as we will later assume, it is more convenient to follow the ISM and
the dust, whereas for different gas and dust velocities it is more
convenient to follow dust and gas as two independent fluids.
Therefore we have:

\vspace{-9pt}
\begin{equation}
\label{dsigmarf1} \left[ \frac{d
\sigma^{\mathcal{C}}r\left(k\right)}{d t} \right]_{rf} = -
\frac{1}{\pi \left( r^2_{k+\frac{1}{2}} - r^2_{k-\frac{1}{2}}
\right) }\Delta F_{\mathcal{C}}\left(r_{k}\right)
\end{equation}
\vspace{-6pt}

\noindent where $\mathcal{C}=\mathcal{M},G,D$ ($\mathcal{C}$ is for
component) means that we have the same formal expression  for ISM
($\mathcal{M}$), gas ($G$) and dust ($D$). The term   $\Delta$  is
$F_{\mathcal{C}}\left(r_{k}\right) =
F_{\mathcal{C}}(r_{k+\frac{1}{2}}) -
F_{\mathcal{C}}(r_{k-\frac{1}{2}})$. The flux across the external
border of the ring/shell $r_{k+\frac{1}{2}}$ is:

\vspace{-8pt}
\begin{flalign}
\label{flux3GS} F_{\mathcal{C}}(r_{k+\frac{1}{2}}) &= 2 \pi
r_{k+\frac{1}{2}} \, v_{k+\frac{1}{2}}^{\mathcal{C}} \left[
\chi(v_{k+\frac{1}{2}}^{\mathcal{C}}) \,
\sigma^{\mathcal{C}}_{K} + \right. \nonumber \\
&+ \left.\chi(-v_{k+\frac{1}{2}}^{\mathcal{C}}) \,
\sigma^{\mathcal{C}}_{k+1} \right]
\end{flalign}
\vspace{-6pt}

\noindent where $\chi(x)$ is the step function: {\mbox{$\chi(x)=1$
or 0}} for {\mbox{$x >$ or $\leq 0$}}, respectively.
Eqns.~(\ref{flux3GS}) are a sort of ``upwind approximation'' for the
advection term to be included in the model equations, describing
either inflow or outflow depending on the sign of the  velocity
$v_{k+\frac{1}{2}}^{\mathcal{C}}$. Clearly, since there are  two
components, gas and dust, it may happen that $v_{k+\frac{1}{2}}^{G}
\ne v_{k+\frac{1}{2}}^{D}$. The simplest case is for
$v_{k+\frac{1}{2}}^{G} = v_{k+\frac{1}{2}}^{D} =
v_{k+\frac{1}{2}}^{\mathcal{M}}$, i.e. when gas and dust are coupled
and move along the same direction.  Analogous expression holds for
the flux $F_{\mathcal{C}}(r_{k-\frac{1}{2}})$ across the inner
border of the ring/shell.

 Let's take the inner edge
$r_{k-\frac{1}{2}}$ at the midpoint between $r_{k-1}$ and $r_k$, and
similarly for the outer edge $r_{k+\frac{1}{2}}$. Writing
Eqn.~(\ref{dsigmarf1}) separately for each chemical species $i$ and
as function of   $G_i$'s,  we may write the radial flow term in eqn. \ref{DUST}
and similar ones for other components as follows:

\vspace{-7pt}
\begin{equation}
\begin{split}
\label{dGirf} \left[ \frac{d}{dt} \mathcal{C}_i(r_k,t) \right]_{rf}
& = \alpha_k^{\mathcal{C}} \, \mathcal{C}_i(r_{k-1},t)
\,-\, \beta_k^{\mathcal{C}} \, \mathcal{C}_i(r_k,t) \,+ \\
 & +\, \gamma_k^{\mathcal{C}} \, \mathcal{C}_i(r_{k+1},t)
\end{split}
\end{equation}
\vspace{-7pt}

\noindent where again $\mathcal{C}=\mathcal{M}/G/D$ and:

\begin{footnotesize}
\begin{flalign}\label{coeffradf}
\alpha_k^{\mathcal{C}} &= \frac{2}{r_k + \frac {r_{k-1} +
         r_{k+1}}{2}} \left[ \chi(v_{k-\frac{1}{2}}^{\mathcal{C}})
         v_{k-\frac{1}{2}}^{\mathcal{C}}
         \frac{r_{k-1}+r_k}{r_{k+1}-r_{k-1}} \right] \cdot \nonumber
         \\
         &\cdot \frac{\sigma_{A (k-1)}}{\sigma_{A k}} \nonumber \\
\beta_k^{\mathcal{C}} &= - \, \frac{2}{r_k + \frac {r_{k-1} +
        r_{k+1}}{2}}
        \cdot \left[ \chi(-v_{k-\frac{1}{2}}^{\mathcal{C}}) v_{k-\frac{1}{2}}^{\mathcal{C}}
        \frac{r_{k-1}+r_k}{r_{k+1}-r_{k-1}} - \right. \nonumber \\
        &\left. - \chi(v_{k+\frac{1}{2}}^{\mathcal{C}})
        v_{k+\frac{1}{2}}^{\mathcal{C}} \frac{r_k+r_{k+1}}{r_{k+1}-r_{k-1}} \right] \nonumber \\
\gamma_k^{\mathcal{C}} &= - \frac{2}{r_k + \frac {r_{k-1} +
         r_{k+1}}{2}}
         \left[ \chi(-v_{k+\frac{1}{2}}^{\mathcal{C}}) v_{k+\frac{1}{2}}^{\mathcal{C}}
         \frac{r_k+r_{k+1}}{r_{k+1}-r_{k-1}} \right] \cdot \nonumber
         \\
         &\cdot \frac{\sigma_{A (k+1)}}{\sigma_{A k}}
\end{flalign}
\end{footnotesize}

\noindent The terms at the r.h.s.  of the eqn.~(\ref{dGirf})
describe the contribution of the 3 contiguous shells to driving the
matter flows: the first term represents the gas being gained by
shell $k$ from the shell $k$--1, the second term is the gas  lost by
the shell $k$ towards the shells $k$--1 and $k$+1, and the third
term is the gas acquired by the shell  $k$ from the shell $k$+1
\citep{Portinari00}. The quantities $\sigma_{A(k-1)}$,
$\sigma_{Ak}$, and $\sigma_{A(k+1)}$ are compact notations for the
present-day surface mass density profile
$\sigma(r_j,t_G)\equiv\sigma_{A}(r_j)$ where $j=k-1,k,k+1$, and $A$
stands for accreted matter. The coefficients given by Eqns.
(\ref{coeffradf}) are all $\geq 0$ and depend only on the shell $k$,
not on the chemical species $i$ considered. They will be different
for gas and dust if a drift between gas and dust is considered. If
there is no drift, then the $\mathcal{M}$ equation can be used
together with one between dust and gas equations. If the velocity
pattern is constant in time, $\alpha_k^{\mathcal{C}}$,
$\beta_k^{\mathcal{C}}$ and $\gamma_k^{\mathcal{C}}$ are also
constant in time. There is an important point to note concerning the
present-day surface mass density profile. In the case of static
models (no radial flows) the present-day surface mass density is
known a  \textit{a priori} and is determined by the mass accretion
law one has assumed, i.e. {\mbox{$\sigma(r_k,t_G) \equiv
\sigma_{A}(r_{k}) \equiv \sigma_{Ak}$}}. In other words, in static
models the radial profile for accretion can be directly chosen so as
to match the observed present--day surface density in the Disk
\citep{Portinari98,Portinari99}. The inclusion of  radial gas flows
changes the expected final density profile (as it is now set up by
infall and radial flows). Therefore, {\mbox{$\sigma(r_k,t_G) \neq
\sigma_A(r_k)$}} and $\sigma(r,t_G)$ cannot be known in advance. It
is known indeed only  {\it a posteriori} \citep[see][for more
details]{Portinari00}.  At the end of each simulation we need to
check how much radial flows have altered the actual density profile
$\sigma(r_k,t_G)$ with respect to the pure accretion profile
$\sigma_A(r_k)$. In any case, if the flow speeds are small (of the
order of $v \sim 1$~km~sec$^{-1}$), the two profiles are expected to
be close each other.

\subsection{Boundary conditions} \label{boundaryCond}

Eqns.~(\ref{dGirf}) need to be modified for the innermost and the
outermost shells, since for these regions at  shells $k$--1 or $k$+1
are not defined.

\subsubsection{The innermost shell}
\label{shell1}

Our model is limited to the Disk of the MW and cannot follow the
innermost regions of the Galaxy where the Bulge dominates. A correct
description of this region would require a more complicated model: a
spherical component simulating the Bulge combined with the Disk each
of which with its own mass distribution and formation history.
Therefore we truncate the Disk to a innermost boundary (ring)
located where the Bulge stars dominating, i.e. at about
$r_1=2\sim2.5$~kpc. At this innermost boundary, the first shell is taken
symmetric with respect to  $r_1$ and its innermost radius is

\vspace{-7pt}
\begin{equation}
r_{\frac{1}{2}} = ( 3 r_1 - r_2 )/2
\end{equation}
\vspace{-7pt}

\noindent At this layer we impose that always
{\mbox{$v_{1/2}^{\mathcal{C}} \leq 0$}}, therefore {\mbox{$k=1$}},
and Eqn.~(\ref{dGirf}) become

\begin{equation} \label{dGirf1}
\left[ \frac{d}{dt}
\mathcal{C}_i(r_1,t) \right]_{rf} = - \beta_{1}^{\mathcal{C}}
\mathcal{C}_i(r_1,t) \, + \, \gamma_{1}^{\mathcal{C}}
\mathcal{C}_i(r_2,t)
\end{equation}
\vspace{-7pt}

\noindent with:

\vspace{-7pt}
\begin{small}
\begin{equation}
\beta_{1}^{\mathcal{C}} = -\frac{1}{2 r_1} \left[
v_{\frac{1}{2}}^{\mathcal{C}}\,\frac{3 r_1-r_2}{r_2-r_1} \,-\,
\chi(v_{\frac{3}{2}}^{\mathcal{C}}) \, v_{\frac{3}{2}}^{\mathcal{C}}
\, \frac{r_1+r_2}{r_2-r_1} \right] \nonumber
\end{equation}
\end{small} \vspace{-7pt}
\begin{small}
\begin{equation}
\label{coeffradf1} \gamma_{1}^{\mathcal{C}} = -
\chi(-v_{\frac{3}{2}}^{\mathcal{C}}) \,
v_{\frac{3}{2}}^{\mathcal{C}} \, \frac{1}{2 r_1} \,
\frac{r_1+r_2}{r_2-r_1}\, \frac{\sigma_{A 2}}{\sigma_{A
1}}~~~~~~~~~~
\end{equation}
\end{small} \vspace{-7pt}

\subsubsection{Boundary conditions at the Disk edge}
\label{boundary}

The Galactic Disk is conceived as made of a inner region, where star
formation and chemical enrichment occurs and stars gas and dust
exist, and a external region in which only gas is present, no stars
can be formed and no chemical enrichment can take place. This
picture is supported both by observational data showing that  in
external spirals HI disks are observed to extend much beyond the
optical disk and theoretical considerations about stability
preventing SF beyond a certain radius \citep{Toomre64,Quirk72}. The
part of the Disk we are interested in is the first one, which may
extend up to about 20--22 kpc. Therefore the mid radius of the most
external shell  ($k=N$) can be located at  $r_{N} = 20-22$~kpc. all
the regions belonging to the Disk external to this limit can be
considered as a large   reservoir of gas with primordial composition
from which gas can flow inside. At the present time  when the
gravitational settling of the proto-galactic cloud is over, the
radial flow of gas from this external region of Disk can be even
more efficient than the classical infall mechanism.  If no star
formation occurs, the evolution of the gas surface density in the
outer disk $\forall \, r>r_{N+1/2}$ can be described by
\citep{Lacey85,Portinari00}:

\vspace{-7pt}
\begin{equation}
\label{radfnoSF} \frac{\partial
\left[G\left(r,t\right)\right]}{\partial t} = A(r) \,
e^{-\frac{\displaystyle t}{\displaystyle \tau(r)}} \,-\, \frac{1}{r}
\, \frac{\partial}{\partial r} [r v \cdot G\left(r,t\right)]
\end{equation}
\vspace{-7pt}

Without star formation, there is no chemical evolution and the
pattern of abundances always remain the primordial one ($X_{i,
inf}$). We assume that this primordial material is also  dust-free,
that is $D(r,t)=0$ for $r>r_{N+\frac{1}{2}}$. As a consequence of
it,   the normalized surface density of the outer Disk is given only
by $\mathcal{M}(r,t)=G(r,t)$ for $r>r_{N+\frac{1}{2}}$ and this
component is the only one to be  considered in Eqn.
$\left(\ref{radfnoSF}\right)$. Other simplifying conditions for the
the most external part of the Disk have been discussed by
\citet{Portinari00}. They all  apply for  $r>r_{N+\frac{1}{2}}$. In
brief, the infall time-scale is uniform $\tau(r) \equiv
\tau(r_{N})$; the inflow velocity is uniform and constant $v(r,t)
\equiv v_{N+\frac{1}{2}}$ and the infall profile is flat $A(r)
\equiv A_{ext}$ in agreement with the observed gas discs in spirals.
With these assumptions, Eqn.~(\ref{radfnoSF}) becomes:

\vspace{-7pt}
\begin{equation}
\label{bordereq} \frac{\partial
\left[G\left(r,t\right)\right]}{\partial t}\,+\,v\,\frac{\partial
\left[G\left(r,t\right)\right]}{\partial r}
 = A \, e^{-\frac{t}{\tau}} \,-\, \frac{v}{r} \,\sigma
\end{equation}
\vspace{-7pt}

\noindent where  $\tau \equiv \tau(r_N)$, $v \equiv
v_{N+\frac{1}{2}}$, and $A \equiv A_{ext}$ to simplify the notation.
Eqn.~(\ref{bordereq}) has a straightforward analytical solution
\citep{Portinari00}:

\vspace{-7pt}
\begin{small}
\begin{equation}
\label{borderconditionTrf}
\begin{array}{l l}
\sigma(r,t) = & A \, \tau \, \times \\
 & \\
\multicolumn{2}{r}{ \times \left[ \left( 1 - e^{-\frac{t}{\tau}}
\right) + \frac{v}{r} \left( \tau \left( e^{-\frac{T_{rf}}{\tau}} -
e^{-\frac{t}{\tau}} \right) - (t - T_{rf}) \right) \right] }
\end{array}
\end{equation}
\end{small}
\vspace{-2pt}

\noindent where $T_{rf} \geq 0$ is the time when radial inflows are
assumed to start. Eqn.~(\ref{borderconditionTrf}) is our boundary
condition at the outermost edge.

Notice that relation (\ref{borderconditionTrf}) is the solution of
Eqn. (\ref{bordereq}) in the idealized case of an infinite, flat gas
layer extending boundless to any $r > r_N$ (see also Appendix~B). Of
course, this does not correspond to gaseous disks surrounding real
spirals. However,  as we are considering  only slow inflow
velocities ($v \sim 1$~km~sec$^{-1}$), with typical values of
$r_N=20-22$~kpc and $t_G=12.8$~Gyr, the gas actually drifting into
the inner  disk shells will be just the gas originally accreted from
regions  within $r \sim 35-40$~kpc. Therefore, the boundary
condition~(\ref{borderconditionTrf}) remains valid as long as the
gas layer extends out to $\sim 35-40$~kpc, a very plausible
assumption since observed gaseous disks extend over a few tens or
even {\mbox{$\sim 100$~kpc}}.

\subsubsection{The outermost shell} \label{shellN}

We take a reference external radius $r_{ext} > r_N$ in the outer
disc where the (total and gas) surface density $\sigma(r_{ext},t)
\equiv \sigma_{ext}(t)$ is given by the boundary condition
(\ref{borderconditionTrf}); typically, $r_N =22$~kpc and $r_{ext}
\sim 24$~kpc. We take the outer edge of the shell at the midpoint:

\vspace{-7pt}
\begin{equation}
 r_{N+\frac{1}{2}} = (r_N + r_{ext})/2
\end{equation}
\vspace{-7pt}

The radial flow term for the $N$-th shell is:

\vspace{-7pt}
\begin{flalign}  \label{dGirfN}
\left[ \frac{d}{dt} \mathcal{C}_i(r_N,t) \right]_{rf} = &
\alpha_N^{\mathcal{C}} \, \mathcal{C}_i(r_{N-1},t) \,-\,
\beta_N^{\mathcal{C}} \, \mathcal{C}_i(r_N,t) \,+ \nonumber \\& +\,
\omega^{\mathcal{C}}_i(t)
\end{flalign}
\vspace{-7pt}

\noindent where:
\begin{equation}
\label{coeffradfN}
\begin{array}{l l}
\omega^{^{\mathcal{C}}}_i(t) = & - X_{i,inf} \,\,\,
\chi(-v_{N+\frac{1}{2}}^{\mathcal{C}}) \,\,
v_{N+\frac{1}{2}}^{\mathcal{C}}
\, \times \\
 & \times \,
\frac{4}{r_{N-1} + 2 r_N +
r_{ext}}\,\,\frac{r_N+r_{ext}}{r_{ext}-r_{N-1}} \,\,
\frac{\sigma_{ext}(t)}{\sigma_{A N}}
\end{array}
\end{equation}

\noindent for $\mathcal{M}$ and $G$, while for dust
$\omega^{\mathcal{C}}_i(t)=0$.

\subsection{The numerical solution}
\label{numericalradf}

Using  Eqns. (\ref{dGirf}), (\ref{dGirf1}) and (\ref{dGirfN}),
neglecting for the moment the terms for dust accretion and
destruction, the basic set of Eqns. (\ref{GISM_B}), (\ref{DUST_B})
and (\ref{GAS_B}) can be written as:

\vspace{-7pt}
\begin{small}
\begin{equation} \label{RFSystem}
\left \{
\begin{aligned}
\frac{d}{dt}\mathcal{C}_i(r_1,t) &= \vartheta^{\mathcal{C}}_1(t)
\mathcal{C}_i(r_1,t)+\gamma_1^{\mathcal{C}}\mathcal{C}_i(r_2,t)+W_{i,\mathcal{C}}(r_1,t)\\
\frac{d}{dt}\mathcal{C}_i(r_k,t) &= \alpha_k^{\mathcal{C}}
\mathcal{C}_i(r_{k-1},t) +\vartheta^{\mathcal{C}}_k(t)\mathcal{C}_i(r_k,t)\, +  \\
&+ \gamma_k^{\mathcal{C}} \mathcal{C}_i(r_k,t) +
W_{i,\mathcal{C}}(r_k,t) \, \, \, \, \\ &k=2,...,N-1 \\
\frac{d}{dt}\mathcal{C}_i(r_N,t) &= \alpha_N^{\mathcal{C}}
\mathcal{C}_i(r_{N-1},t) + \vartheta^{\mathcal{C}}_N(t) \, \mathcal{C}_i(r_N,t)\, +  \\
&+W_{i,\mathcal{C}}(r_N,t)+\omega^{\mathcal{C}}_i(t)
\end{aligned}
\right.
\end{equation}
\end{small}
\vspace{-3pt}

\noindent where we have introduced:

\vspace{-7pt}
\begin{align}
\vartheta^{\mathcal{C}}_k(t) \equiv - \left(
\eta^{\mathcal{C}}(r_k,t) + \beta^{\mathcal{C}}_k \right) \leq 0
\end{align}
\vspace{-13pt}
\begin{align}
\eta^{\mathcal{C}}(r_k,t) \equiv  \frac{\Psi}{^{\mathcal{C}}}(r_k,t)
\end{align}
\vspace{-7pt}

The terms $W_{i,\mathcal{C}}(r_k,t)$ are defined as in Eqns.
(\ref{GISM_B}), (\ref{DUST_B}) and (\ref{GAS_B}). We refer to
\citet{Portinari98} for further details on the quantity $\eta$.
Neglecting, for the time being, that  $\eta$'s and  $W_i$'s contain
the $\mathcal{C}_i$'s themselves, we are dealing with a linear,
first order, non homogeneous system of differential equations with
non constant coefficients, of the kind:
\begin{equation}
\label{systemradf} \frac{d \vec{\mathcal{C}}_i}{dt} \,=\,{\cal A}(t)
\, \vec{\mathcal{C}}_i(t) \,+\, \vec{W}_i(t)
\end{equation}
There is a system like (\ref{systemradf}) for each chemical species
$i$, but the matrix of the coefficients ${\cal A}(t)$ is independent
of $i$. The way by which this system is solved and then dust
evolution is calculated depends on the specific hypotheses made for
gas and dust.

(1) \textsf{Radial flows are neglected but dust is included}. Since
the radial flows are not started from the beginning of the evolution
of the galaxy, this case usually happens in the early stages of the
evolution of the MW and it lasts some Gyrs \citep{Portinari00}. In
this case the evolution of the $\mathcal{M}$ and $D$ components
should be followed. The classical static system for the ISM in which
the different rings evolve independently (\ref{GISM_B}) (without
flowing term) is solved according to the classical Talbot \& Arnett
technique \citep{Portinari98}. Once this system is solved the
accretion/destruction terms for dust can be taken into account (see
\citet{Piovan11b}) is solved to find the evolution of the dust. The
presence of the accretion/destruction terms does no longer allow to
use the standard resolving method and the new system of ordinary
differential equations must be integrated with a different
technique. We adopt a combination of the ODEINT routine (4-th order
Runge-Kutta) with controlled step-size \citep{Press92} and the
powerful DOP853 Runge-Kutta routine (method of Dormand \& Prince) of
order 8 with a 5th order estimator and a 3rd order correction
\citep{Hairer10a,Hairer10b}.

(2) \textsf{Radial flows and dust are both included}. The equations
to be solved depend on whether or not some velocity drift exists
between dust and gas. The simplest and most widely adopted
hypothesis  is that not only dust is homogeneously mixed to the gas,
but both  radially flow at the same velocity. In this case, the two
systems of equations to be solved are (\ref{GISM_B}) and
(\ref{DUST_B}). The system (\ref{GISM_B}) is reduced to the form
(\ref{RFSystem}) by including the radial flows and solved according
to the technique developed by \citet{Portinari00} to whom the reader
should refer for all  details. Once the  evolution of the ISM is
computed,  Eqn. (\ref{DUST_B}) for dust  is setup by means of
accretion/destruction terms \citep{Piovan11b}. The system of
ordinary differential equations is now solved with the ODEINT and
DOP853 integrators. The reason to calculate first the evolution for
the ISM is that the equations for the accretion/destruction
processes regulating the dust evolution include the total amount of
interstellar matter as an ingredient.

(3)  \textsf{Velocity drift between the ISM components}. In such a
case, the global system of differential equations, now including gas
and dust together, made by
 Eqns. (\ref{GAS_B}) and (\ref{DUST_B}), with the radial flow term as
in Eqn. (\ref{dGirf}) and the accretion/destruction terms for dust
must be solved. In such a case it is not possible to obtain a priori
the evolution of the ISM. Again the ODEINT and DOP853 integrators
are used to evolve the system as a function of time.

Some general considerations can be made: rather small time-steps are
needed to keep stable  the numerical model; the required time-steps
get smaller and smaller at increasing the flow velocities  and
making thinner the shells. The  integrators of ordinary differential
equations  self-regulate the internal time-step according to the
required precision once the external time-step is chosen. To
describe the gas flows in a disk with an exponential density
profile, the shells are equi-spaced in the logarithmic, rather than
linear, scale (so that they roughly have the same mass, rather than
the same width). We model the Galactic Disk using 33 shells from 2
to 21 kpc, equally spaced in the logarithmic scale; their width
ranges from $\sim 0.15$~kpc for the inner shells to $\sim 1-1.5$~kpc
for the outermost ones. With such a grid spacing, and velocities up
to {\mbox{$\sim 1$~km~sec~$^{-1}$}}, suitable time-steps are of
$10^{-4}$~Gyr. This means that roughly {\mbox{$1.5 \times 10^5$}}
time-steps, times 35 shells, are needed to complete each model,
which would translate in excessive computational times. This
drawback is avoided by separating the time-scales in the code
following the suggestion by \citep{Portinari00}.

%%%%%%%%%%%%%%%%%%%%Figure 1
\begin{figure}
\centerline{\hspace{-15pt}
\includegraphics[height=11cm,width=8.5cm]{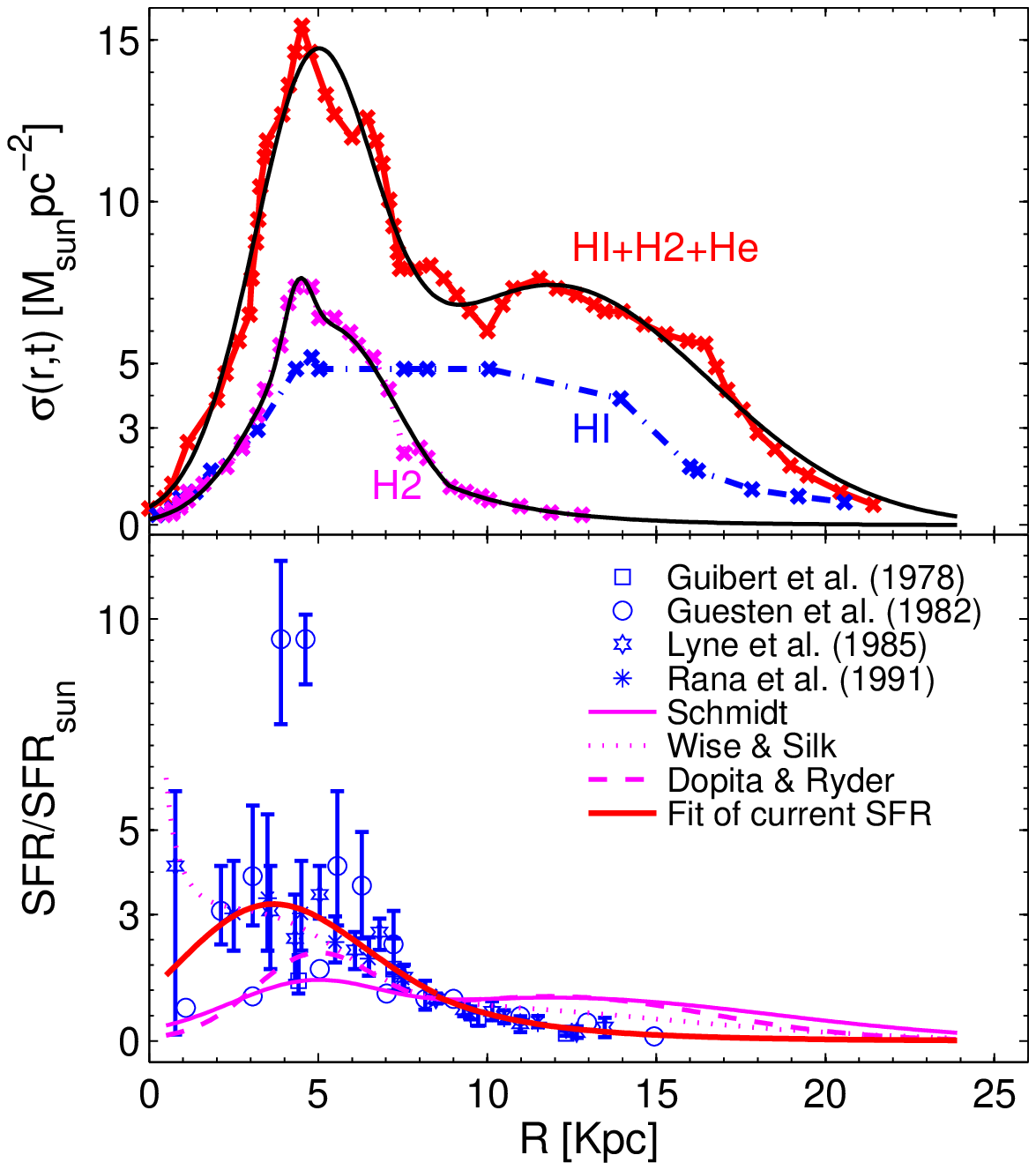}\vspace{-25pt}}
\caption{\textbf{Upper panel}: present-day radial profiles across
the MW Disk for  molecular hydrogen
$\sigma_{H_{2}}\left(r,t_{G}\right)$,  neutral hydrogen
$\sigma_{HI}\left(r,t_{G}\right)$, and  total amount of gas
$\sigma^{G}\left(r,t_{G}\right)$ obtained taking into account helium
correction \citep{Boissier99,Yin09}. \textbf{Lower panel}:
present-day  radial profile of star formation across the MW Disk,
according to the data by \citet{Guibert78} (squares),
\citet{Guesten82} (circles), \citet{Lyne85} (hexagons),
\citet{Rana91} (stars). Superposed to the data are three theoretical
SF laws (Schmidt - continuous line, Wyse \& Silk - dotted line,
Dopita \& Ryder - dashed line) and an exponential best fit, all
referring to the present age $t_{G}$ of the MW.} \label{GASeSFRMW}
\end{figure}

\section{The Radial variation of the fraction of MCs} \label{RadialMCs}

The accretion of dust in the MW Disk can be simulated in our model
by means of two possible choices: (1) a simple model that adopts
typical accretion timescales to simply determine the amount of each
element embedded into dust; (2) a more refined model with variable
timescales that follows the accretion of dust grains in some typical
grain families inside the cold regions of the Disk. This latter
description is the one included in our reference model and is
adopted here to examine the radial behaviour of dust formation and
evolution. Since in our  one-phase chemical model no description is
available for a multi-phase ISM, it is unavoidable to estimate from
time to time (possibly related to the physical properties of the
system) the amount mass  in cold molecular clouds (MCs) in each
ring. The partition between MCs and diffuse ISM (the remaining part
of the ISM) crucially enters the accretion term in eqns.
\ref{GISM_B}, \ref{DUST_B} and \ref{GAS_B} and drives the evolution
of the dust budget. In this formulation the fraction, $\chi_{MC}$,
of ISM locked up in MCs is a parameter. For this reason we have
included two possibilities for $\chi_{MC}$. In the first one, from
the present data for the Solar Vicinity \citet{Zhukovska08} estimate
$\chi_{MC}=0.2$, and assume that it has remained constant from the
formation of the MW to the present. Therefore, a constant value is
adopted during the entire evolution of the Solar Neighbourhood. The
second description for $\chi_{MC}$ is presented below and it stands
on the data for the MW Disk. In practice, it correlates the fraction
of MCs to the SFR and the local gas density. In \citet{Piovan11b} we
observed as both recipes produced similar results for the Solar
Vicinity, but in the case of a constant fraction $\chi_{MC}$ the
amount of dust could be underestimated and overestimated in the
innermost and outermost regions of the MW disk, respectively. If
compared to other more physically sounded prescriptions, a constant
$\chi_{MC}$ would give rise to a flatter "gradient in dust". We
expect however that  the local physical conditions affect the
accretion process and the fraction of MCs varies with time and
space. This strongly favors the second prescription for a varying
fraction of MCs.

Fig. \ref{GASeSFRMW}  shows  the radial dependence across the MW
Disk of  the current surface density profiles of $\mathrm{H}_{2}$,
$\mathrm{HI}$ and total gas (upper panel) of the MW and  the current
SFR normalized to the solar value
$\textrm{SFR}/\textrm{SFR}_{\odot}$ (lower panel)indicated by the
observational data. Superposed to the data we show some typical SF
laws \citet[See][for more details about these analytical
laws]{Piovan11b}. It is worth recalling that the total gas surface
density  includes also the contribution by helium
\citep{Boissier99,Yin09}. With the aid of this,  we look at the
correlations between $\textrm{SFR}/\textrm{SFR}_{\odot}$,
$\sigma^{G}\left(r_{k},t\right)$ (one of quantities followed by the
chemical model), and $\sigma_{H_{2}}\left(r_{k},t\right)$ (related
to the parameter $\chi_{MC}$).  \\

%%%%%%%%%%%%%%%%Figure 2
\begin{figure*}
\centerline{ \hspace{-30pt}
\includegraphics[height=6cm,width=6.5truecm]{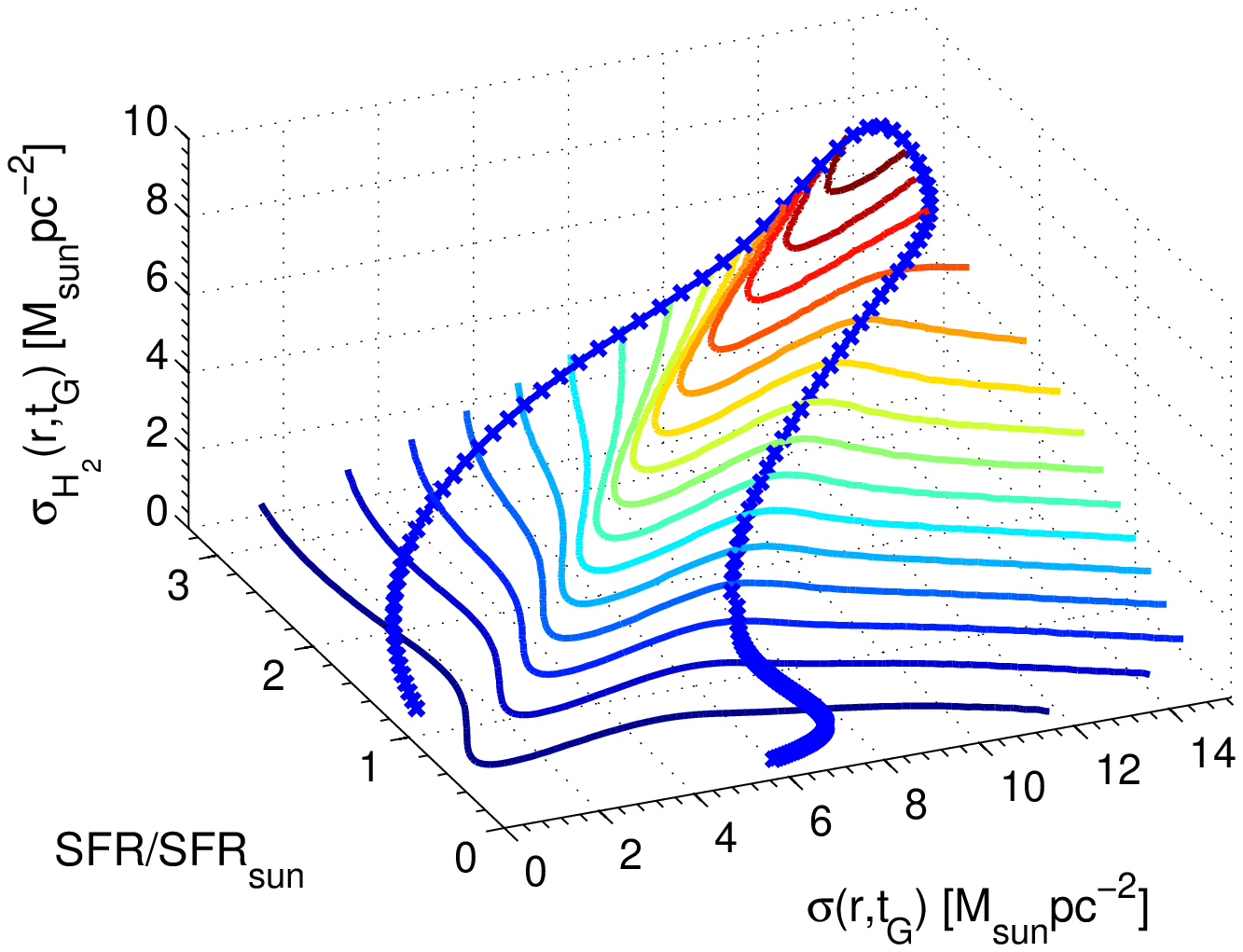}
\hspace{-23pt}
\includegraphics[height=6cm,width=6.5truecm]{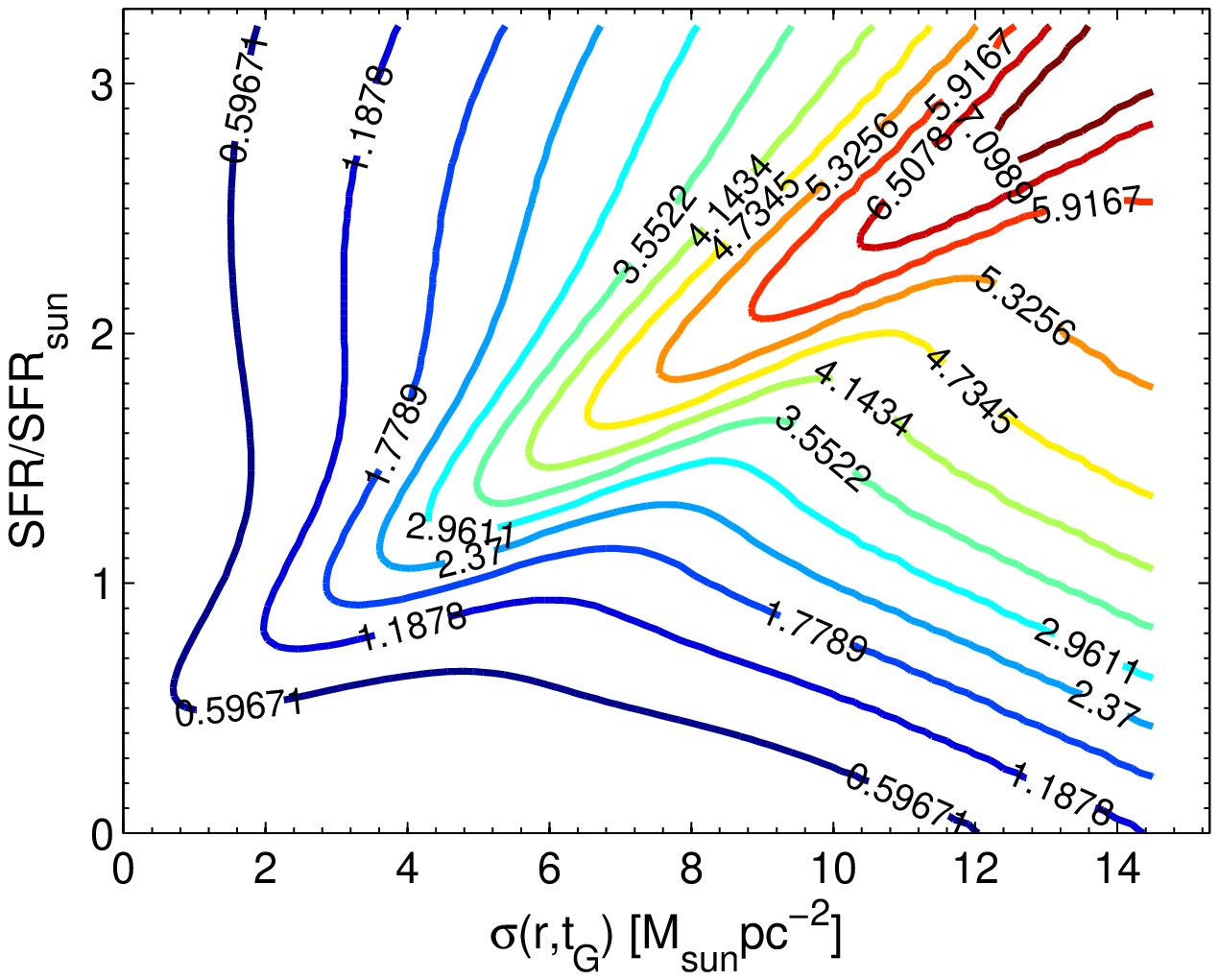}
\hspace{-23pt}
\includegraphics[height=6cm,width=6.5truecm]{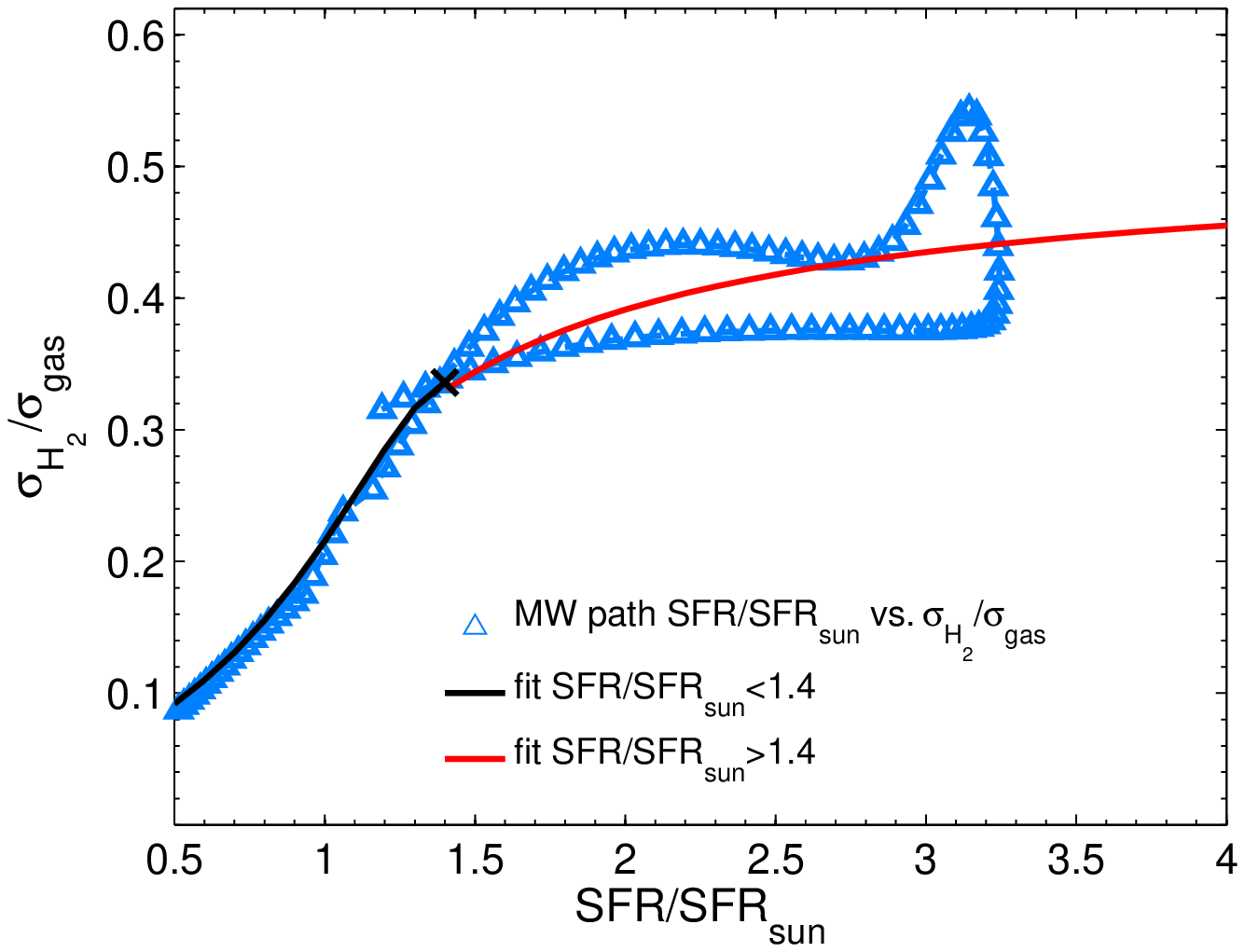}
\hspace{-30pt}} \caption{\textbf{Left Panel}: 3D plot of
$\sigma_{H_{2}}(r,t_{G})$ as a function of
$\textrm{SFR}/\textrm{SFR}_{\odot}$ and $\sigma^{G}(r,t_{G})$ for
the present-day radial profiles across the MW Disk  (crosses). The
solid line superposed to the data is the fit from the  ANNs. Contour
isolines obtained by extrapolating the ANNs weights to the regions
uncovered by the data are also shown. \textbf{Middle Panel}: The
same as in the Left Panel but showing only the 2D contour levels of
 $\sigma_{H_{2}}(r,t_{G})$ in the
$\textrm{SFR}/\textrm{SFR}_{\odot}$ vs. $\sigma^{G}(r,t_{G})$ plane.
\textbf{Right Panel}: 2D plot with the radial
$\textrm{SFR}/\textrm{SFR}_{\odot}$ vs
$\sigma_{H_{2}}(r,t_{G})/\sigma^{G}(r,t_{G})$ relationship for the
MW (triangles) and a two-piece fit of the data. The region of
interest is the one for SFR higher than solar.}\label{3Dand2DH2}
\end{figure*}

%%%%%%%%%%%%%%%%%%%%%%%Figure 3
\begin{figure*}
\centerline{
\includegraphics[height=6.5cm,width=8.0cm]{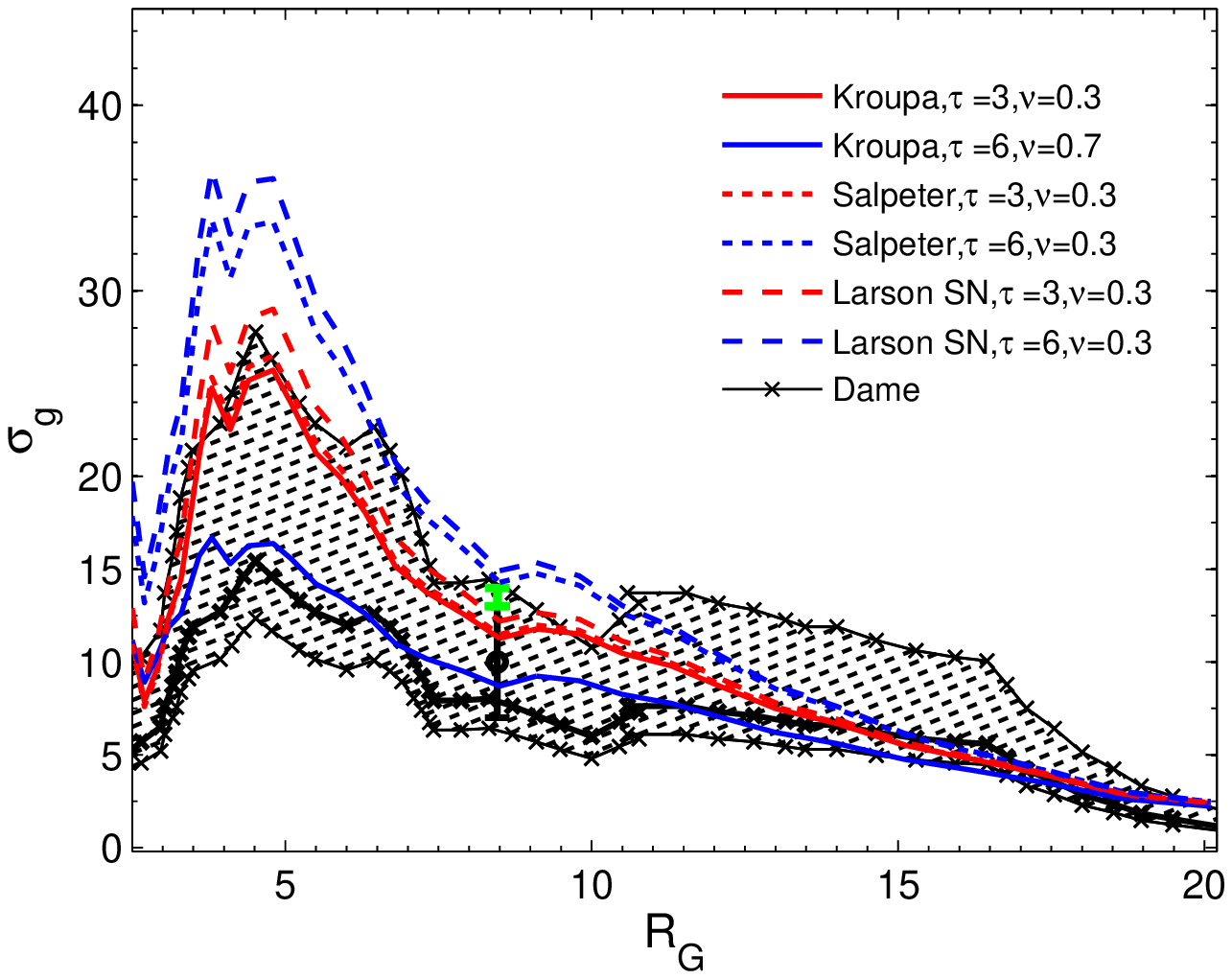}
\hspace{-20pt}
\includegraphics[height=6.5cm,width=8.0cm]{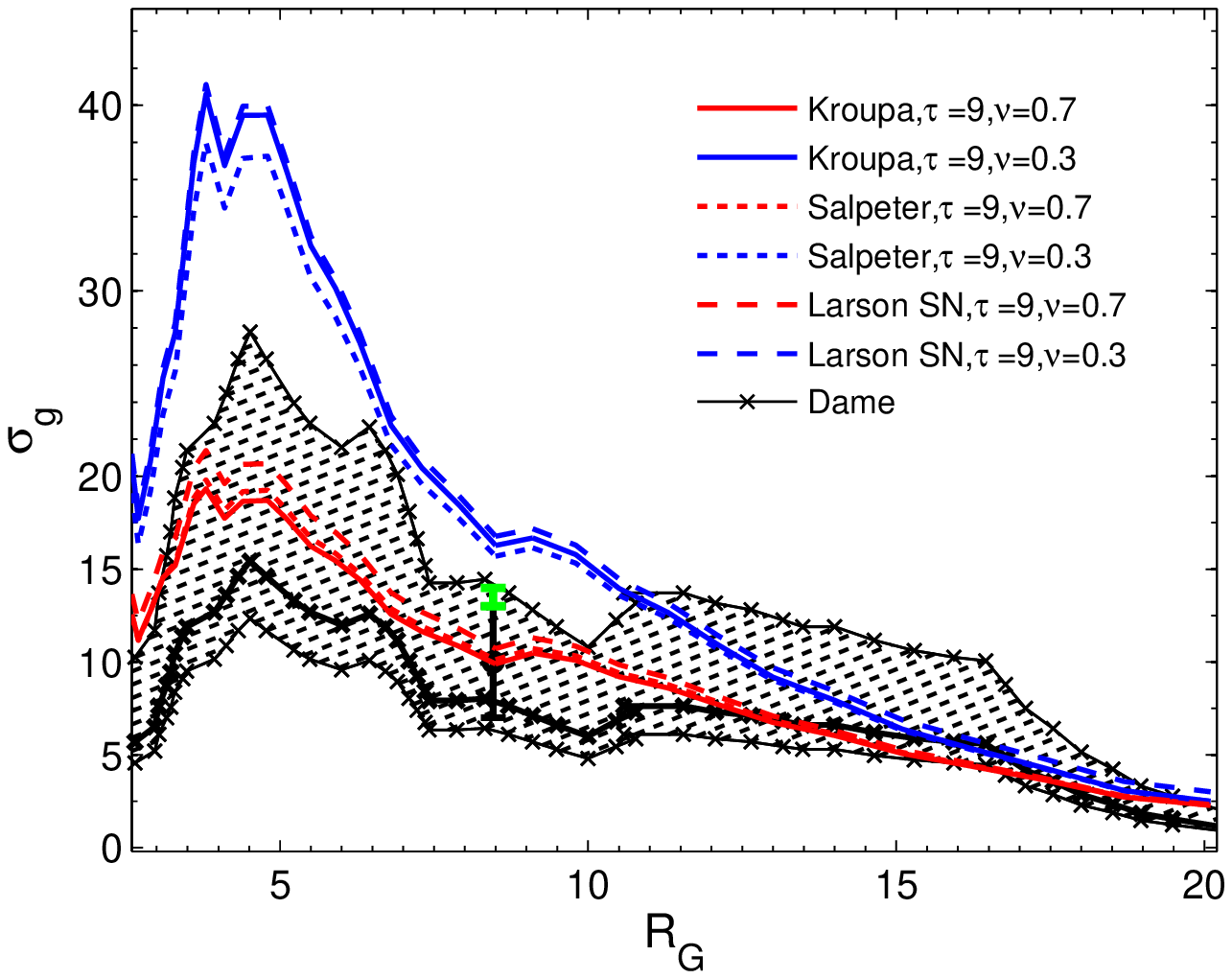}}
\caption{Radial surface density profile of the gas in the MW for
twelve models taken from \citet{Piovan11b}. The models differ for
the parameters:  IMF (Larson SoNe, Salpeter, Kroupa multi power
law), infall timescale $\tau $ (3, 6 and 9 Gyr),  the efficiency
$\nu$ (0.3 and 0.7). Not all the combinations have been represented,
but only the most significant ones. Data for the SoNe  are taken
from \citet{Dame93} and the hatched area represents the confidence
region taking into account the errors in the observational
estimates. Data for the SoNe gas mass density are from
\citet{Dickey93,Dame93} and \citet{Olling01}} \label{DensityProfile}
\end{figure*}

%%%%%%%%%%%%%%%%%%%%%%%Figure   4
\begin{figure*}
\centerline{\hspace{-30pt}
\includegraphics[height=6cm,width=5.8truecm]{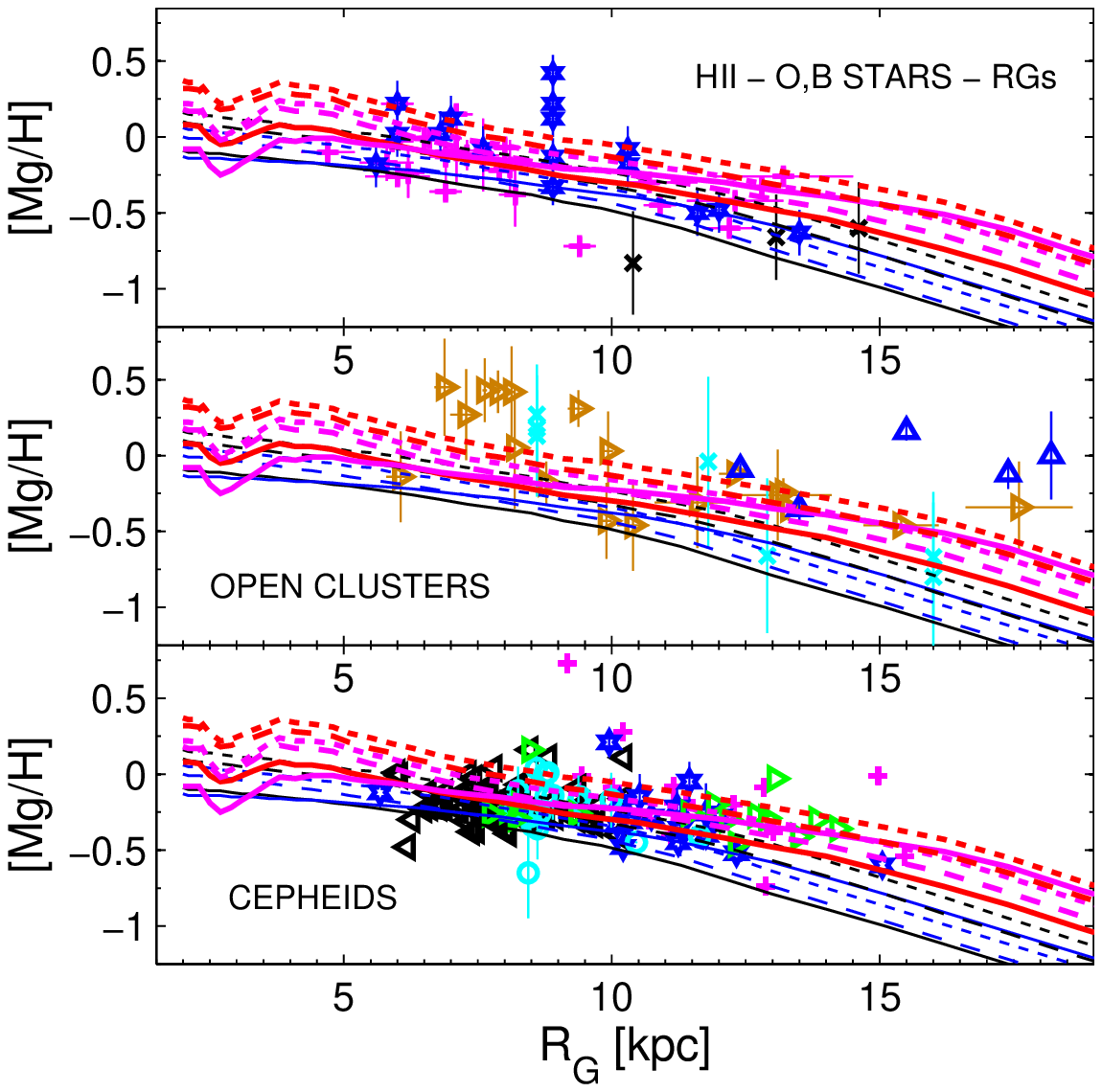}
\includegraphics[height=6cm,width=5.8truecm]{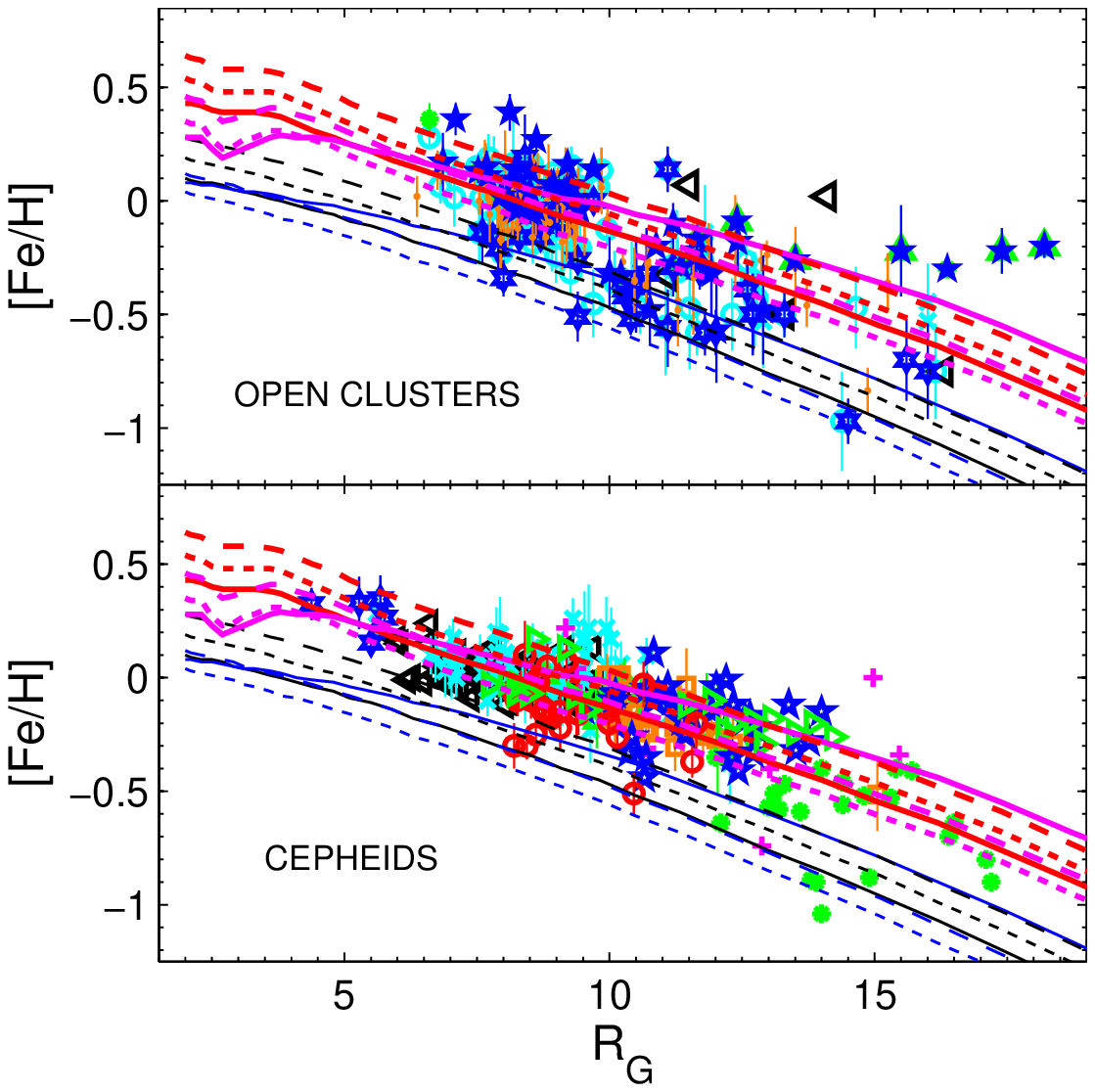}}
\centerline{\hspace{-20pt}
\includegraphics[height=6cm,width=5.8truecm]{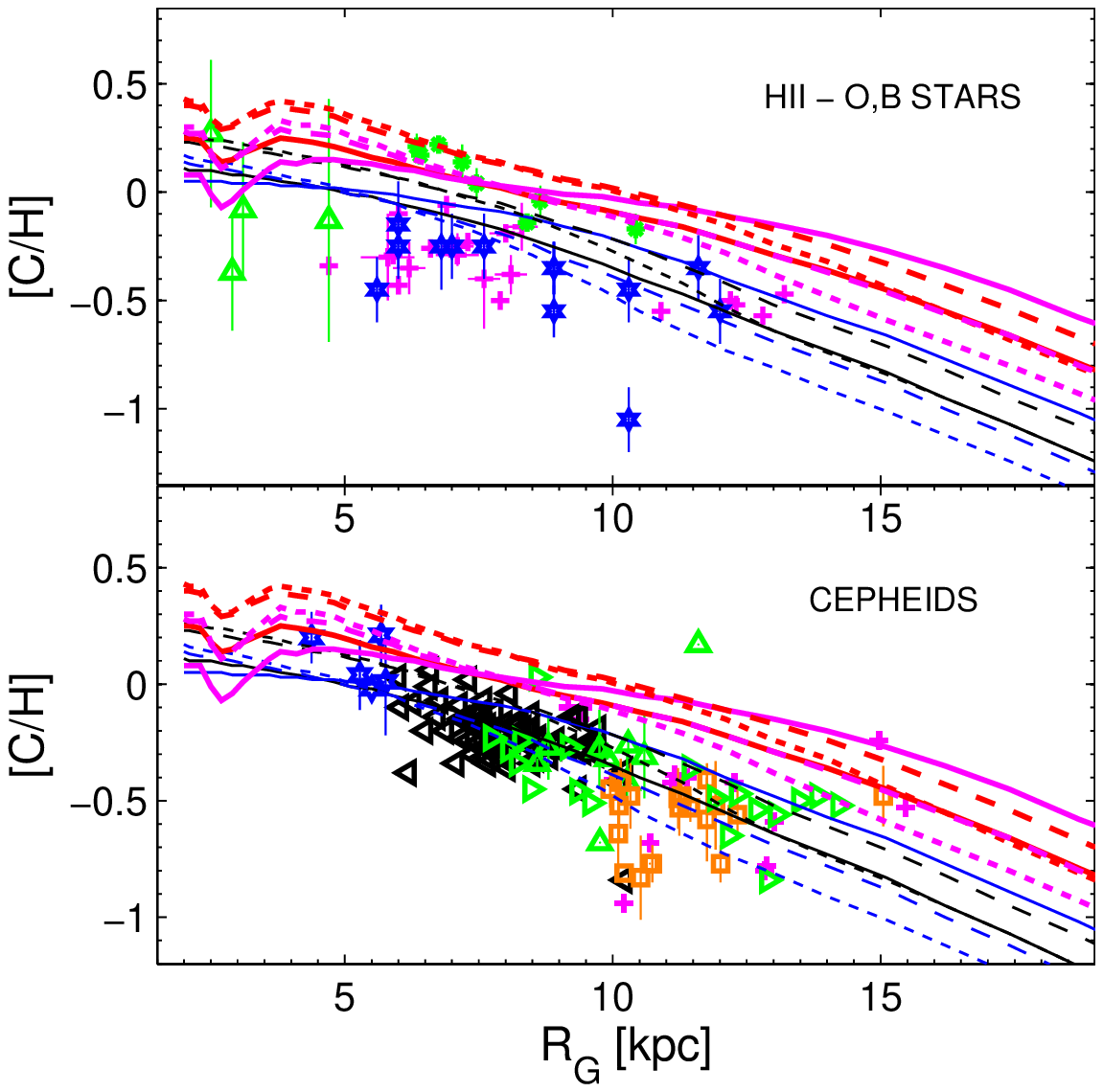}
\hspace{-2pt}
\includegraphics[height=6cm,width=5.8truecm]{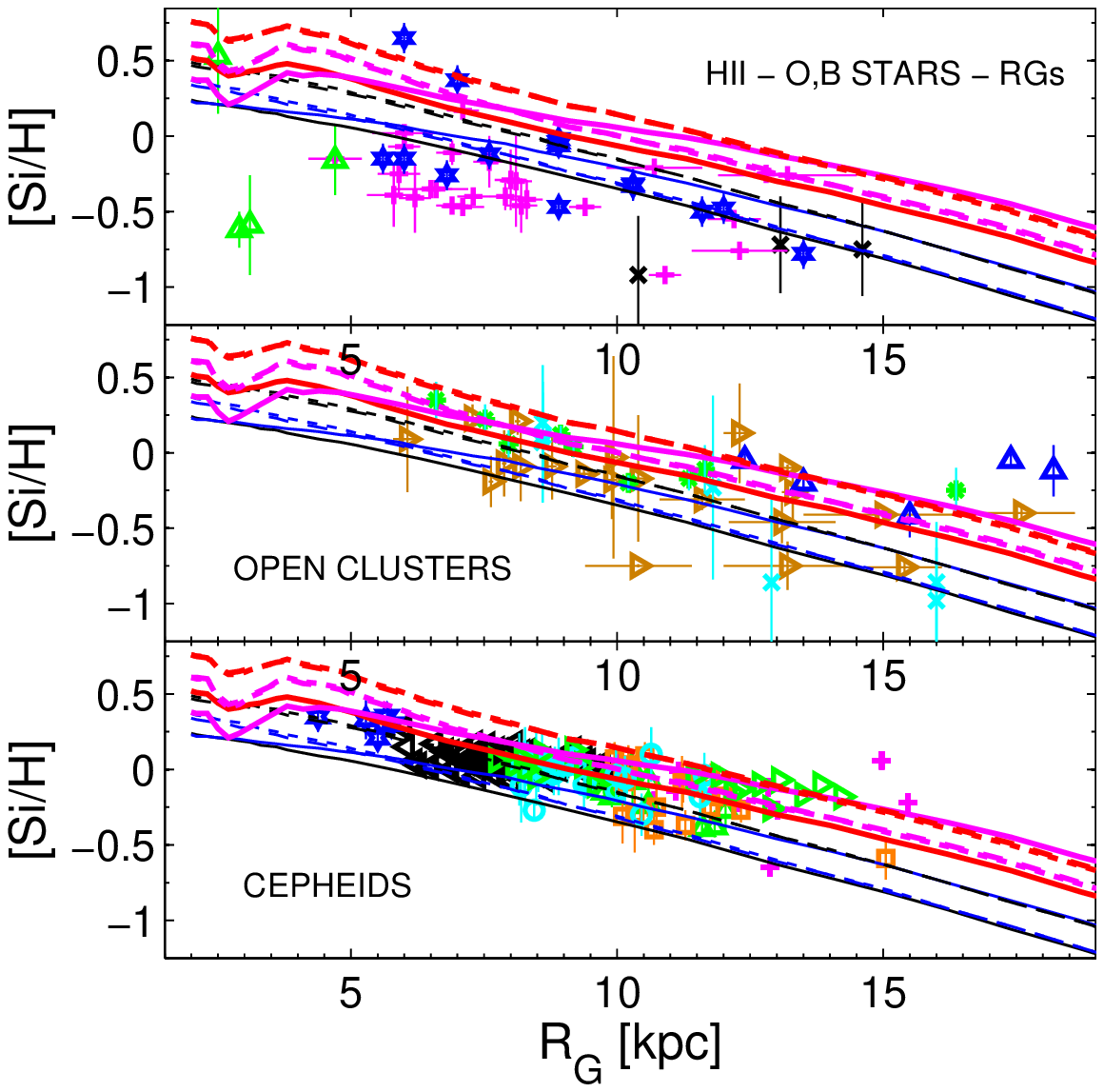}
\hspace{-2pt}
\includegraphics[height=6cm,width=5.8truecm]{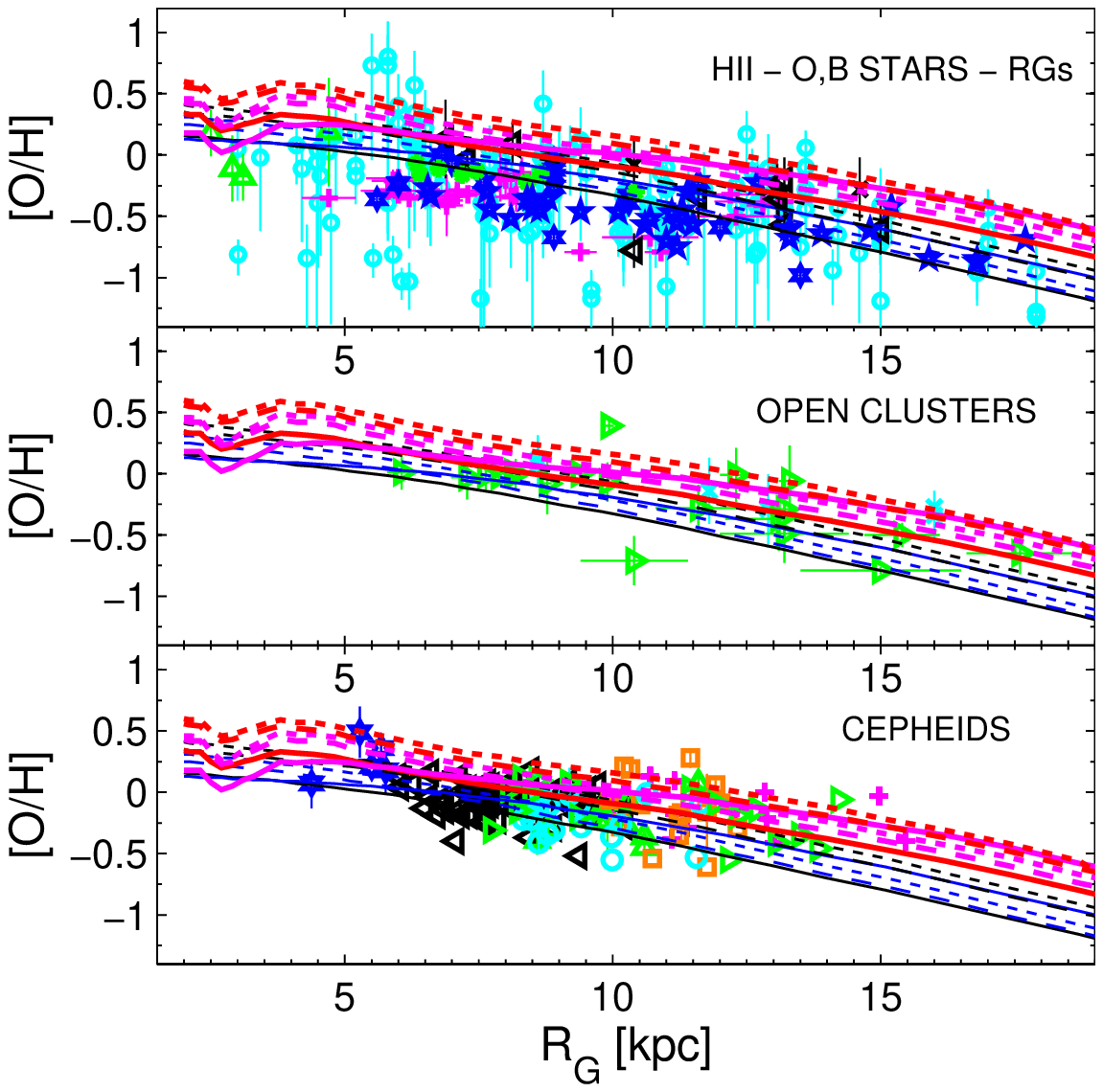}
\hspace{-12pt}}
\centerline{\hspace{-20pt}
\includegraphics[height=6cm,width=5.8truecm]{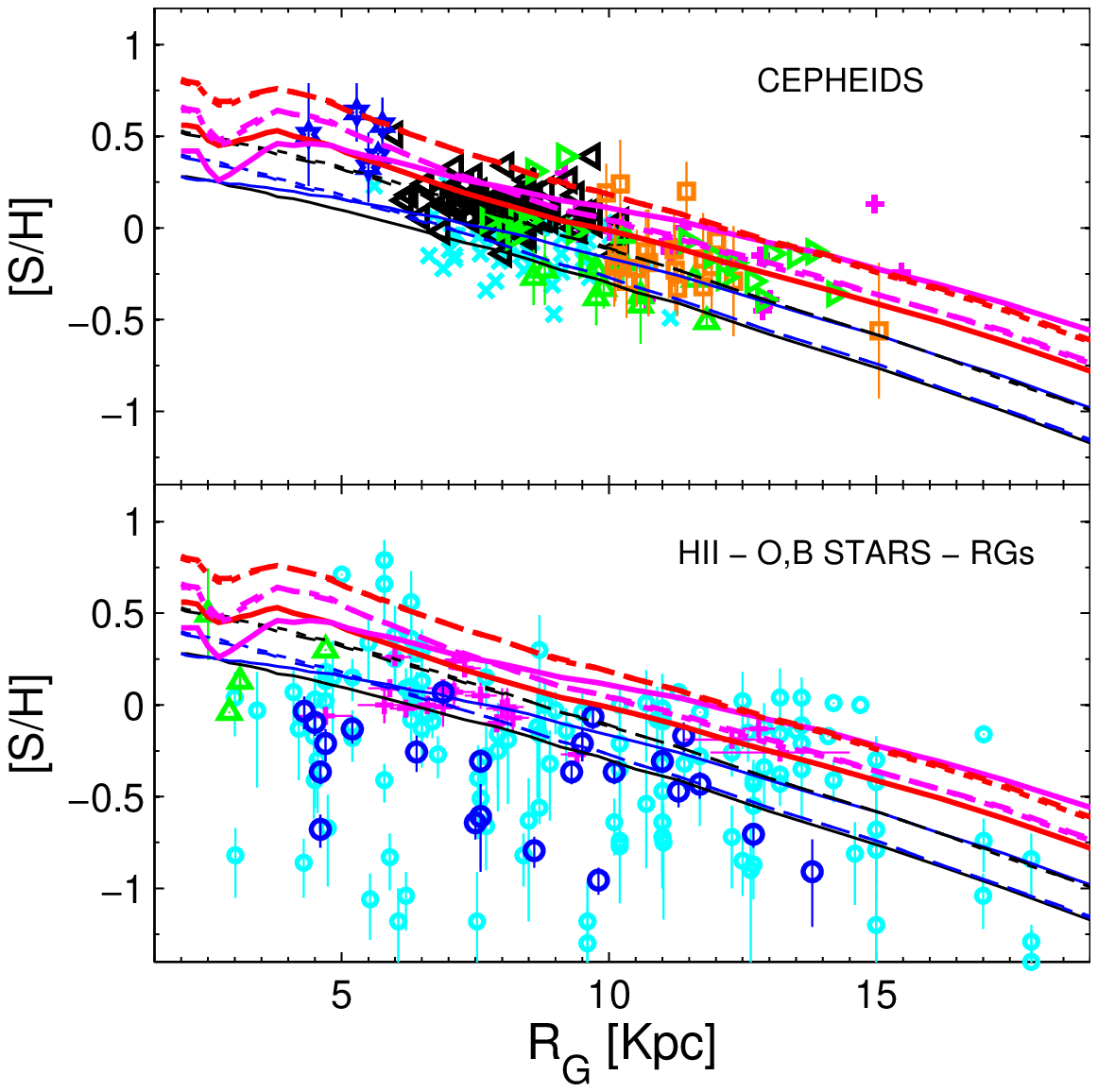}
\hspace{-2pt}
\includegraphics[height=6cm,width=5.8truecm]{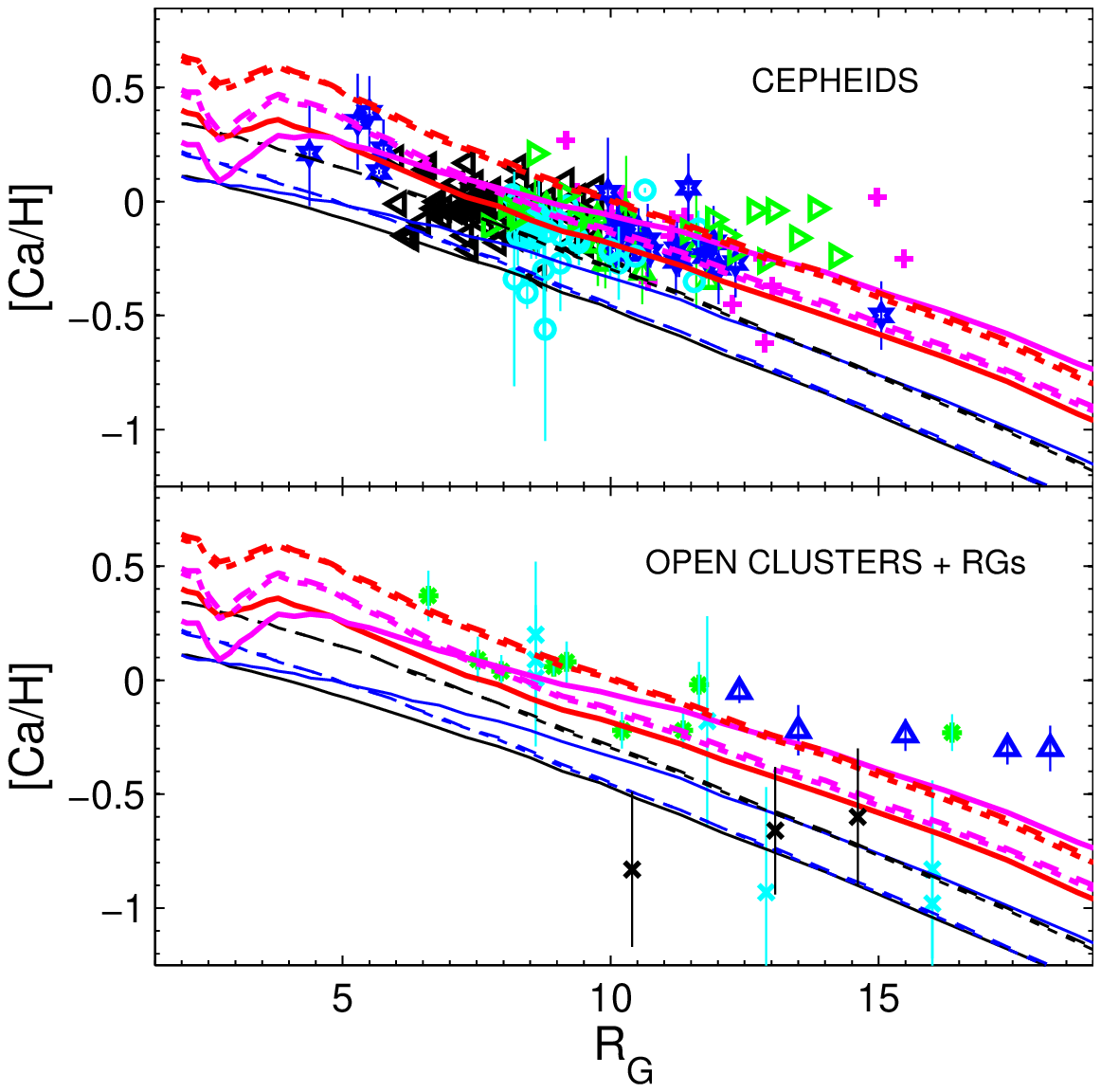}
\hspace{-2pt}
\includegraphics[height=6cm,width=5.8truecm]{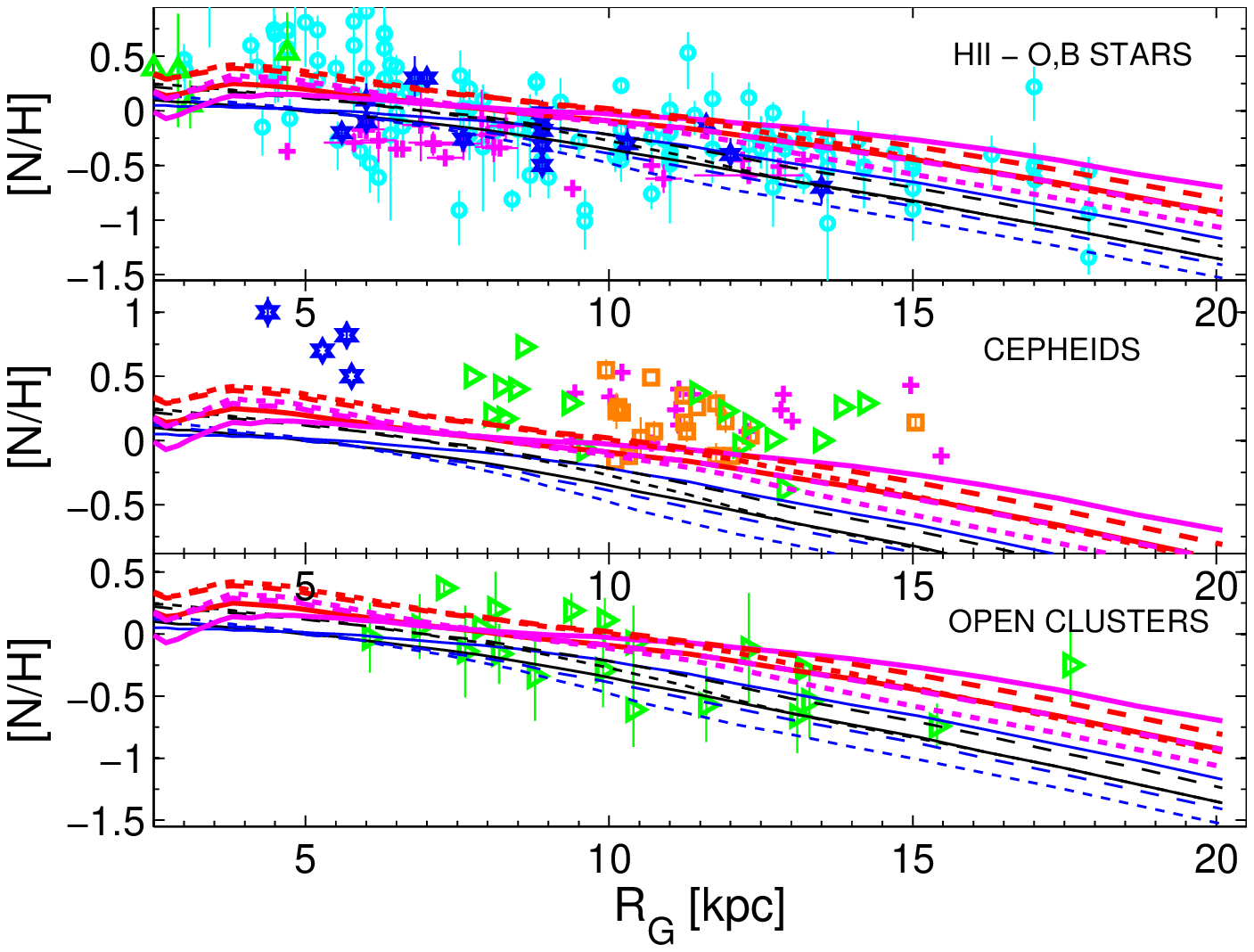}
\hspace{-12pt}} \caption{Evolution of the Galactic  abundances of
$[\textrm{Mg}/\textrm{H}]$, $[\textrm{Fe}/\textrm{H}]$,
$[\textrm{C}/\textrm{H}]$, $[\textrm{Si}/\textrm{H}]$,
$[\textrm{O}/\textrm{H}]$, $[\textrm{S}/\textrm{H}]$,
$[\textrm{Ca}/\textrm{H}]$ and $[\textrm{N}/\textrm{H}]$ as a
function of the galacto-centric distance for different types of
abundance indicators. In all the diagrams we present the evolution
of the radial gradient from the age at which the Sun was formed
(thin lines) to the current age (thick lines). The elemental
abundances are measured in the following sources: O and B stars,
Field Red Giants (RGs), HII regions, open clusters and Cepheid
variables. The sources have been divided into two or three
groups/panels, depending on the available ones for that element, in
such a way to examine the gradients from different sources. The
following models are considered: Kroupa IMF, $\tau =3$ and $\nu
=0.3$ (continuous thin black and thick red lines); Kroupa IMF, $\tau
=6$ and $\nu =0.7$ (continuous thin blue and thick magenta lines);
Salpeter IMF, $\tau =3$ and $\nu =0.3$ (dotted thin black and thick
red lines); Salpeter IMF, $\tau =6$ and $\nu =0.3$ (dotted thin blue
and thick magenta lines); Larson SoNe IMF, $\tau =3$ and $\nu =0.3$
(dashed thin black and thick red lines), and finally Larson SoNe
IMF, $\tau =6$ and $\nu =0.3$ (dashed thin blue and thick magenta
lines).} \label{EvolRadialAbunGAS}
\end{figure*}

\footnotetext[7]{\textbf{Magnesium}. O and B stars:
\citet{Gummersbach98} (Gu98), \citet{Smartt01} (Sm01),
\citet{Daflon04} (Da04); Field Red Giants: \citet{Carney05} (Ca05);
Open Clusters: \citet{Rolleston00} (Ro00), \citet{Carraro04} (Ca04),
\citet{Carraro07} (Ca05), \citet{Yong05} (Yo05); Cepheid stars:
\citet{Andrievsky02a} (An02a), \citet{Andrievsky02b} (An02b),
\citet{Andrievsky02c} (An02c), \citet{Andrievsky02d} (An02d),
\citet{Luck03} (Lu03), \citet{Kovtyukh05} (Ko05), \citet{Luck06}
(Lu06), \citet{Yong06} (Yo06), \citet{Lemasle07} (Le07).
\textbf{Iron}. Open Clusters: \citet{Twarog97} (Tw97),
\citet{Carraro98b} (Ca98), \citet{Hou02} (Ho02), \citet{Chen03}
(Ch03), Ca04, Yo05, Ca07, \citet{Sestito08} (Se08),
\citet{Magrini09} (Ma09); Cepheid stars: An02a, An02b, An02c, An02d,
Lu03, Ko05, Lu06, Yo06, Le07 and \citet{Lemasle08} (Le08).
\textbf{Carbon}. O and B stars (Gu98, Sm01, Da04); HII regions:
\citet{Esteban05} (Es05); Cepheid stars: An02a, An02b, An02c, An02d,
Lu03, Ko05 and \citet{Andrievsky04} (An04). \textbf{Silicon}. O and
B stars Gu98, Sm01, Da04; Field Red Giants: Ca05; Open Clusters:
Ro00, Ca04, Ca07, Yo05, Se08; Cepheid stars: An02a, An02b, An02c,
An02d, Lu03, An04, Ko05, Lu06, Yo06, Le07. \textbf{Sulfur}. Cepheid
stars: An02a, An02b, An02c, An02d, Lu03, An04, Ko05, Lu06; O and B
stars: Sm01, Da04; HII regions \citet{MartinHernandez02} (Ma02),
\citet{Rudolph07} (Ru07); Open Clusters: Ca04, Ca07, Yo05, Se08;
Field Red Giants: Ca05; Cepheid stars: An02a, An02b, An02c, An02d,
Lu03, An04, Ko05, Lu06, Yo06, Le07.  \textbf{Calcium}. Open
Clusters: Ca04, Ca07, Yo05, Se08; Field Red Giants: Ca05; Cepheid
stars: An02a, An02b, An02c, An02d, Lu03, An04, Ko05, Lu06, Yo06,
Le07. \textbf{Oxygen}: HII regions \citet{Deharveng00} (De00), Es05,
Ru07; O and B stars: \citet{Smartt97} (Sm97), Da04, Sm01; Field Red
Giants: Ca05; Open Clusters: Ro00, Ca04, Yo05; Cepheid stars: An02a,
An02b, An02c, An02d, Lu03, An04, Ko05, Le07. \textbf{Nitrogen}: HII regions:
Ru07; O and B stars: Gu98, Da04, Sm01; Open Clusters: Ro00; Cepheid stars:
An02b, An02c, Lu03, Ko05.}

%%%%%%%%%%%%%%%%%%Figure 5
\begin{figure*}
\centerline{\hspace{-30pt} \hspace{+30pt}
\includegraphics[height=6cm,width=6.0truecm]{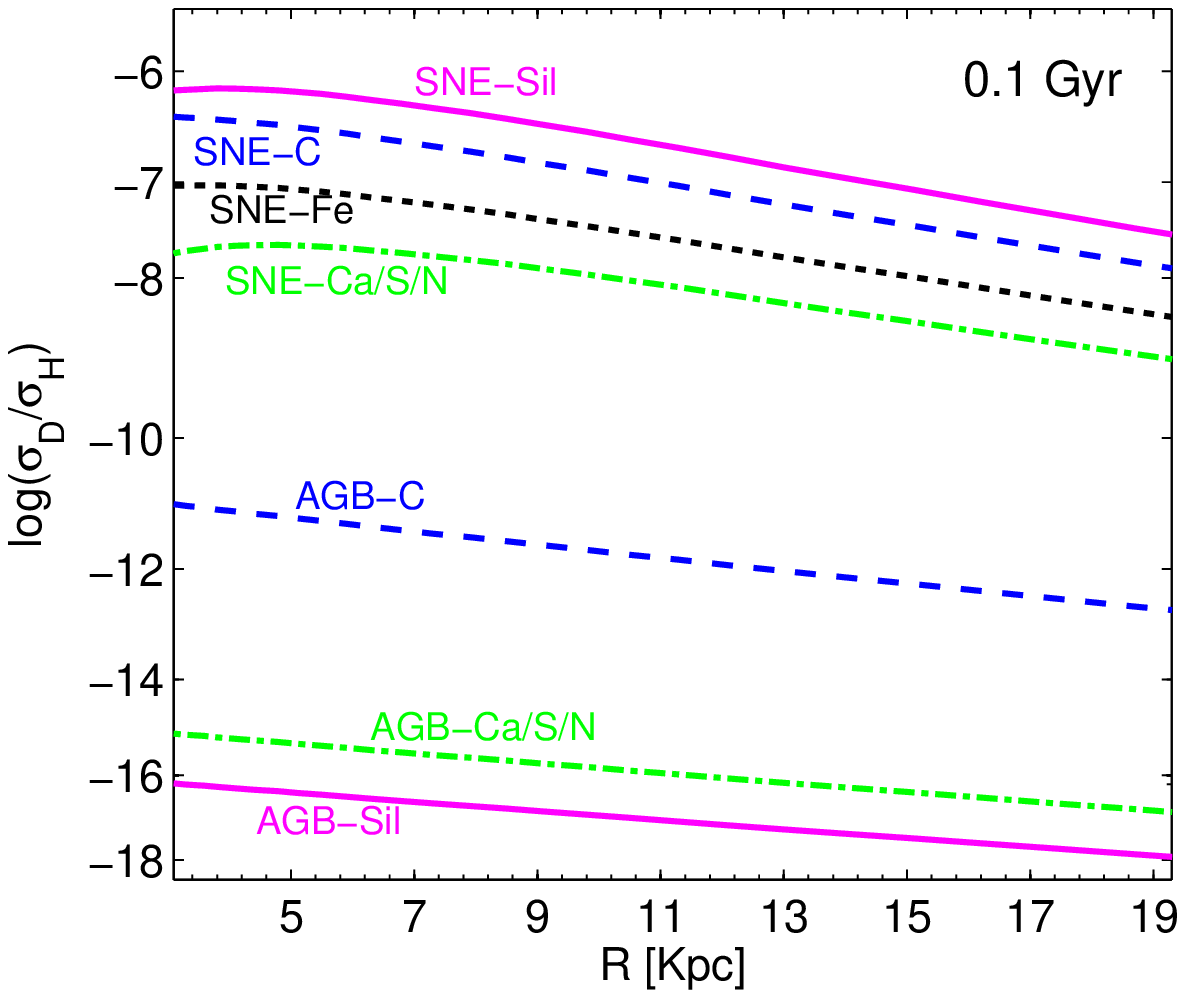}
\includegraphics[height=6cm,width=6.0truecm]{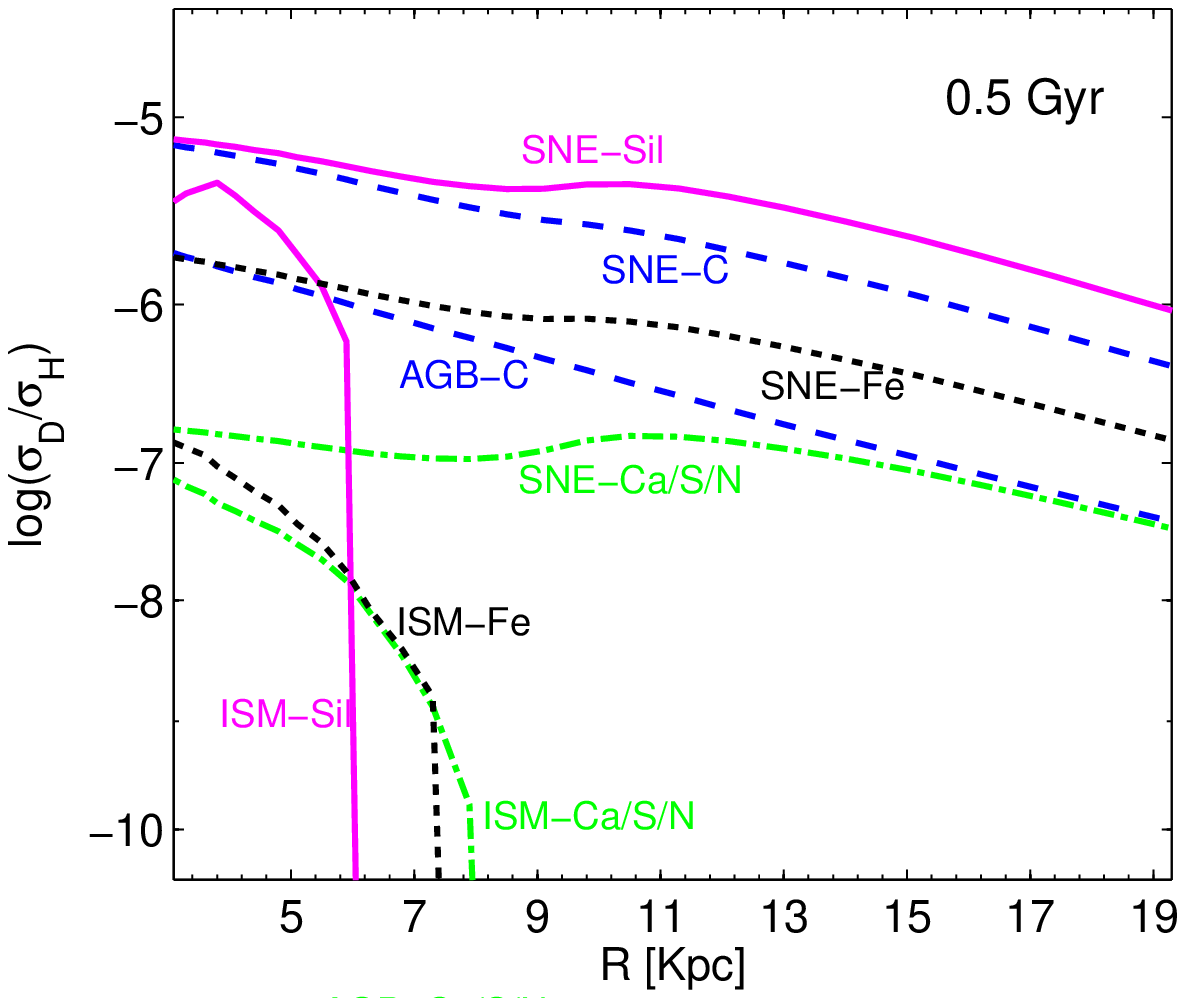}
\includegraphics[height=6cm,width=6.0truecm]{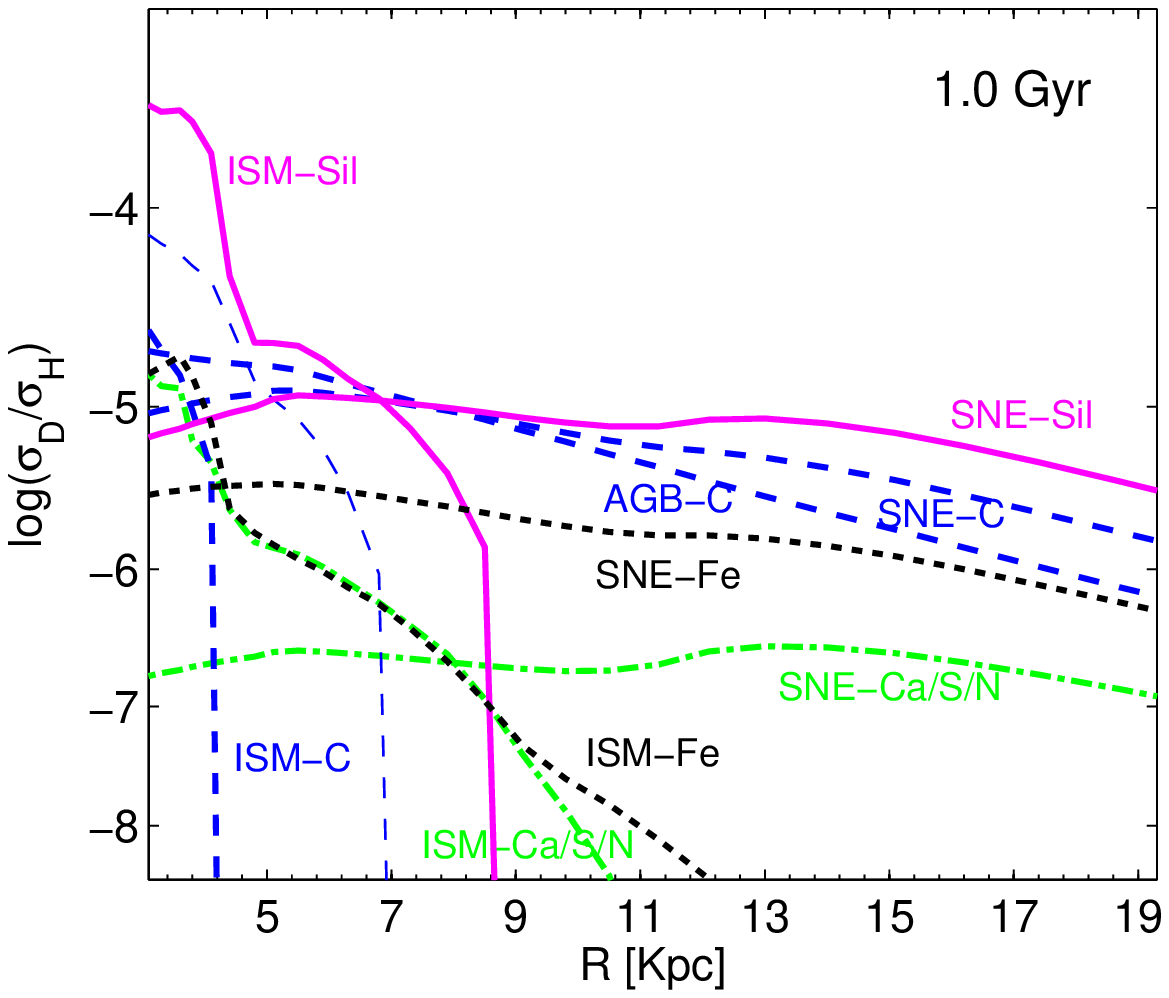}}
\centerline{\hspace{-12pt}
\includegraphics[height=6cm,width=6.5truecm]{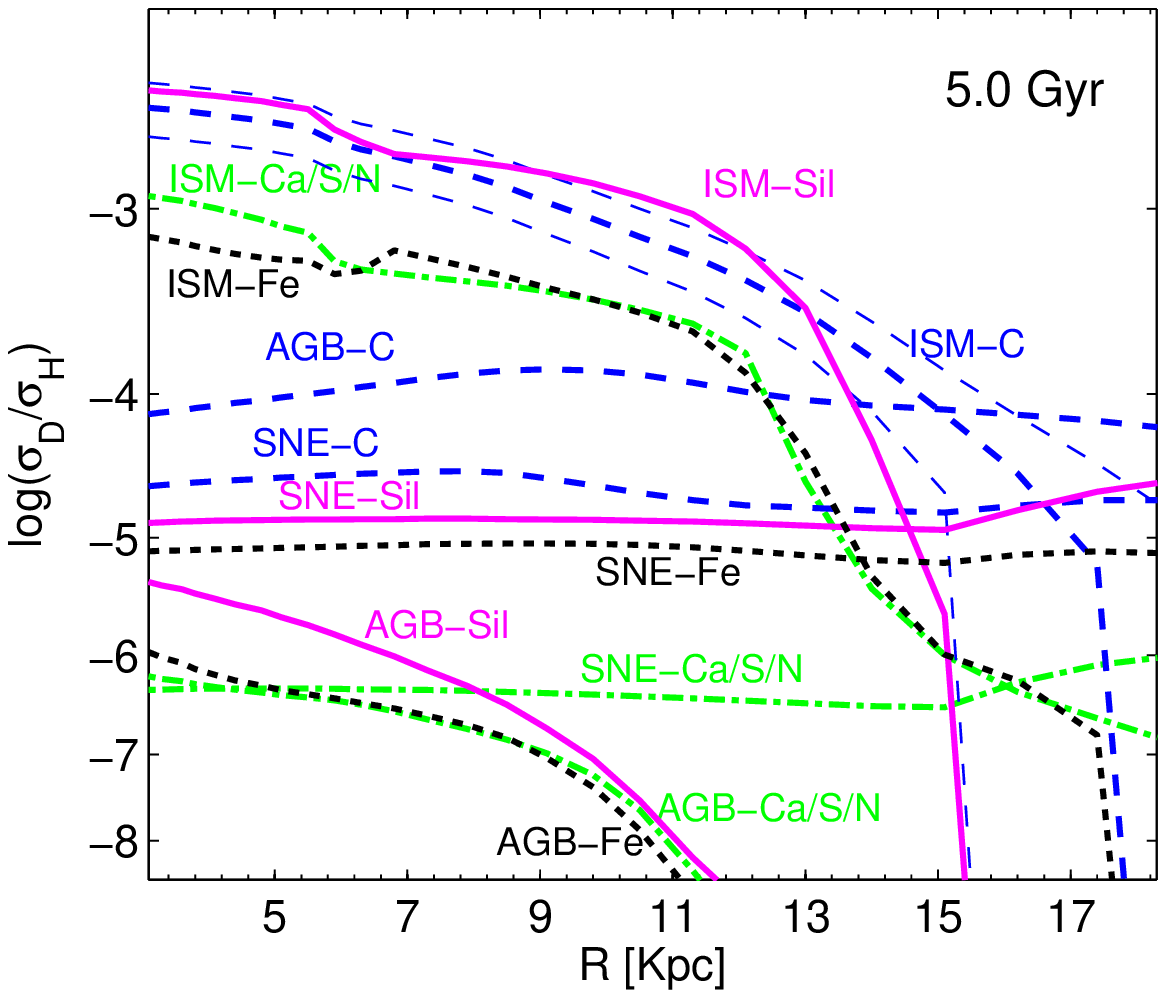}
\includegraphics[height=6cm,width=6.5truecm]{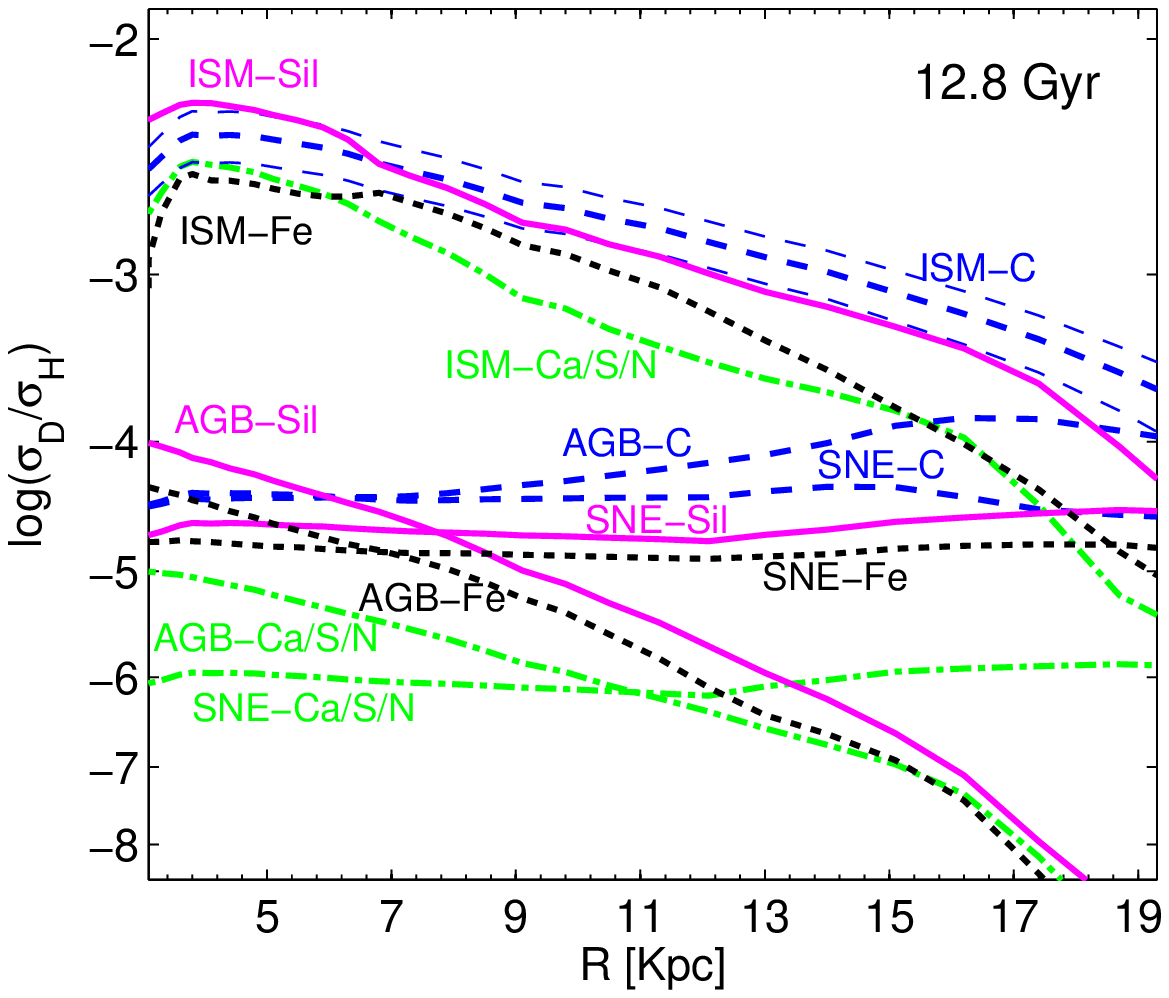}
\hspace{-12pt}} \caption{Temporal evolution of the radial
contribution to the abundance of dust by the four types of grain
among  which we distributed the single elements (that is silicates,
carbonaceous grains, iron dust grains and generic grains embedding
Ca, N or S - see \citet{Zhukovska08} and \citet{Piovan11b} for more
details) and the three sources of dust, namely  SN{\ae}, AGB stars
and accretion in the ISM. All the contributions have been properly
corrected for the destruction of dust. Five ages are represented
from the early stages to the current time. \textbf{Upper left
panel}: the radial contribution of the four types of dust grains to
the total dust budget at 0.1 Gyr. We show: silicates (continuous
lines), carbonaceous grains (dashed lines), iron dust (dotted lines)
and, finally, generic grains containing  $\mathrm{S}$, $\mathrm{Ca}$
and $\mathrm{N}$ (dot-dashed lines). For each group we distinguish
the net contributions from the ISM accretion, AGB and SN{\ae}, that
is: ISM-C, AGB-C and SN{\ae}-C for carbonaceous grains, ISM-Sil,
AGB-Sil and SN{\ae}-Sil for silicates, ISM-Fe, AGB-Fe and SN{\ae}-Fe
for the iron dust and finally, ISM-Ca/S/N, AGB-Ca/S/N and
SN{\ae}-Ca/S/N for the other grains. In all cases the abundance of
CO is fixed to $\xi_{\mathrm{CO}}=0.30$ taken as the reference
value. For the last two ages we also show for the ISM the results
for $\xi_{\mathrm{CO}}=0.15$ and $\xi_{\mathrm{CO}}=0.45$.
\textbf{Upper Central panel}: the same as in the upper left panel
but for 0.5 Gyr. \textbf{Upper Right panel}: the same as in the
upper left panel but for 1.0 Gyr. \textbf{Lower Left panel}: the
same as in the upper left panel but for 5.0 Gyr. \textbf{Lower Right
panel}: the same as in the upper left panel but for 12.8 Gyr. Since
the ratio $\sigma_{D}/\sigma_{H}$ grows at varying  age, the scale
of the y-axis is continuously shifted according to the value of the
plotted data.} \label{EvolRadialAbunGrains}
\end{figure*}

%%%%%%%%%%%%%%%%%%%%%%%%%%Figure 6
\begin{figure*}
\centerline{\hspace{-30pt}
\includegraphics[height=6.2cm,width=6.5truecm]{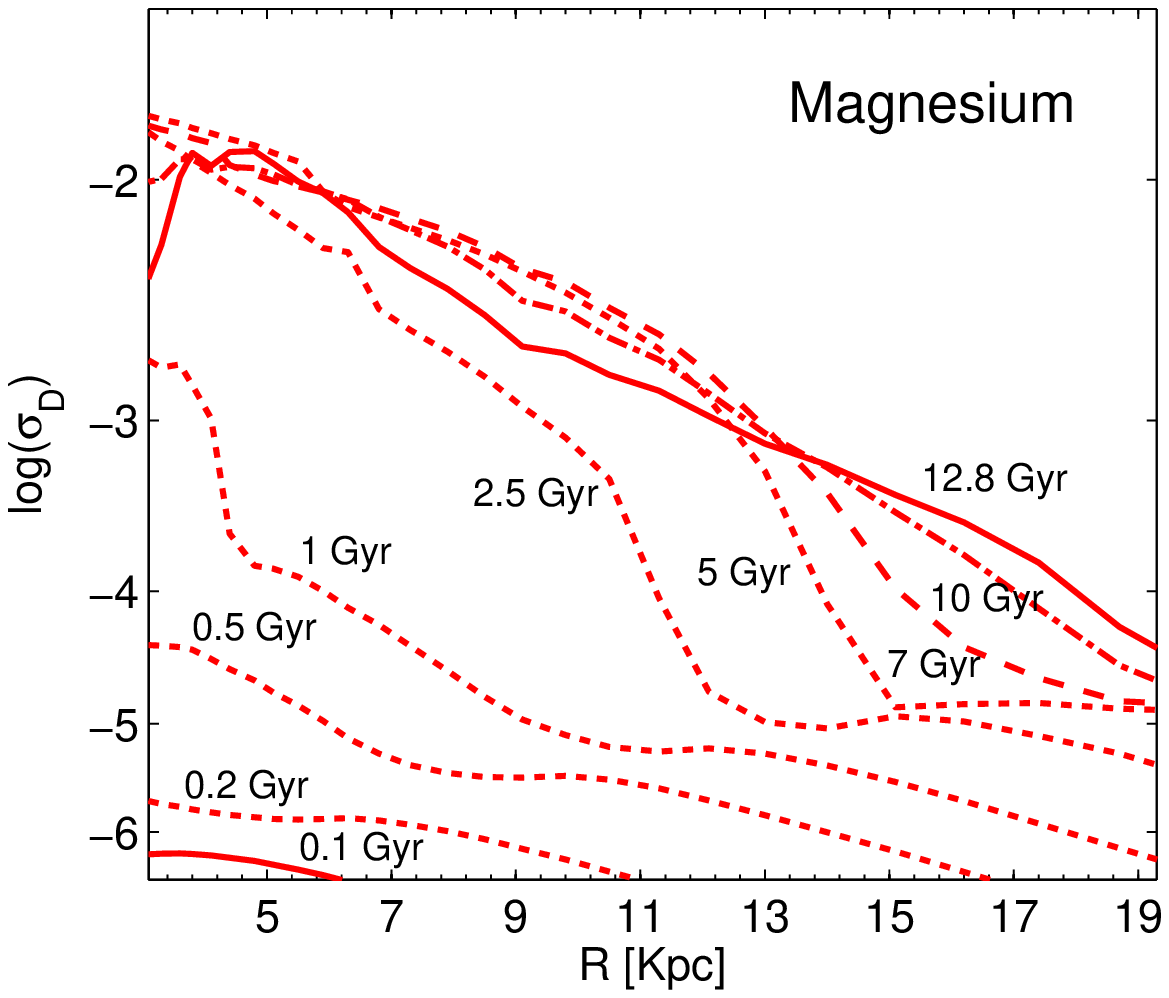}
\includegraphics[height=6.2cm,width=6.5truecm]{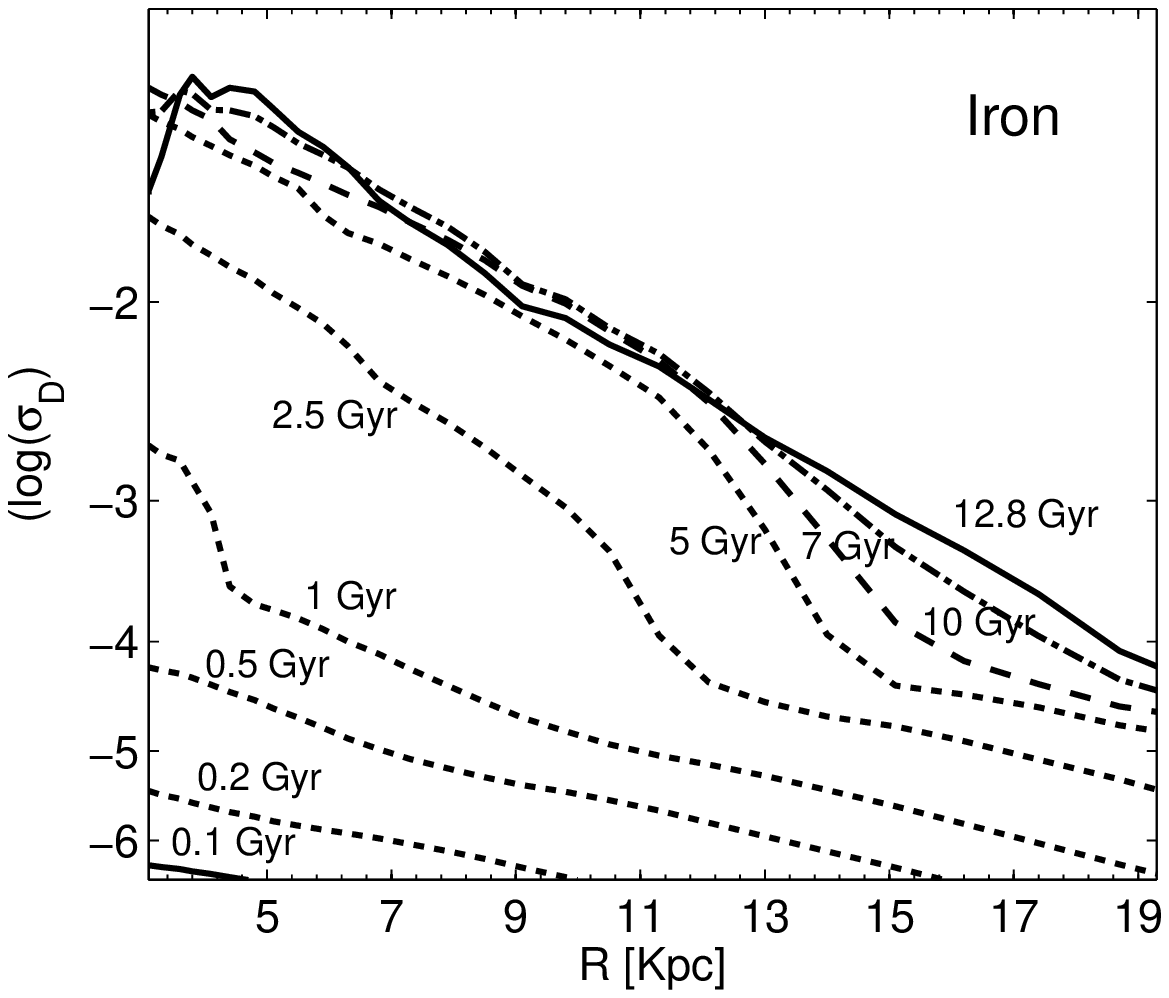}}
\centerline{\hspace{-12pt}
\includegraphics[height=6.2cm,width=6.5truecm]{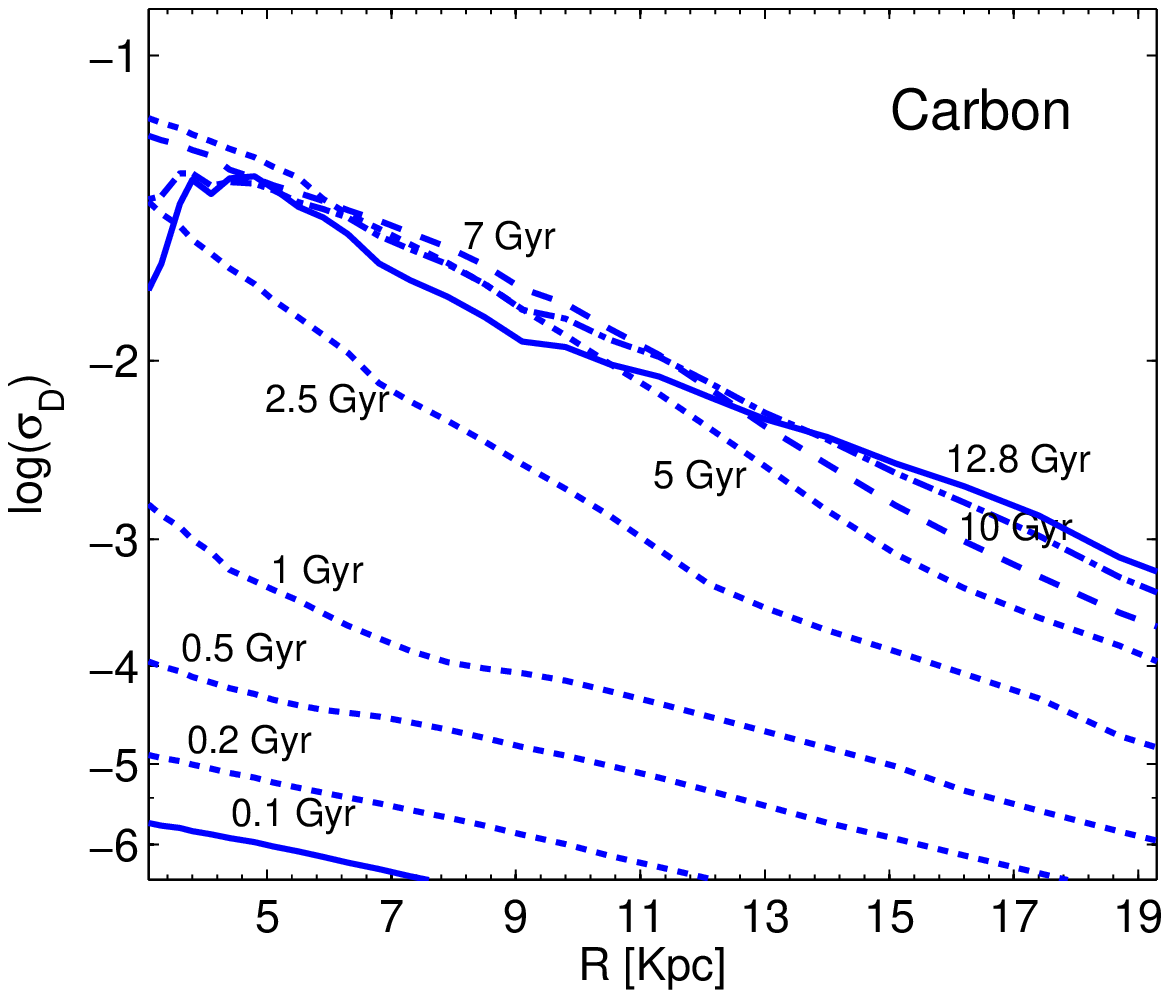}
\hspace{-15pt}
\includegraphics[height=6.2cm,width=6.5truecm]{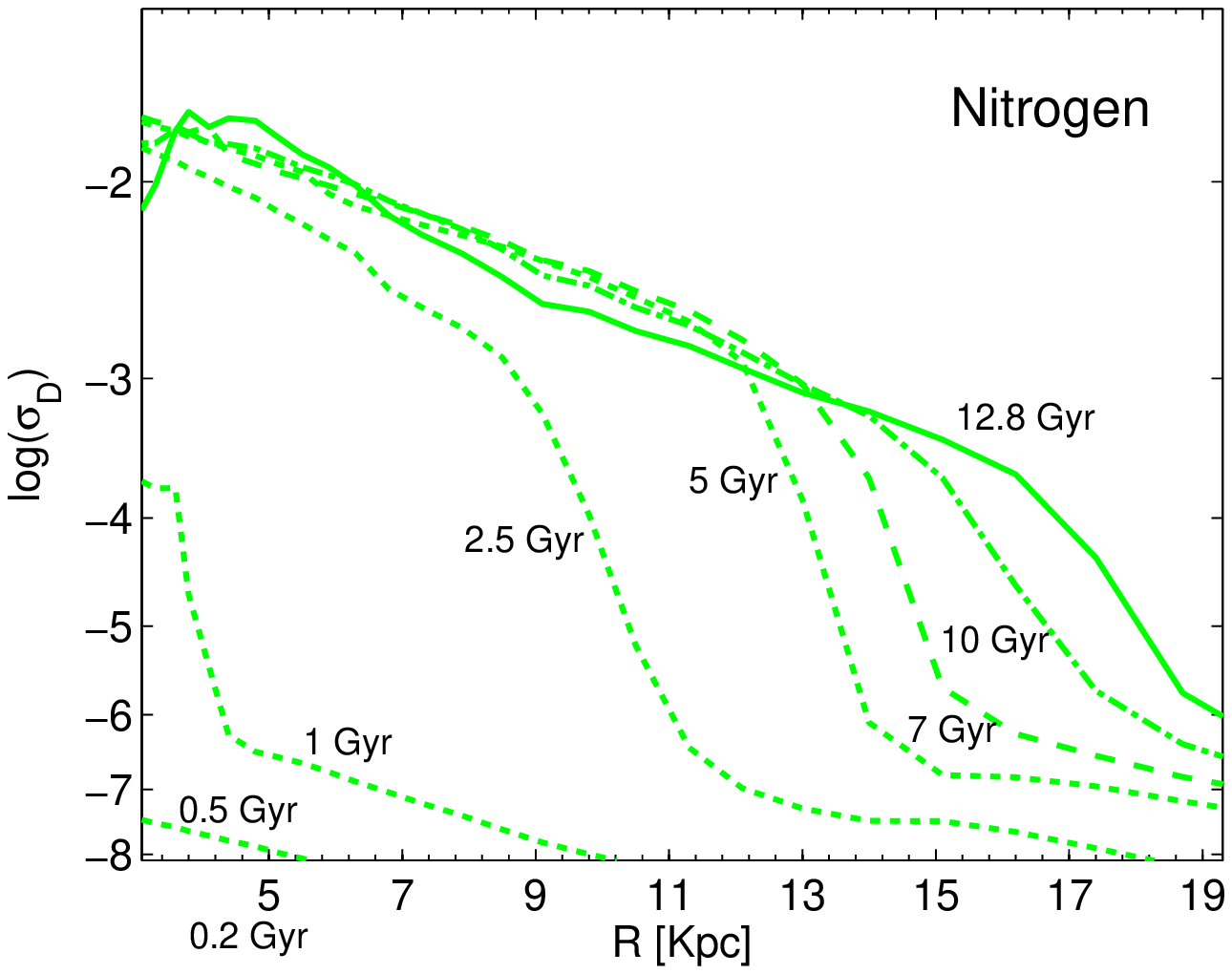}
\hspace{-15pt}
\includegraphics[height=6.2cm,width=6.5truecm]{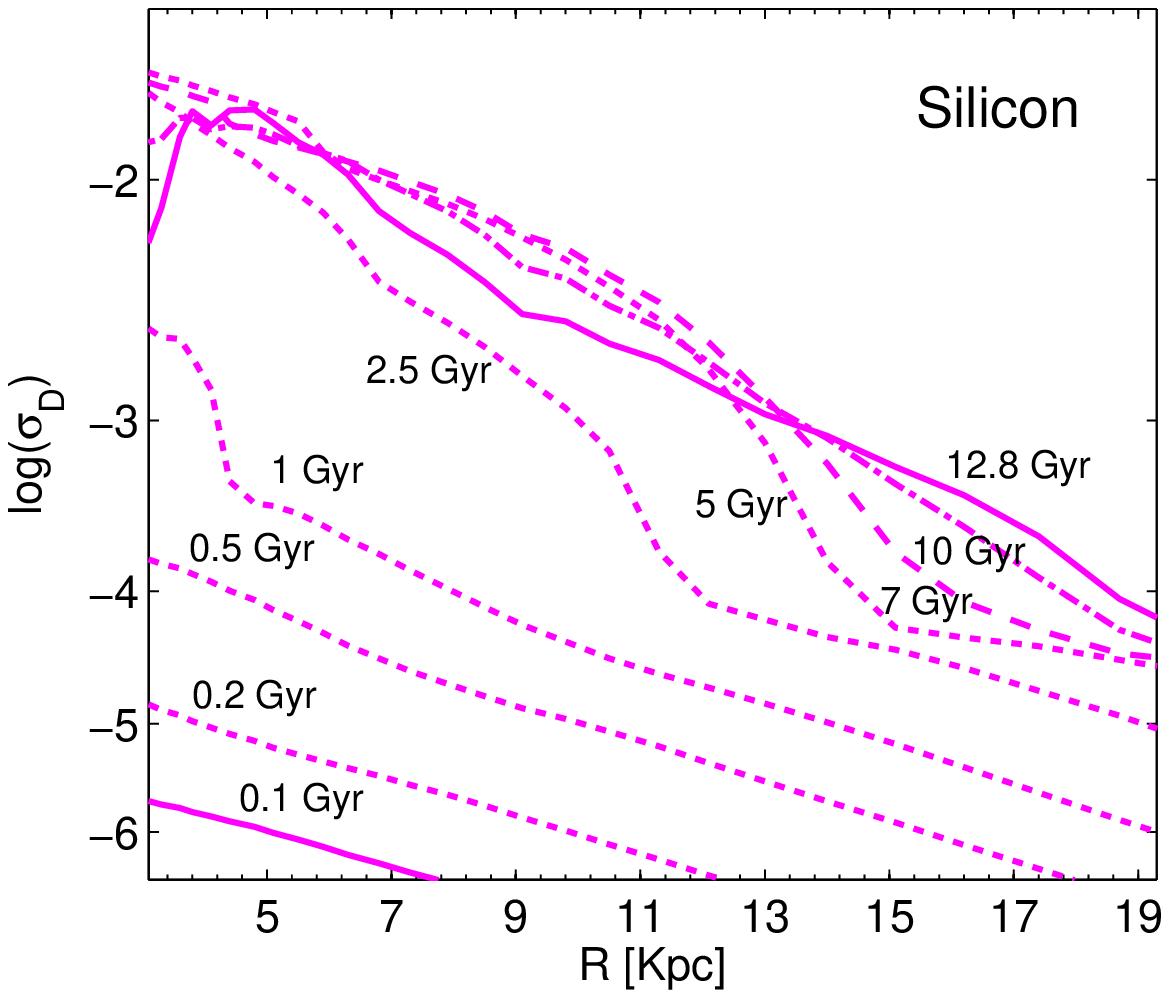}}
\centerline{\hspace{-12pt}
\includegraphics[height=6.2cm,width=6.4truecm]{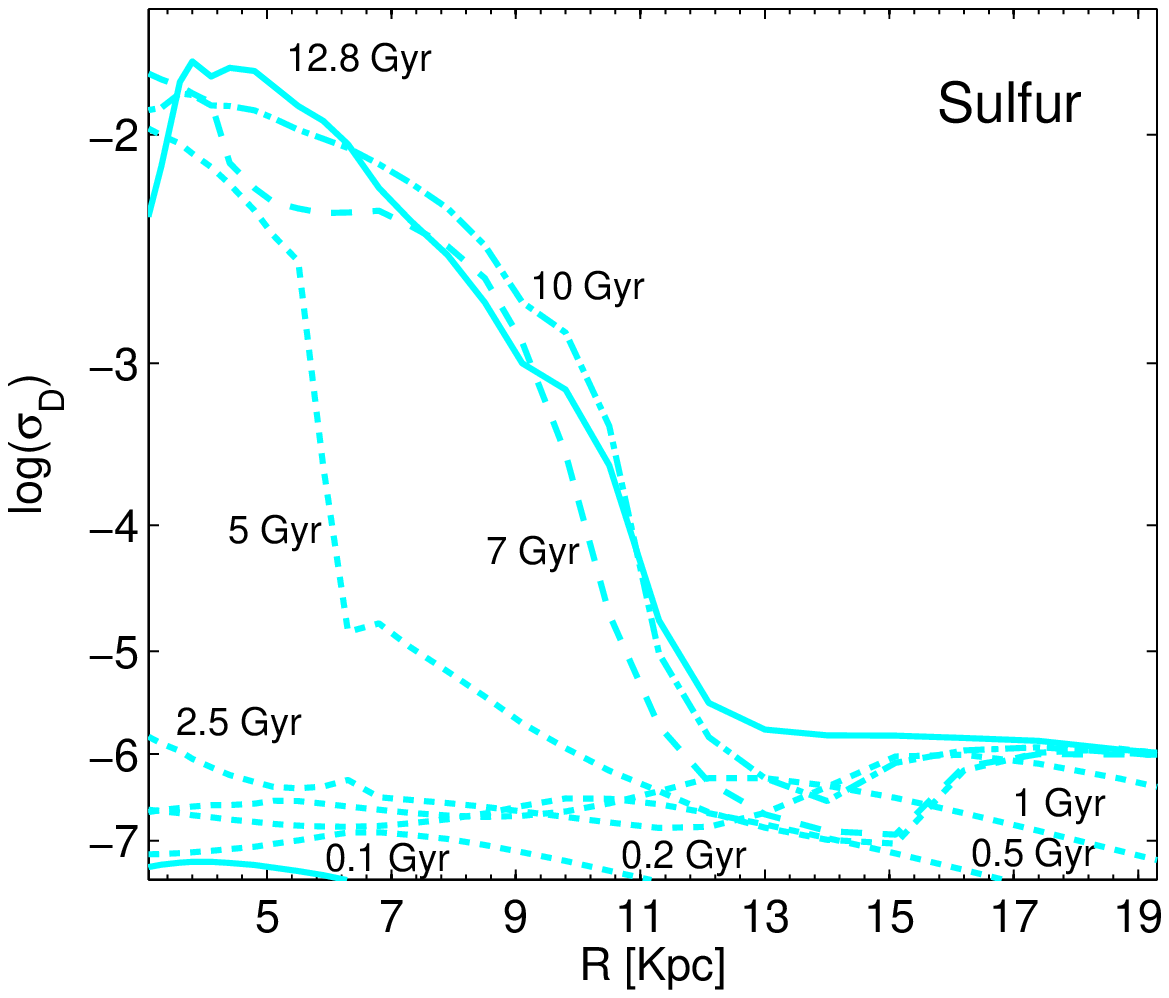}
\hspace{-15pt}
\includegraphics[height=6.2cm,width=6.4truecm]{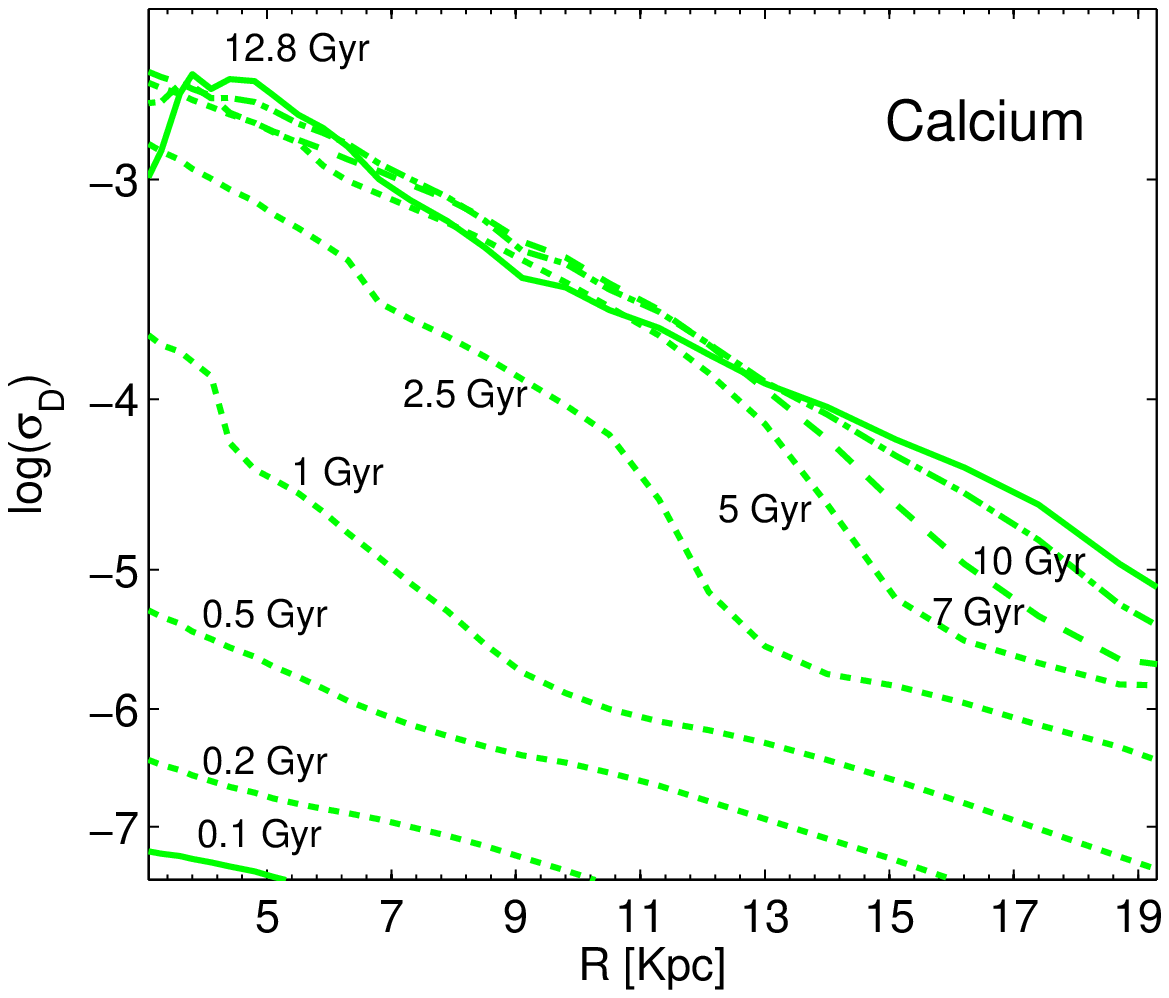}
\hspace{-15pt}
\includegraphics[height=6.2cm,width=6.4truecm]{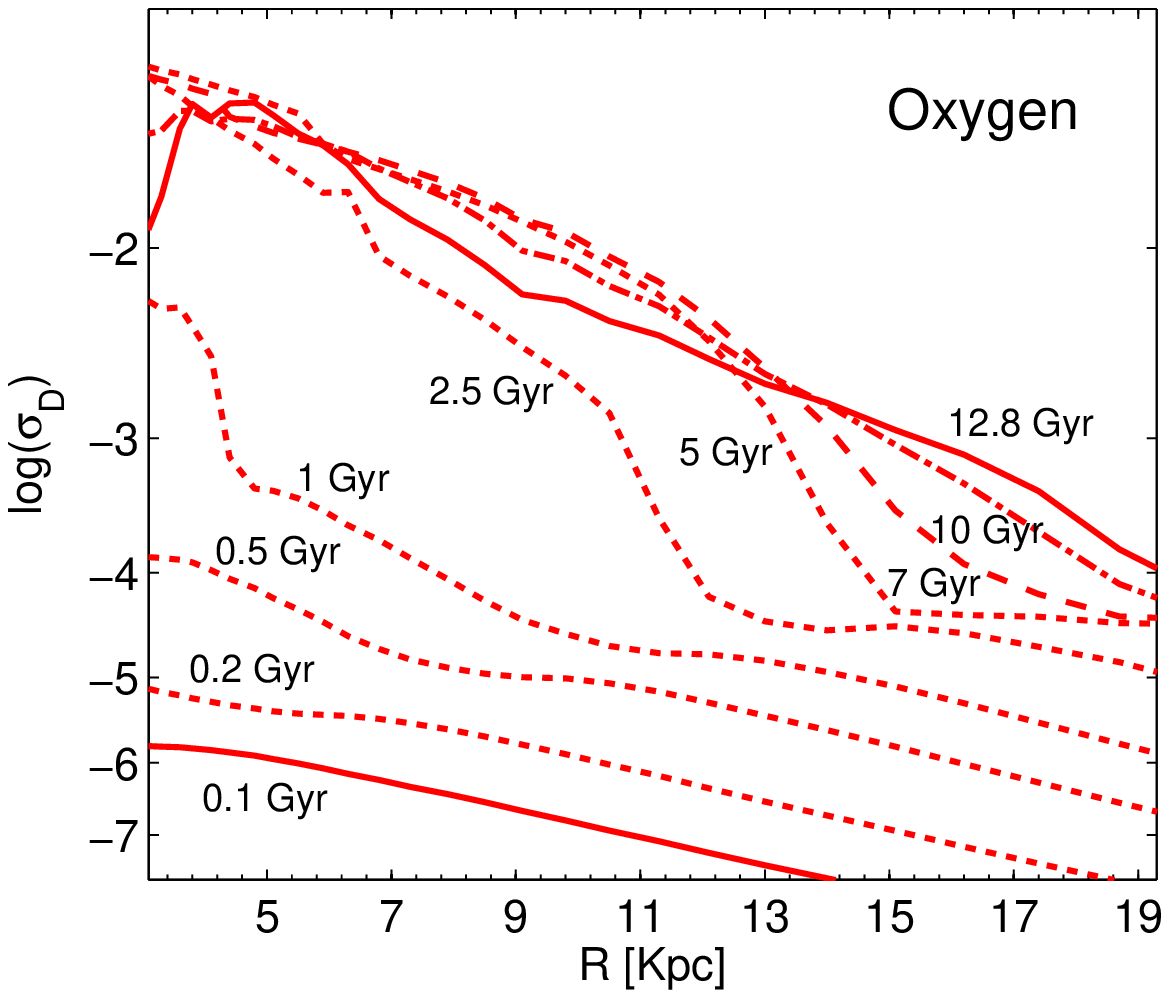}
\hspace{-12pt}} \caption{Temporal evolution of the logarithmic
radial abundance of dust $\sigma^{D}_{i}\left(r_{k},t\right)$, in
[M$_{\odot}$pc$^{-2}$], for all the elements belonging to our set
and involved into the process of dust formation, that is C, N, O,
Mg, Si, S, Ca and Fe (see also \citet{Piovan11b} for more details).
All the contributions have been properly corrected for the dust
destruction. Nine ages are represented from the early stages to the
current time, that is 0.1, 0.2, 0.5, 1, 2.5, 5, 7, 10 and 12.8 Gyr,
assuming that the formation of the MW started when the Universe was
$\sim$0.9 Gyr old \citep{Gail09}. \textbf{Upper panels}: from left
to right the time evolution of the radial abundance $\sigma^{D}$ for
Mg and Fe. \textbf{Central panels}: the same as in the upper panels
but for C, N and Si, from left to right. \textbf{Lower panels}: the
same as in the central panels, but for S, Ca and O. Since the
abundance $\sigma^{D}_{i}$ changes at varying the age in a different
range for each element, the scale of the y-axis is continuously
shifted according to the represented range of values.}
\label{EvolRadialAbunElementsDust}
\end{figure*}

%%%%%%%%%%%%%%%%%%%%%%%%%%%Figure 7
\begin{figure*}
\centerline{\hspace{-30pt}
\includegraphics[height=6.cm,width=6.5truecm]{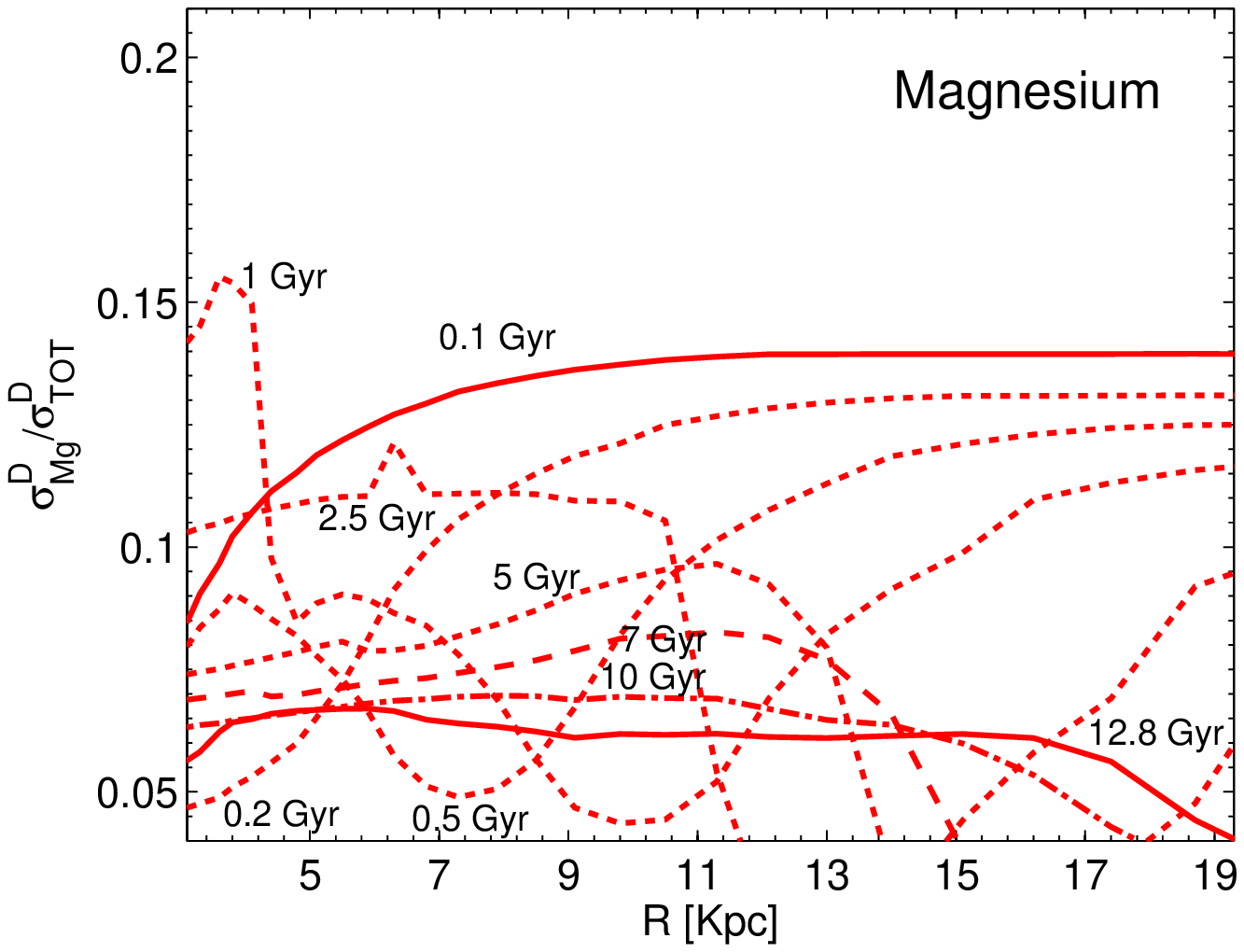}
\includegraphics[height=6.cm,width=6.5truecm]{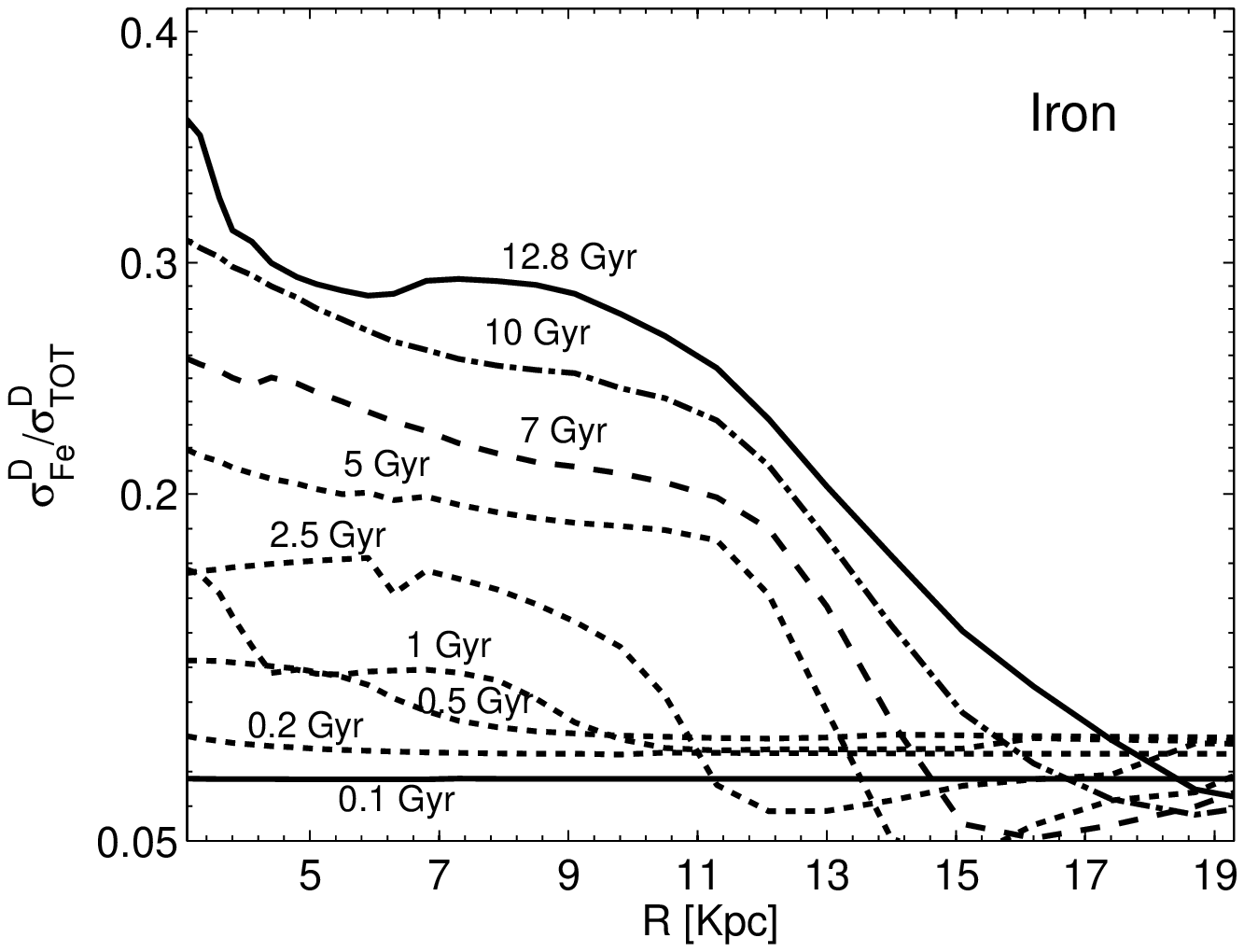}}
\vspace{-3pt}
\centerline{\hspace{-30pt}
\includegraphics[height=6.cm,width=6.5truecm]{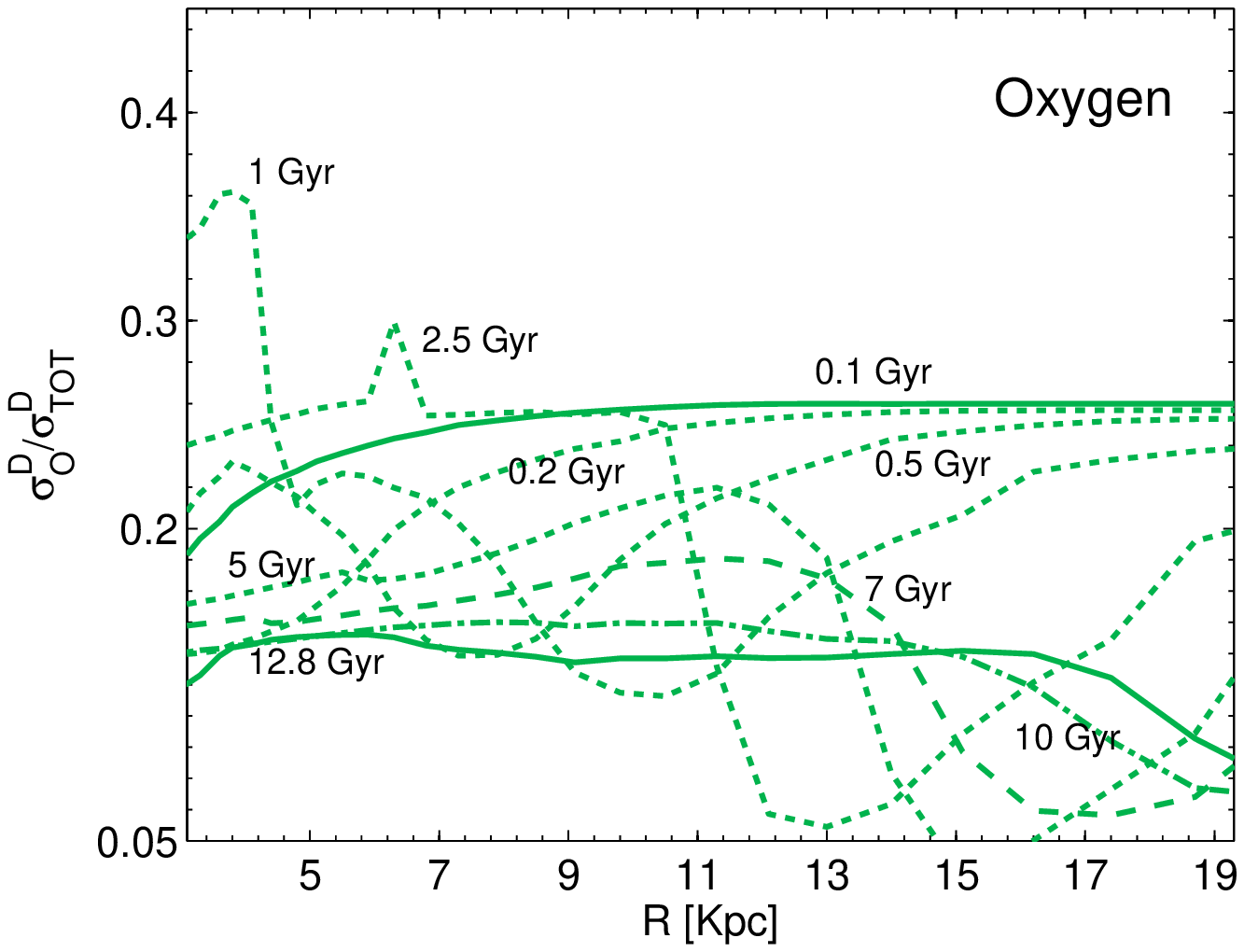}
\includegraphics[height=6.cm,width=6.5truecm]{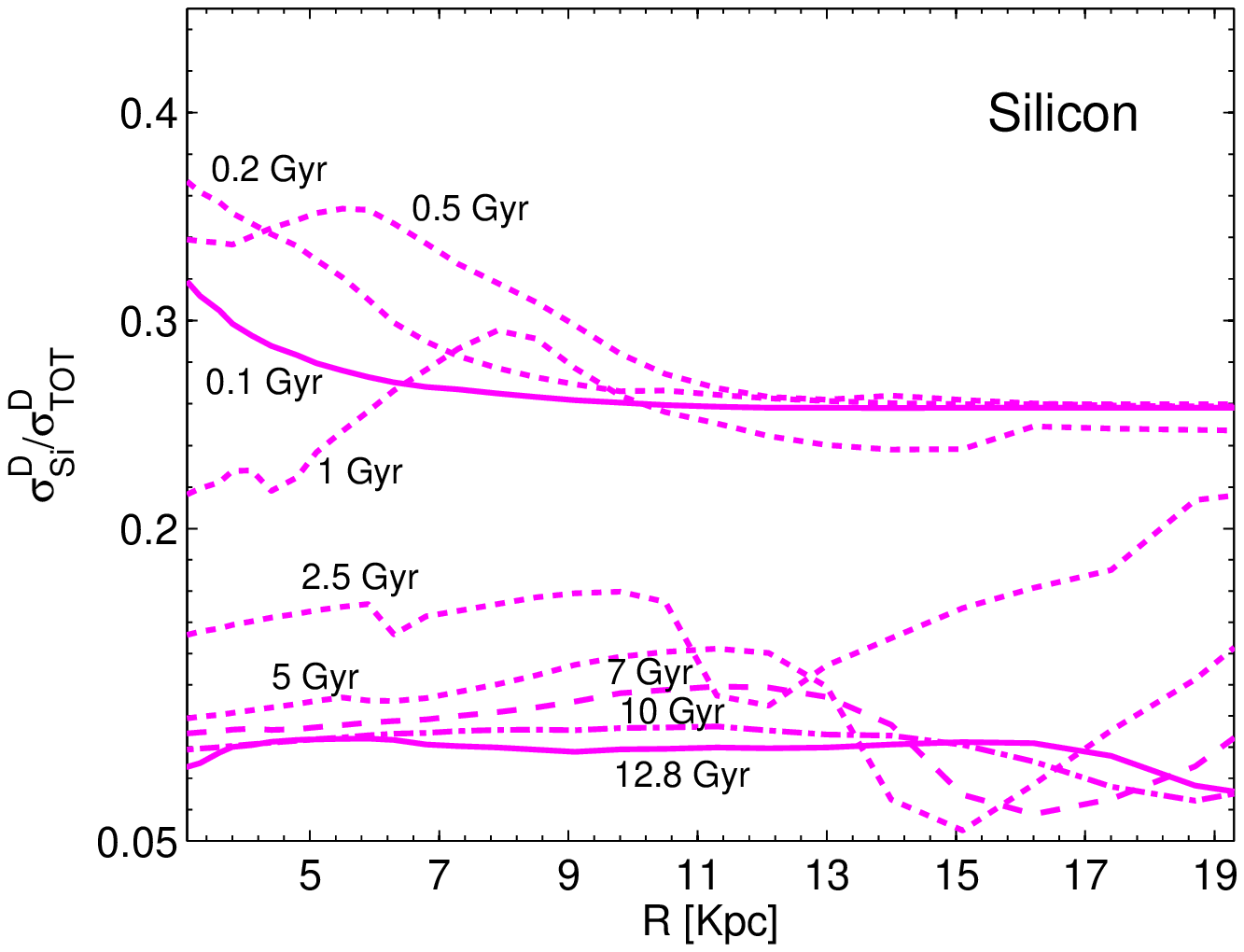}}
\vspace{-3pt}
\centerline{\hspace{-12pt}
\includegraphics[height=6.cm,width=6.4truecm]{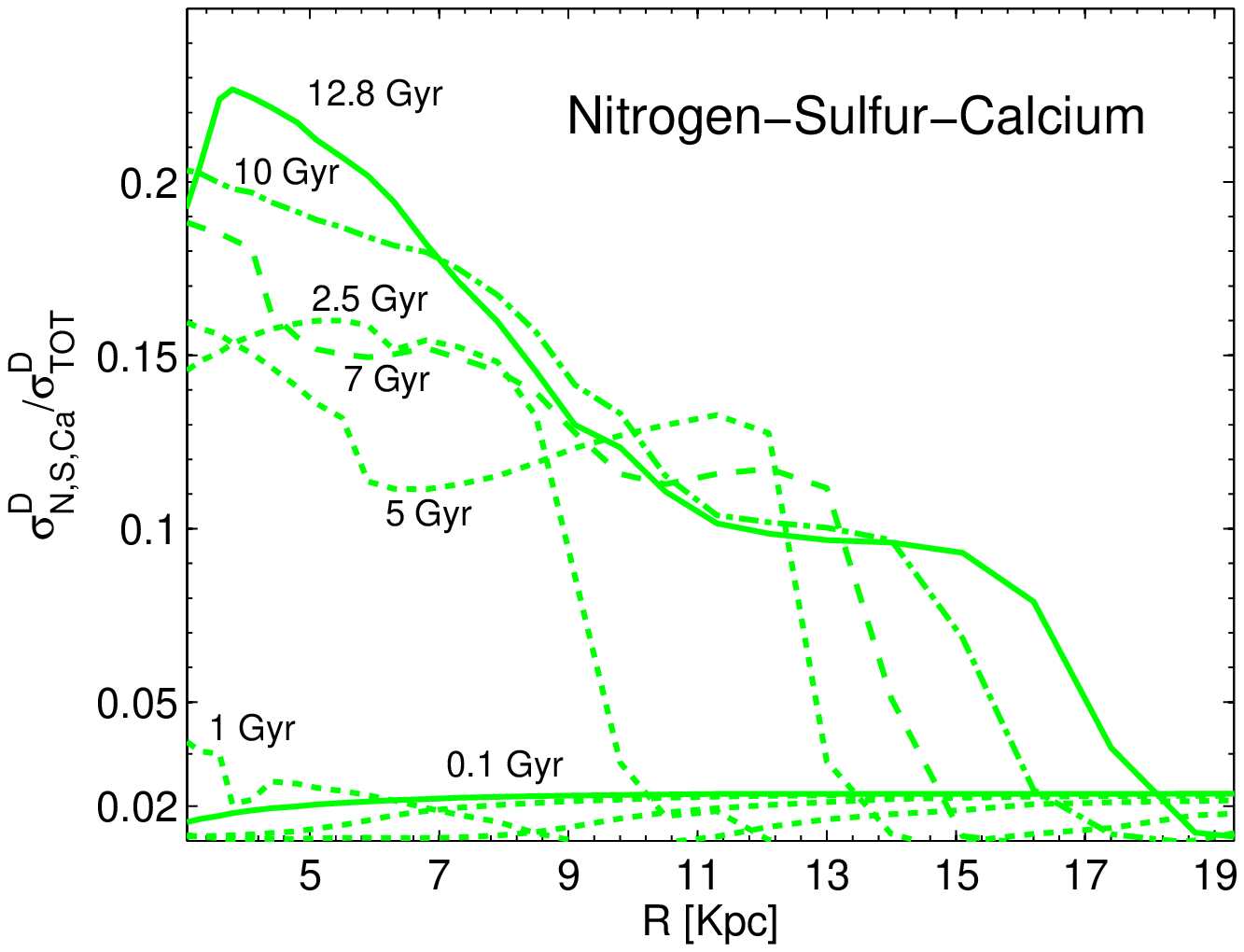}
\hspace{-15pt}
\includegraphics[height=6.cm,width=6.4truecm]{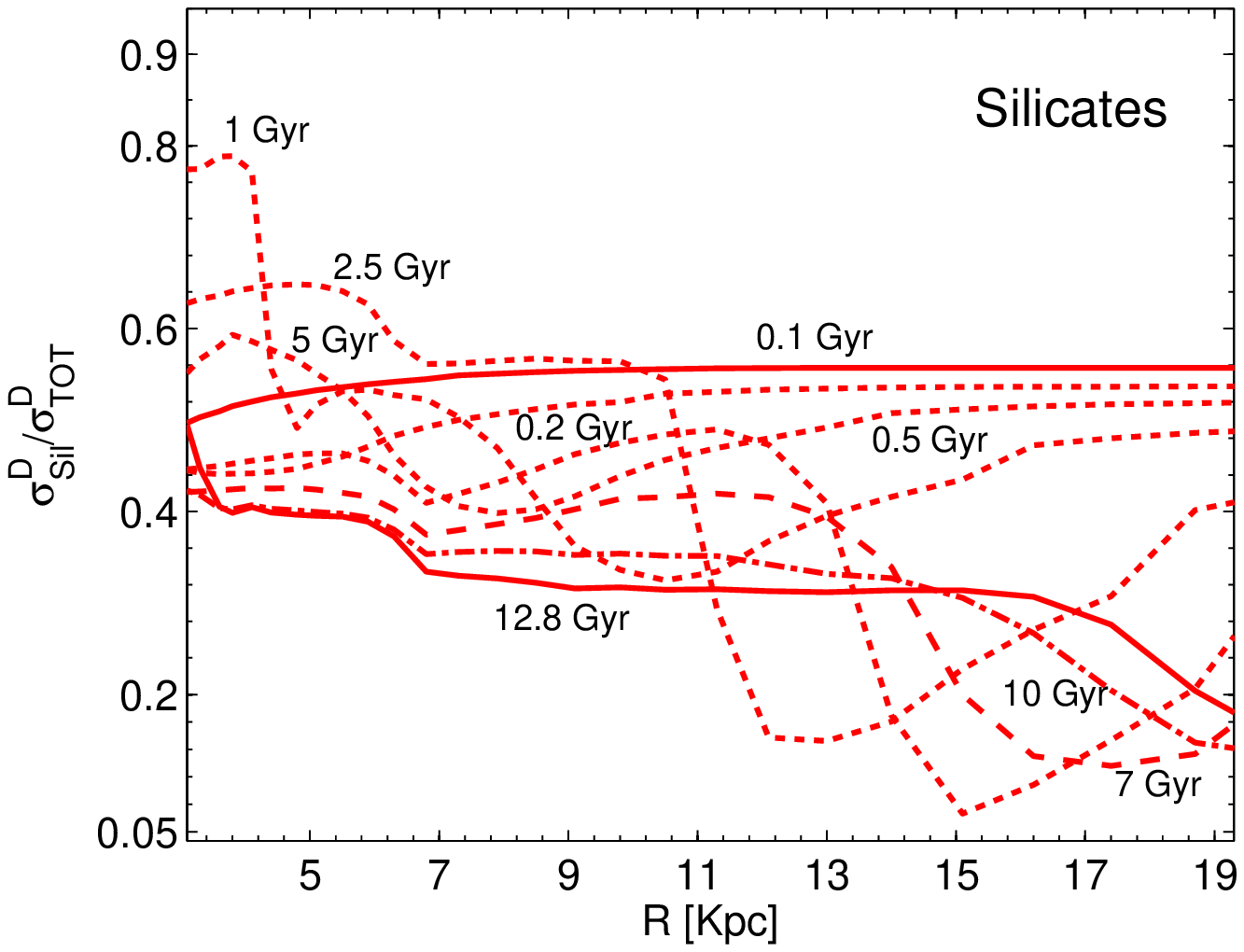}
\hspace{-15pt}
\includegraphics[height=6.cm,width=6.4truecm]{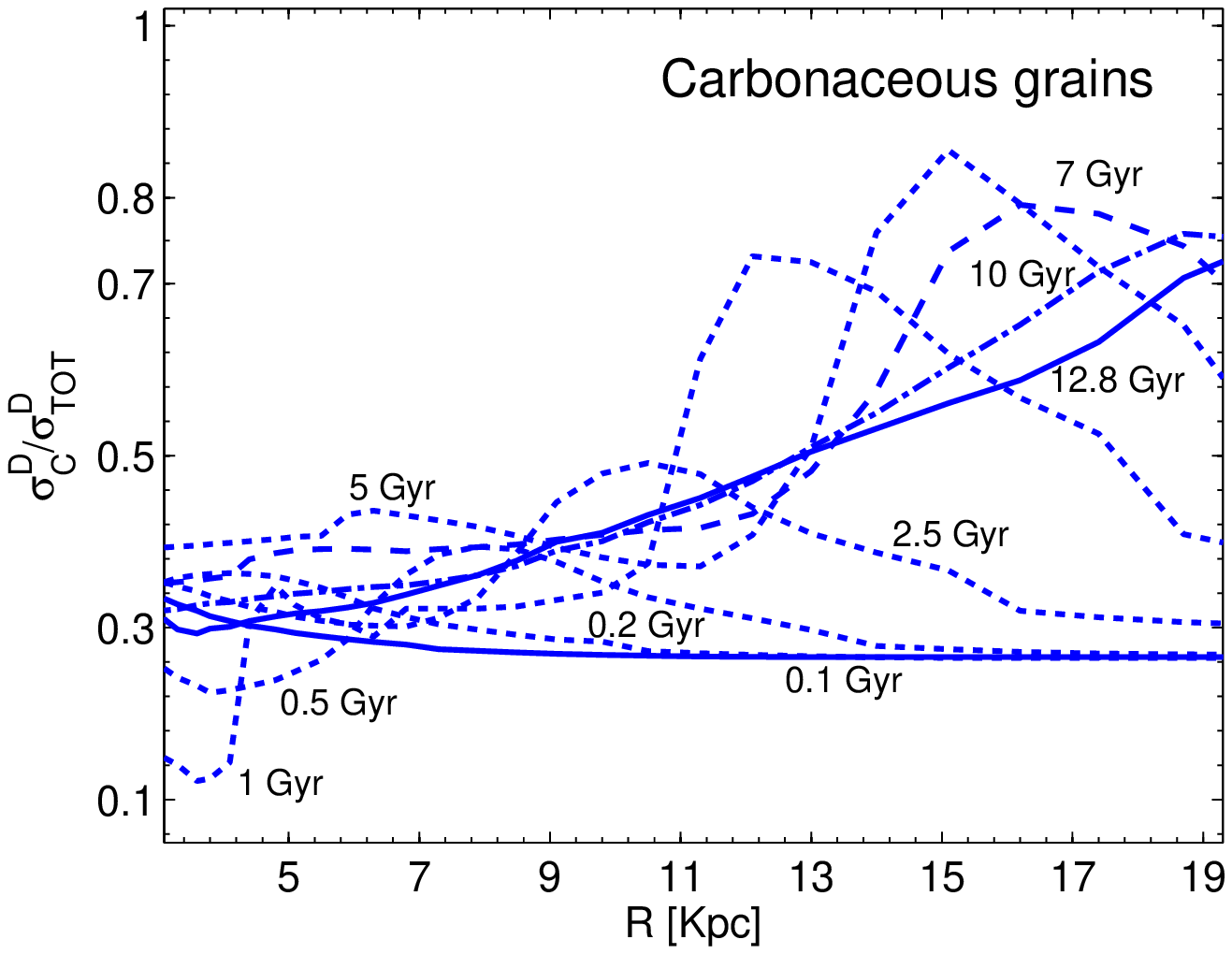}
\hspace{-12pt}} \caption{Temporal evolution of the radial
\textit{mass fraction of single elements or grain families embedded
into dust}
$\sigma^{D}_{i}\left(r_{k},t\right)/\sigma^{D}\left(r_{k},t\right)$,
\textit{normalized to the total dust budget}, for some of the
elements belonging to our set and involved in the process of dust
formation (that is C, N, O, Mg, Si, S, Ca and Fe)  and some grain
types (silicates, carbonaceous grains, iron grains and Ca/N/S - see
also \citet{Piovan11b} for more details). All the contributions have
been properly corrected for the dust destruction. Nine ages are
represented as in Fig. \ref{EvolRadialAbunElementsDust}, from the
early stages to the current time, that is 0.1, 0.2, 0.5, 1, 2.5, 5,
7, 10 and 12.8 Gyr, assuming that the formation of the MW started
when the Universe was $\sim$0.9 Gyr old \citep{Gail09}.
\textbf{Upper panels}: from left to right the time evolution of the
radial mass fraction of Mg and Fe, normalized to the total dust
mass. \textbf{Central panels}: the same as in the upper panels but
for O and Si, from left to right. \textbf{Lower panels}: the same as
in the central panels, but for S, N and Ca (lumped together)  and O,
silicates (that is the mass fraction of Mg, Si, Fe and O involved
into quartz/pyroxenes/olivines) and C (that in practice, once
subtracted the carbon embedded in SiC, forms the carbonaceous
grains).} \label{EvolRadialFracElementsDust}
\end{figure*}

%%%%%%%%%%%%%%%%%%%%%%%Figure 8
\begin{figure*}
\centerline{\hspace{-40pt}
\includegraphics[height=6.cm,width=16.0cm]{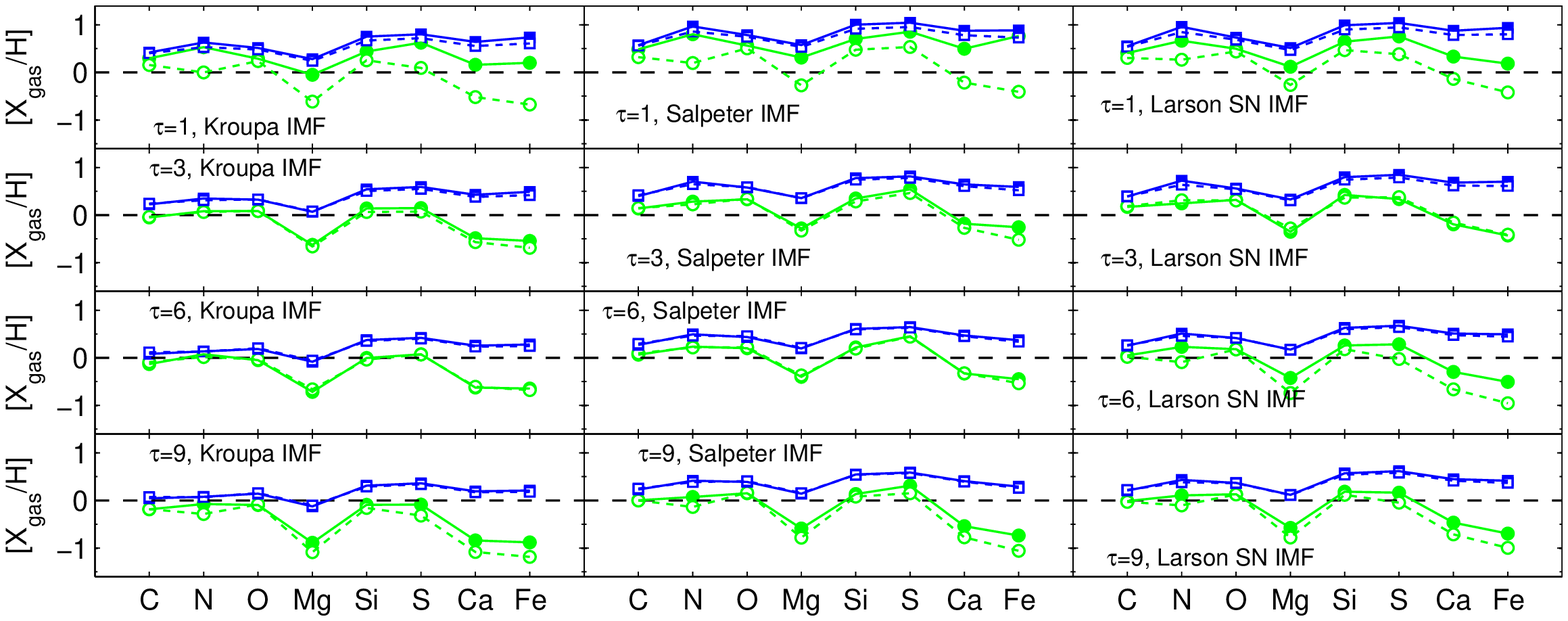}}
\vspace{-3pt}
\centerline{\hspace{-40pt}
\includegraphics[height=6.cm,width=16.0cm]{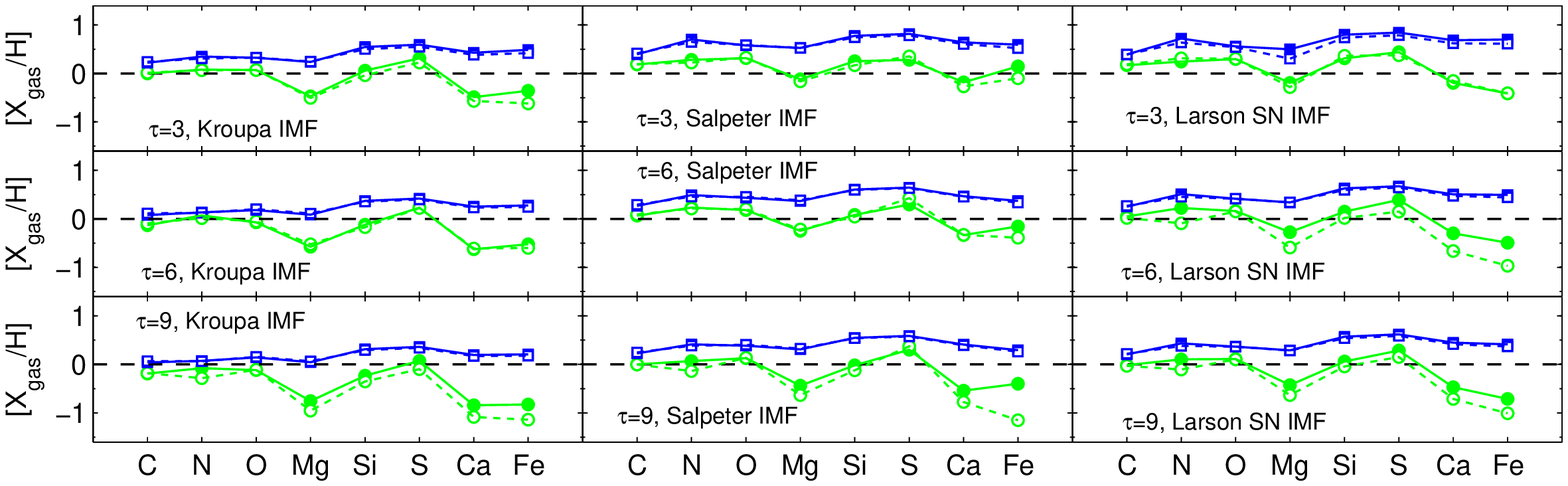}}
\caption{Depletion of the elements C, N, O, Mg, Si, S, Ca and Fe in
the ISM for an \textit{inner ring} of the MW centered at 2.3 Kpc.
\textbf{Top panels}: the element depletions are obtained using the
original yields, \textit{no correction for the Mg under-abundance is
applied}. The same combinations of the parameters adopted by
\citet{Piovan11b} to simulate the depletion in the SoNe are
considered, that is three IMF (Salpeter, Kroupa multi power-law and
Larson adapted to the SoNe) and four infall time scales ($\tau =1$,
$\tau =3$, $\tau =6$ and $\tau =9$). In each panel the global ISM
abundance (gas+dust) is shown. The values of the abundances are
connected with continuous lines for the case $\nu = 0.7$ (filled
squares and circles), while for $\nu = 0.3$ (open squares and
circles) we use dashed lines. The squares represent the global ISM
abundance, while the circles the depleted gas abundance. For each
panel there is a label showing the combination IMF/infall timescale
we have adopted. \textbf{Bottom panels}: the same as in the top
panel but with the original yield corrected for the Mg
under-abundance. Only three infall timescales are shown ($\tau =3$,
$\tau =6$ and $\tau =9$).} \label{DepletionINNER_CORR}
\end{figure*}

%%%%%%%%%%%%%%%%%%%%%%%Figure 9
\begin{figure*}
\centerline{\hspace{-40pt}
\includegraphics[height=6.cm,width=16.0cm]{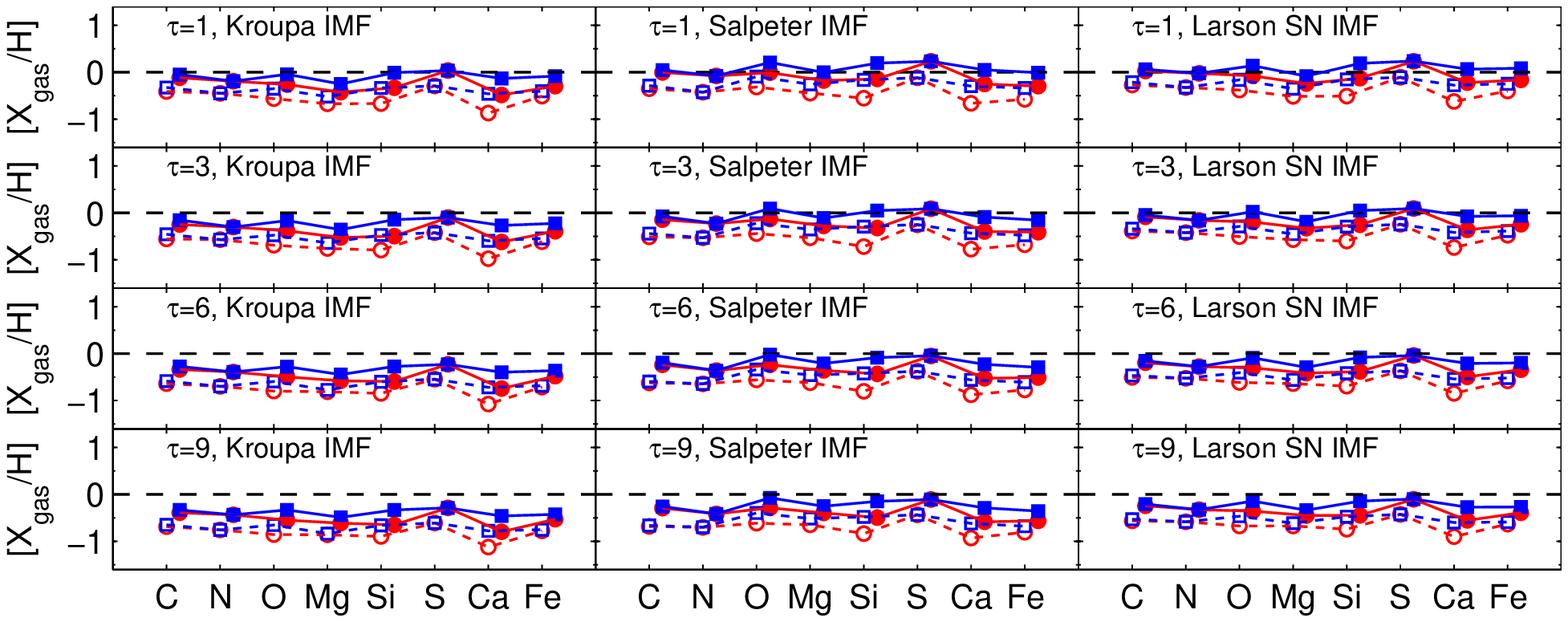}}
\vspace{-3pt}
\centerline{\hspace{-40pt}
\includegraphics[height=6.cm,width=16.0cm]{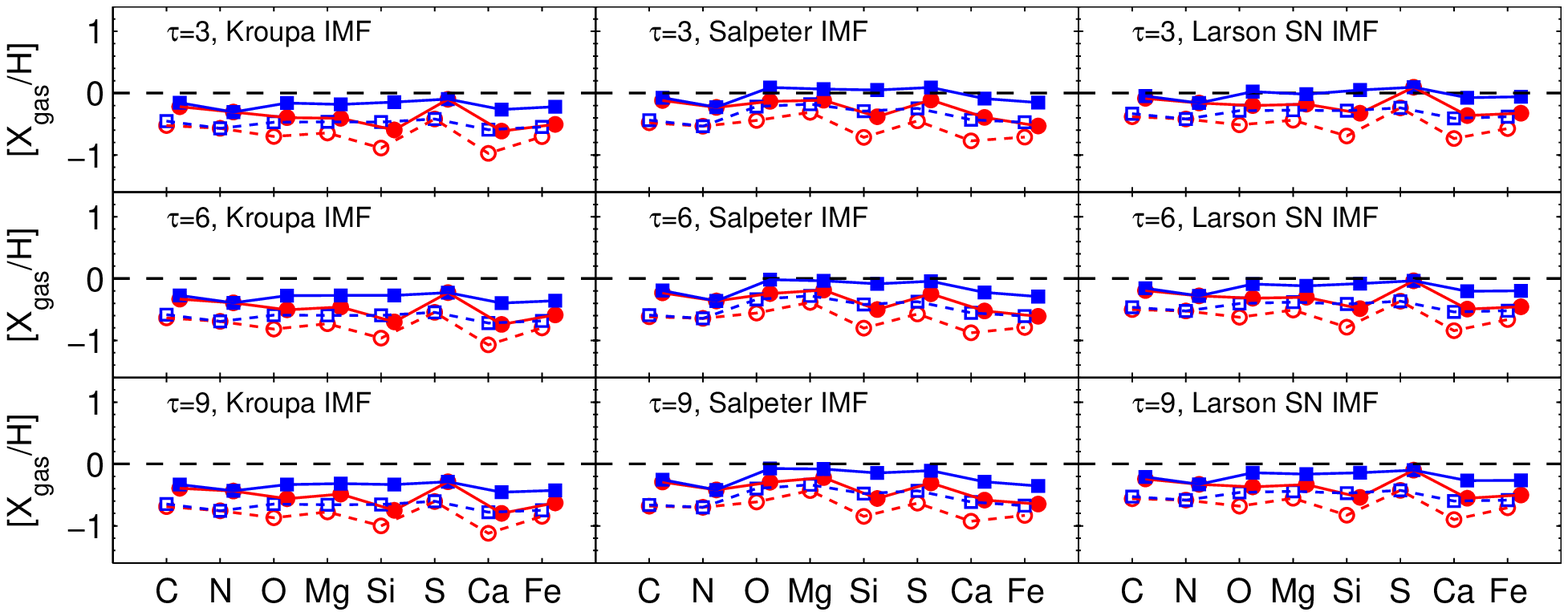}}
\caption{Depletion of the elements C, N, O, Mg, Si, S, Ca and Fe in
the ISM for an \textit{outer ring} of the MW at 15.0 Kpc.
\textbf{Top panels}: the element depletions are obtained using the
original yields, \textit{no correction for the Mg under-abundance is
applied}. The choice of the parameters and the meaning of the
symbols are the same as for Fig. \ref{DepletionINNER_CORR}. For the
sake of a better representation of the data, the case with $\nu =
0.7$ \textit{is slightly shifted to the right} in such a way to
avoid the superposition with $\nu = 0.3$. As in Fig.
\ref{DepletionINNER_CORR}, the dashed lines connect the un-depleted
and depleted gas abundances for the $\nu = 0.3$ case. \textbf{Bottom
panels}: the same of the top panel, but with the original yield
corrected for the Mg under-abundance. The same timescales of Fig.
 \ref{DepletionINNER_CORR} are used.}
\label{DepletionOUTER_CORR}
\end{figure*}

To this aim, first we derive the best fit for the three quantities:
the results (obtained using the MATLAB \textit{Curve Fitting
Toolbox}) are indicated by the solid lines in the top panel and the
thick line in the bottom panel of Fig. \ref{GASeSFRMW}. Second,
using the best fits we derive the 3D correlation between the
$\mathrm{H}_{2}$ mass, the star formation rate, and total gas mass.
They are shown in the three panels of Fig. \ref{3Dand2DH2}. To
reproduce the 3D data and extrapolate the results to unknown regions
of the parameter space, we use  the Artificial Neural Networks
technique (hereafter ANNs) modelled from biological nervous systems:
ANNs are like a network of neurons linked together by synapses
(numerical weights). The network in usage  here is taken from
\citet{Grassi11b} to whom the reader should refer for all details.
It  contains two layers of hidden neurons whose number is $n_{h}=5$
and $n_{g}=25$, respectively and has  two input neurons, i.e.
($\textrm{SFR}/\textrm{SFR}_{\odot}$ and and
$\sigma^{G}\left(r_{k},t_{G}\right)$) (plus the bias input that is
always set to a constant value) and one output neuron, i.e.
$\sigma_{H_{2}}\left(r_{k},t_{G}\right)$. The evolution of the RMS
is monotonic and decreases to approximately $\mathrm{RMS}=10^{-3}$
after $10^6$ iterations. In the left panel of  Fig. \ref{3Dand2DH2}
we show  the fit found by the ANNs (solid line) compared with the
data (crosses). We suppose now that the correlation between the
three parameters established from the present-day data remains the
same as a function of the Galaxy age and therefore we drop the
dependence on time. Furthermore, adopting the same ANN weights, we
extrapolate the ANN predictions to the whole space of the parameters
$\textrm{SFR/SFR}_{\sun}$, $\sigma^{G}(r,t)$ and
$\sigma_{H_{2}}(r,t)$. The contours and isolines describing the
extrapolated regions are shown in left and middle panels of Fig.
\ref{3Dand2DH2}.\\
\indent Some comments are worth being made here. First, we have no
warranty that the behaviour in the extrapolated region is so regular
as predicted by the ANNs. In the huge gaps uncovered by the
experimental data for the MW, we could for example have valleys and
peaks so that to cast light on this problem a complete coverage of
the parameter space with observational data for spiral galaxies in
general would be required. However, this goes beyond the aims of
this study.  It suffices here to  check that extrapolating   the MW
data does not lead to  un-physical results. The predictions seem to
be reasonable: for low SFR or low gas amount, the H$_{2}$ mass is
low. For high SFR and gas mass  we have a reasonable amount of
H$_{2}$. Even if we look only at the region where the SFR is high
but the total amount of gas is low, the prediction for H$_{2}$ is
also reasonable. Even if the correlation between H$_{2}$ and SFR is
not clear and somewhat in contradictions with the expectations,
stars are seen to form from the cold molecular gas
\citep{Kennicutt98} and a significant amount of H$_{2}$ is
physically acceptable when SFR is significant.\\
\indent Second, the region covered by the ANN is bounded  by the
limits
$[(\textrm{SFR}/\textrm{SFR}_{\sun})_{min},(\textrm{SFR}/\textrm{SFR}_{\sun})_{max}]$
and
$[\sigma^{G}\left(r,t\right)_{min},\sigma^{G}\left(r,t\right)_{max}]$.
This may be a problem in particular with high SFRs, because the
limits for the ANN have been set basing upon data for the SFR in the
innermost regions of the Disk where the uncertainty is high. To cope
with this point of embarrassment, when the SFR is higher than about
$3\times \textrm{SFR}_{\sun}$, we replace the predictions based on
the ANNs with a simpler fit based on $\sigma_{H_{2}}(r,t_{G})$
versus SFR data and we keep the fraction  $\chi_{MC}$ in the
interval $0.4-0.5$. All of this is shown in the right panel of Fig.
\ref{3Dand2DH2}.

\section{Radial gradients in element abundances} \label{RadialAbundances}

Studying the properties of dust in the SoNe is by far easier than
across the Disk of the MW: indeed the data to our disposal for the
Disk are much fewer and of poorer quality  than those for the SoNe.
For instance, we cannot estimate the depletion of elements at
different distances from the SoNe toward the Galactic Center and/or
the outermost regions of the Disk. However, we can compare the
radial gradients in element abundances, as indicated by Cepheid
stars, Open Clusters, HII regions and O,B stars. In the case of a
satisfactory agreement between the observations and theoretical
predictions for the element abundances, we may argue with some
confidence that the predictions for radial variation of the dust
properties are reasonably good.  In relation to this see the case of
the relative abundances of Mg and Si and how they would affect the
formation of silicates in the dust and depletion of these elements
as thoroughly discussed by \citet{Piovan11b}. \\
\indent In the following, we examine  the radial variations of the
elemental abundances to check if our multi-shells model, in presence
of radial flows of gas and  inner bar, is able to reproduce in a
satisfactory way at least the main radial properties of the MW. As
already said we are not looking for the best model in absolute sense
but only the a general agreement with observational data so that our
predictions for dust formation and evolution stand on solid ground. \\
\indent Two important observational  constraints must be satisfied
and reproduced by the models,  namely: (i) the radial distribution
of the gas and (ii) the radial gradients  in those elements that are
involved in the dust formation. In Fig. \ref{DensityProfile},
starting from the reference model introduced in  Sect.
\ref{chemical}, we compare the density profiles for models
calculated with different IMF, efficiency of the star formation
$\nu$ and the infall timescale $\tau$. Twelve models are shown. This
sample is taken from the wider grid of models calculated by
\citet{Piovan11b} to reproduce the depletion of the elements
observed in the SoNe \footnote{ The whole sample contains model for
four v$\tau$  ($\tau = 1$, $\tau = 3$, $\tau = 6$ and $\tau = 9$
Gyr), three IMFs (Salpeter, Kroupa multi-power law and Larson
adapted to the SN), four values of the efficiency of the star
formation $\nu$ ($\nu = 0.3$, $\nu = 0.7$, $\nu = 1.1$ and $\nu =
1.5$), and  two cases for the stellar yields of elements, the
original ones by \citet{Portinari98} and a revised version taking
into account a correction for a slight under-abundance in Mg.}. The
twelve models are shown in Fig. \ref{DensityProfile}. They are with
original stellar yields by \citet{Portinari98} (no correction to the
Mg abundance is applied). Since we have no data for the element
depletion as a function of the radial distance, the slight
under-abundance of Mg is less of a problem here.  Furthermore,
selecting  the twelve cases, very short infall timescales have been
excluded because the models would not reproduce the  element
depletions observed in  the SoNe. Finally, since we have no hints on
the observational error associated to the mass density profile
$\sigma_{g}(r)$ as a function of the galacto-centric distance, at
each distance we apply an error derived from the ratio $\Delta
\sigma_g/\sigma_g$  for the SoNe, and assumed to remain constant.
Indeed, already in the SoNe, there is a significant uncertainty in
the local gas density, the  estimates of which go from 7-13
M$_{\odot}$pc$^{-2}$ \citep{Dickey93}, to $\sim
8$ \citep{Dame93} up to 13-14 \citep{Olling01}. \\
\indent It is soon evident that while some models fairly agree with
the observational data for $\sigma_{g}$, others significantly
deviate from it  in particular for the innermost regions: models
with long infall timescale and low star formation efficiency (such
as the model with  $\tau = 9$ Gyr  and $\nu = 0.3$) do not consume
enough gas. In literature the problem is often solved either
introducing a suitable radial dependence for $\tau$ or changing the
IMF in such a way that the low-mass long-lived stars are favored.
However, for the purposes of the present study, we prefer to keep
both $\tau$ and IMF constant with the galacto-centric distance.
Finally, it is worth noting the crucial contribution of the bar
whose presence nicely reproduces the gas density distribution in the
innermost regions,
simulating the bump at about 5 kpc.\\

\indent In Fig. \ref{EvolRadialAbunGAS} we show  the gradients
across the Galactic Disk in the abundance ratios
$[\textrm{Mg}/\textrm{H}]$, $[\textrm{Fe}/\textrm{H}]$,
$[\textrm{C}/\textrm{H}]$, $[\textrm{Si}/\textrm{H}]$,
$[\textrm{O}/\textrm{H}]$, $[\textrm{S}/\textrm{H}]$,
$[\textrm{Ca}/\textrm{H}]$ and $[\textrm{N}/\textrm{H}]$. All these
elements are also involved in dust formation and evolution. For each
element we group the observational data according to the source of
the abundances, namely O and B stars, field Red Giants (RGs), HII
regions, Cepheid variables and open clusters. The  references for
the observational data are listed in the footnote\footnotemark[7].
Only models with  intermediate values of $\tau$ ($ 1 < \tau < 9$)
 are compared with the data. Since most likely different sources
trace the gradients at different galactic ages, we plot the
gradients as they were at the time the Sun was born (about 4.5 Gyr
ago, the thin blue and black lines) and at the present age (thick
red and magenta lines). Most objects under examination should fall
into the region spanned by the gradients during this time interval.
In general, theoretical predictions and data  fairly agree with
minor  discrepancies that can be, in some cases, attributed to the
yields, such as the slight Mg under-abundance and/or C and S
over-abundance \citep{Portinari98}. Other major discrepancies
between the theoretical and observational abundance gradients that
depend on the type of source under consideration are expected and
can be taken to indicate the performance of the models. For
instance, following \citet{Kovtyukh05}, in intermediate mass stars
the C under-abundance (carbon is deficient at about
$[\mathrm{C/H}]=-0.4$) and the N over-abundance (nitrogen is
enhanced at about $[\mathrm{N/H}]=+0.4$) has been predicted by
\citet{Luck78} and \citet{Luck81} and confirmed in studies of
Cepheids and non-variable super-giants
\citep{Andrievsky96,Kovtyukh96}. This is more or less what we
observe when the evolution of theoretical element abundances C and N
along the Disk are superposed to the measurements of C and N
abundances in Cepheids (intermediate mass stars of  3 to 9
M$_{\odot}$). Finally, comparing theory and data, one should keep in
mind that models do not take star dynamics and kinematics into
account and the stars always remain in the same region
where they are born: the radial flows act only on gas and dust.\\

\textsf{Radial gradients in dust}. Assessed that the radial
gradients, in the gaseous element abundances and total gaseous
density, reasonably agree with the observational data, at least for
some combinations of the input parameters, we examine the radial
formation and evolution of dust in the MW Disk being confident that
it is grounded on a realistic evolution of the global abundances in
the ISM. In Fig. \ref{EvolRadialAbunGrains} we show for five
selected ages, from the early stages to the current time, the time
evolution of the radial abundance of four typical grain families,
that is silicates, carbonaceous grains, iron grains and generic
grains formed with Ca, N and S. We grouped these last three elements
into a single group for the sake of simplicity  because of their low
abundance. The contributions to each one of the four groups have
been split among the three sources of  dust, that is star-dust from
SN{\ae} and AGB stars \citep{Piovan11a} and accretion in the cold
regions of the ISM \citep{Piovan11b}. The amount of dust in all the
regions continuously grows  following the  enrichment in metals
produced by the  many  generations of stars. In the early stages of
the evolution, at 0.1 Gyr, star-dust from SN{\ae} is the main source
of dust across the whole disk. At 0.5 Gyr the accretion of dust in
the ISM starts to be significant in the innermost regions and slowly
becomes the dominating source in the whole Disk, earlier for the
innermost regions and later for the outermost ones. AGB stars
contribute mostly with carbon-rich grains in the early evolution at
low metallicity Z and only when  a significant metallicity is
reached they start enriching the ISM in  silicates. It is worth
noting here that the evolutionary scheme across the Disk we have
just outlined is a consequence of the role played by the star-dust
from SN{\ae} in the early stages. Reducing the yields of dust by
SN{\ae} would obviously change the the above results for the early
stages.\\

\indent In Fig. \ref{EvolRadialAbunElementsDust} we present the
evolution with time of the radial dependence of the logarithm of the
dust abundance  $\sigma^{D}_{i}\left(r_{k},t\right)$ for all the
elements in our list taking part to dust  formation, namely C, N, O,
Mg, Si, S, Ca and Fe. Nine ages are represented from the early
stages to the current time. All the elements more or less follow the
same evolutionary trend, and eventually reach a  typical profile
with a bump at  4-5 kpc in agreement with the gas density profile
(as expected). It is interesting to note that while for the
innermost-central rings of the disk the enrichment in dust already
reaches a maximum at about  5 Gyr, for the outermost regions the
dust budget keeps growing until the present age. The reason for that
can be easily attributed to the delayed contribution by accretion in
the ISM  occurring in the outer regions with low star formation
\citep[see][for more details]{Piovan11b}. In brief, as accretion of
dust grains in the ISM  gets important  only when and if  some
enrichment in metals has taken place, in the outer regions of low
star formation this is inhibited by the low metal
content that can be reached over there. \\

It is interesting to examine not only the  abundance evolution of
each element in the dust, but also the relative abundance  of each
element embedded into dust (or the relative fraction of each dust
grain family) with respect to the total dust budget. For this reason
in Fig. \ref{EvolRadialFracElementsDust} we show the evolution of
the radial \textit{mass fraction of each element embedded into dust}
$\sigma^{D}_{i}\left(r_{k},t\right)/\sigma^{D}\left(r_{k},t\right)$,
\textit{normalized to the total dust budget}, for some of the
elements in our list and/or some  families of grains. In particular,
we show the fraction of silicates (which  means Mg, Si, Fe and O
embedded into quartz/pyroxenes/olivines by accretion in the ISM or
star-dust) and carbonaceous grains (this latter is nearly coincides
with the amount of C, once subtracted the amount of C belonging to
the silicon carbide). Let us examine the behaviour of silicates and
carbonaceous grains in some detail. At the beginning of the
evolution, where only the most massive SN{\ae} contribute, the mass
of dust in silicates is a bit larger than in carbon based grains.
This is in agreement with the recent observations of SN 1987A with
the Herschel Space Observatory \citep{Matsuura11}: the significant
emission by a population of cold dust grains observed by PACS and
SPIRE can be reproduced only with a mixture of both carbonaceous
grains and silicates with  these latter  more abundant than the
C-based dust grains. This is indeed what is shown in Fig.
\ref{EvolRadialAbunGrains} at the age of  0.1 Gyr, when SN{\ae} play
the major role. As soon as AGB stars start to contribute (at
low/sub-solar metallicities a 5-6 solar masses has a lifetime of the
order of 0.1 Gyr) the mixture of
 SN{\ae} and low-Z AGB stars slowly increases the fraction of
carbonaceous grains and reduces that of  silicates. Indeed, low-Z
AGB stars are easily C-star and mainly inject carbon in the ISM. At
ages in the range  0.5-1 Gyr, at least in the innermost regions, the
formation of dust grains in the cold ISM starts to be a very
efficient process and strongly influences the total budget. Our
silicates are made by Mg, Si, O and Fe: the timescale for the
formation of silicates in the early evolution, when SN{\ae} enriched
the ISM with all of these elements, is therefore shorter than the
one for carbonaceous grains (carbon is also
partially locked in the unreactive CO).\\

\indent Carbon is  injected by  SN{\ae} (plus a significant
contribution by low-Z AGB stars), but in any case carbon  is less
than the sum of all the other refractory elements involved in
silicates. Their total number density in the ISM, compared to the
one of carbon atoms, favors the formation of silicates during the
early evolution. This can be seen in the upper right panel of Fig.
\ref{EvolRadialAbunGrains}; the fraction of carbon locked up in  CO
also plays an important role. In the panels of Fig.
\ref{EvolRadialFracElementsDust} for the carbonaceous grains and
silicates, one can note that, as a consequence of the behaviour
described above, the fraction of silicates in the innermost regions
grows until the age of about 1 Gyr is reached, while the fraction of
C-rich dust decreases. Afterwards, as  AGB stars join  SN{\ae} to
increase the number  density of Carbon atoms in the ISM,  the
accretion of carbon in cold MCs becomes also efficient, and the
silicates/carbonaceous grains ratio slowly decreases again. A
different behaviour takes place in the outermost regions: at the
beginning there is a ratio similar to the inner regions, but the
fraction silicates/carbonaceous grains keeps decreasing because (i)
the accretion in the ISM has long timescales and starts late due to
the low densities and Z; (ii) the strong contribution of C-based
dust, due to the low-Z AGB stars, dominates in a significant window
between SN{\ae} and start of the ISM accretion. The final result is
the high fraction of carbon dust with respect to silicates in the
outer region. It is interesting now to compare our results with the
evolution of the gradient in Carbon  predicted by
\citet{Zhukovska09}. Since in their case SN{\ae} are very poor
factories of silicates, but significant injectors of Carbon-rich
dust, \citet{Zhukovska09} adopt for the  early evolutionary stages
in the inner regions, a mixture almost totally made by Carbon-based
dust. In our case, on the contrary, we obtain in the inner Disk that
the partition between the two main dust types is more balanced and
both play a comparable role. In any case, in the outer regions we
find results very close to those by \citet{Zhukovska09}.  In brief,
at the current age we all  find that the final gradient in relative
mass fraction of carbonaceous grains has a similar positive slop
reaching in the outermost region a value as high as 70$\%$ of
the dust mass embedded in carbonaceous grains. \\

\indent The behaviour of iron dust has the opposite trend: it starts
very flat and at the current time we have a negative gradient, with
iron dust significantly contributing to the dust mass in the center
of the MW. In the early stages, we have some contribution to the
iron budget by SN{\ae} (however less than for silicates and
carbonaceous grains, see top panels in Fig.
\ref{EvolRadialAbunGrains}) and because of the low metallicities a
negligible contribution by low-Z AGB stars. As time goes on, the sum
of the contributions from (i) CCSN{\ae}, (ii) accretion in the ISM
that for iron dust starts to be significant at t$>1-2$ Gyr, (iii)
AGB stars that reach solar metallicities in the inner regions and
start to inject iron dust,  and, finally, (iv) type Ia SN{\ae}
\citep{Piovan11a,Piovan11b}. The sum
of all of these contributions yields  a gradient with negative slope. \\

\textsf{Radial gradients in element depletion}. Once examined the
results for the radial gradients in element abundances both in the
gas and dust, we discuss the corresponding gradient in element
depletion due to dust formation. First, we concentrate on the
innermost and outermost regions of the Disk. In Fig.
\ref{DepletionINNER_CORR} we show the results for the depletion of r
C, N, O, Mg, Si, S, Ca and Fe in the ISM in an inner ring of the MW
Disk centered at 2.3 Kpc. This is a region of high density and high
star formation in turn. The results are shown for both the original
(upper panel) and corrected Mg yields  (lower panel). Indeed, as
described in \citet{Piovan11b}, a small correction to  the original
yields for the Mg under-abundance leads to different (better)
results for the dust mixture as the Mg to Si ratio affects the
formation of silicates  and the final results  for the  SoNe. The
same group of models presented in \citet{Piovan11b} and used to
study the depletion in the SoNe are shown for the inner and outer
regions, but limited to those with the coefficient of star formation
efficiency $\nu = 0.3$ and $\nu = 0.7$. The infall time scale is
varied between 1 and 9 Gyr (or between 3 and 9 Gyr in the
Mg-corrected models) and three IMFs are examined, namely the Kroupa
multi power-law IMF, the Larson IMF adapted to the SoNe and,
finally, the Salpeter one for the sake of comparison. We also show
the un-depleted (no dust formation) abundances of the gas (i.e. the
gas+dust total abundance). The abundances referring to the same
models are connected by a line (dashed or continuous for $\nu = 0.3$
and $\nu =0.7$ respectively) to better show how the elemental
abundances fall down because of the dust formation. As expected, all
the elements exhibit a lower abundance in the gas. Some common
features are: (i) models with corrected yields of Mg (bottom panels)
show a stronger depletion because they form more silicates, due to
the more balanced ratio between Mg and Si; (ii) Mg is much more
depleted than Si because of the higher abundance $\Delta
[\textrm{X}_{\textrm{gas}}/H]$ of Si (actually $\Delta
[\textrm{X}_{\textrm{gas}}/H]$ of Si and Mg are nearly the same but
the total starting abundance of Si is higher); (iii) compared to the
depletion in the SoNe  examined by \citet{Piovan11b}, there is a
less significant difference between the models with $\nu = 0.3$ and
$\nu = 0.7$, with the exception of the case with $\tau = 1$ Gyr. The
reason for this can be tracked in the fast onset of the ISM
accretion phase in both cases.  Very early on, accretion drives the
dust budget and tends to smooth differences due to injections of
stardust; (iv) the IMFs with higher number of SN{\ae} produce a
stronger enrichment than the case with a smaller population of
SN{\ae}, but the $\Delta [\textrm{X}_{\textrm{gas}}/H]$
caused by the depletion is quite similar. \\

\indent The situation for a companion outermost region of low
density and low star formation centered at 15 kpc is presented in
Fig. \ref{DepletionOUTER_CORR}. The meaning of the symbols is the
same as before. In this Figure we applied a small to the right for
the case $\nu = 0.7$ to avoid superposition of the models. Comparing
the data displayed in  Fig. \ref{DepletionINNER_CORR} and Fig.
\ref{DepletionOUTER_CORR} we note that: (i) in general the innermost
regions form more dust with a higher reduction of the gas abundance
$\Delta [\textrm{X}_{\textrm{gas}}/H]$. This is the result of
forming dust in a medium richer in metals and denser in which the
accretion process is favored; (ii) the effect of the different
efficiencies of star formation (parameterized by $\nu$) is more
evident in the low density/low SFR regions because it makes more
evident the effect of metals and ISM enrichment on dust formation;
(iii) in the outermost regions almost all the elements suffer a
comparable depletion $\Delta [\textrm{X}_{\textrm{gas}}/H]$, simply
because the accretion process is very weak over there.

%%%%%%%%%%%%%%%%%%%%%%%Figure 10
\begin{figure*}
\centerline{\hspace{-30pt}
\includegraphics[height=6.cm,width=7.5truecm]{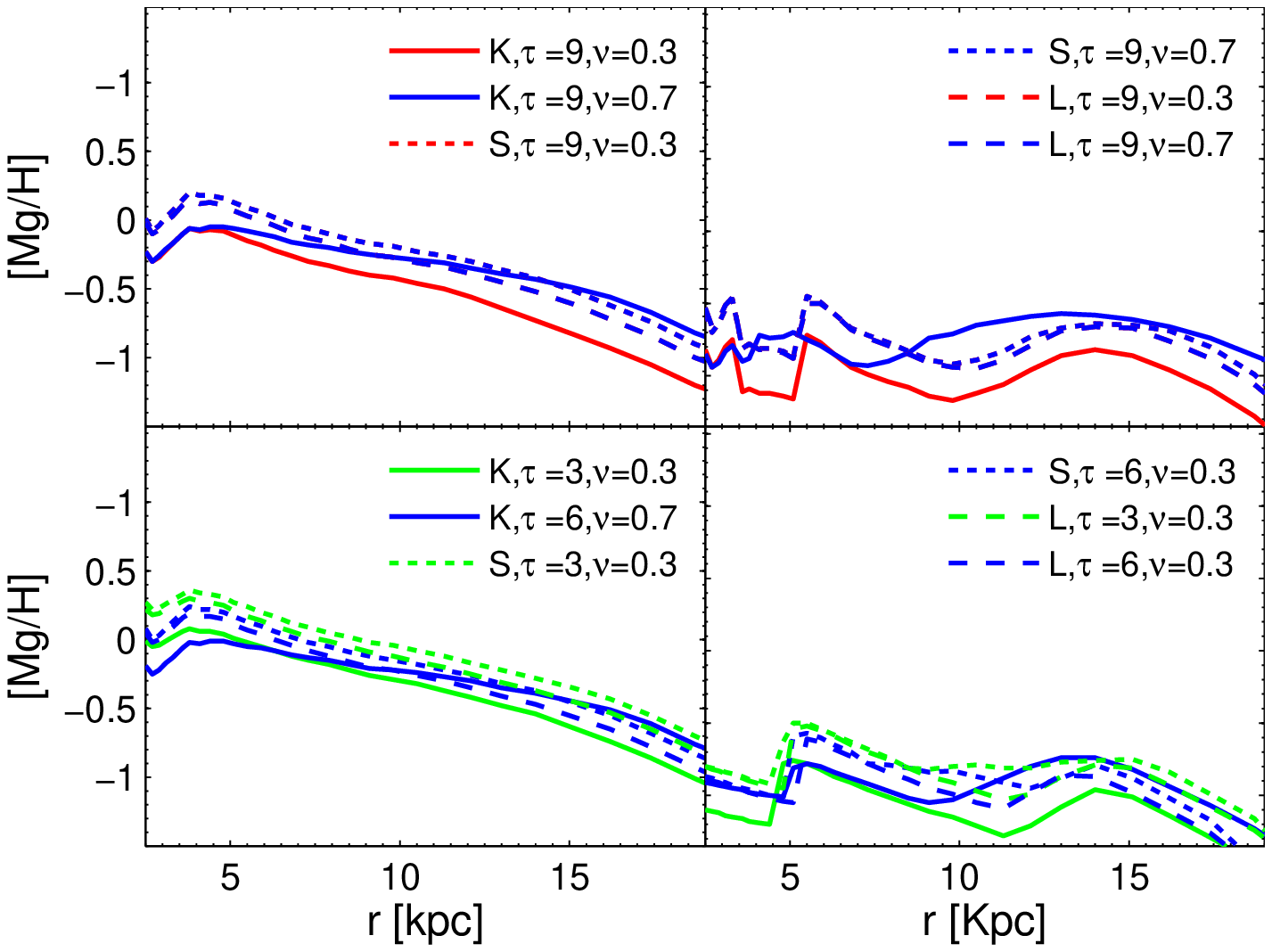}
\includegraphics[height=6.cm,width=7.5truecm]{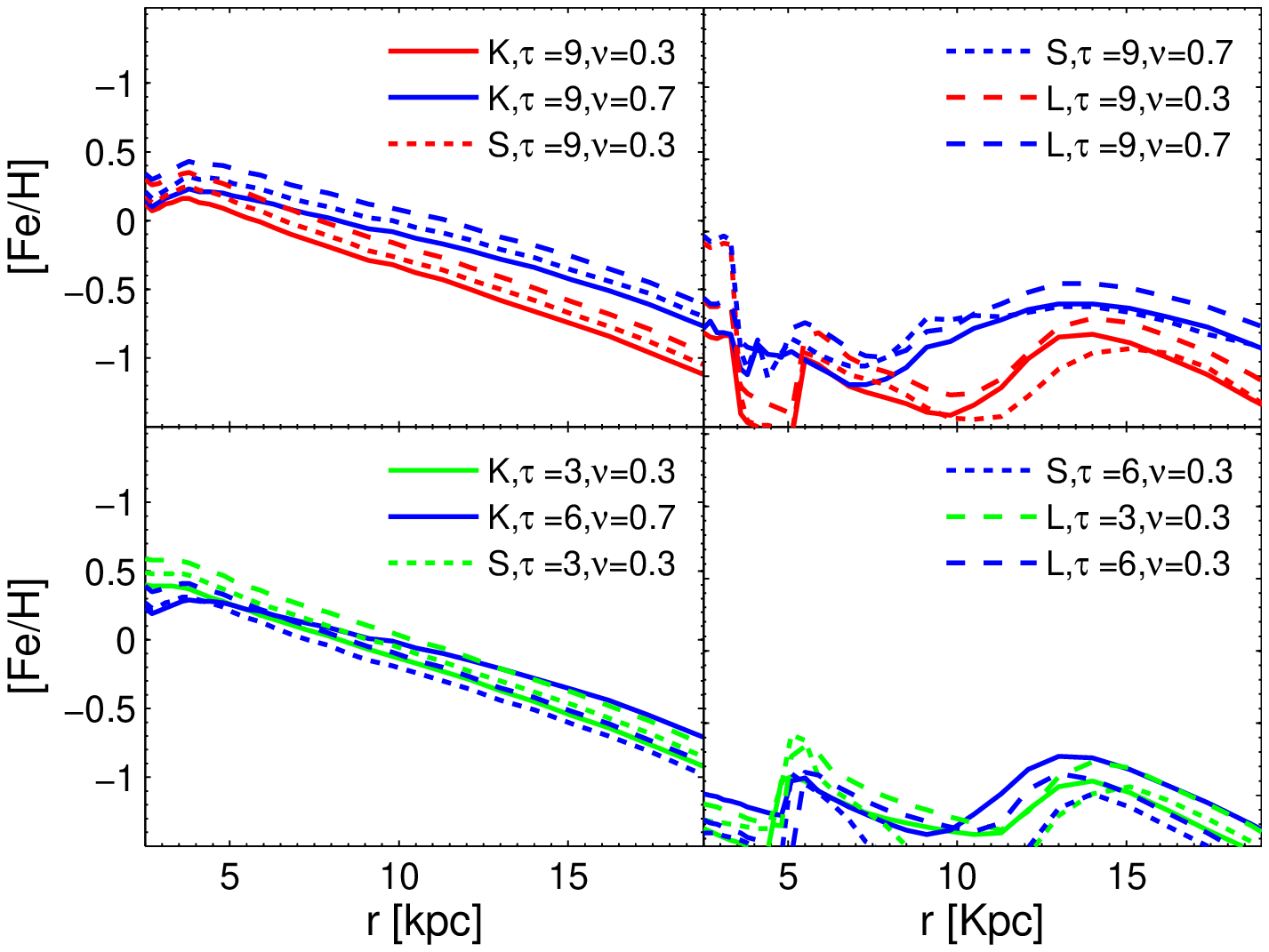}}
\vspace{-3pt}
\centerline{\hspace{-30pt}
\includegraphics[height=6.cm,width=7.5truecm]{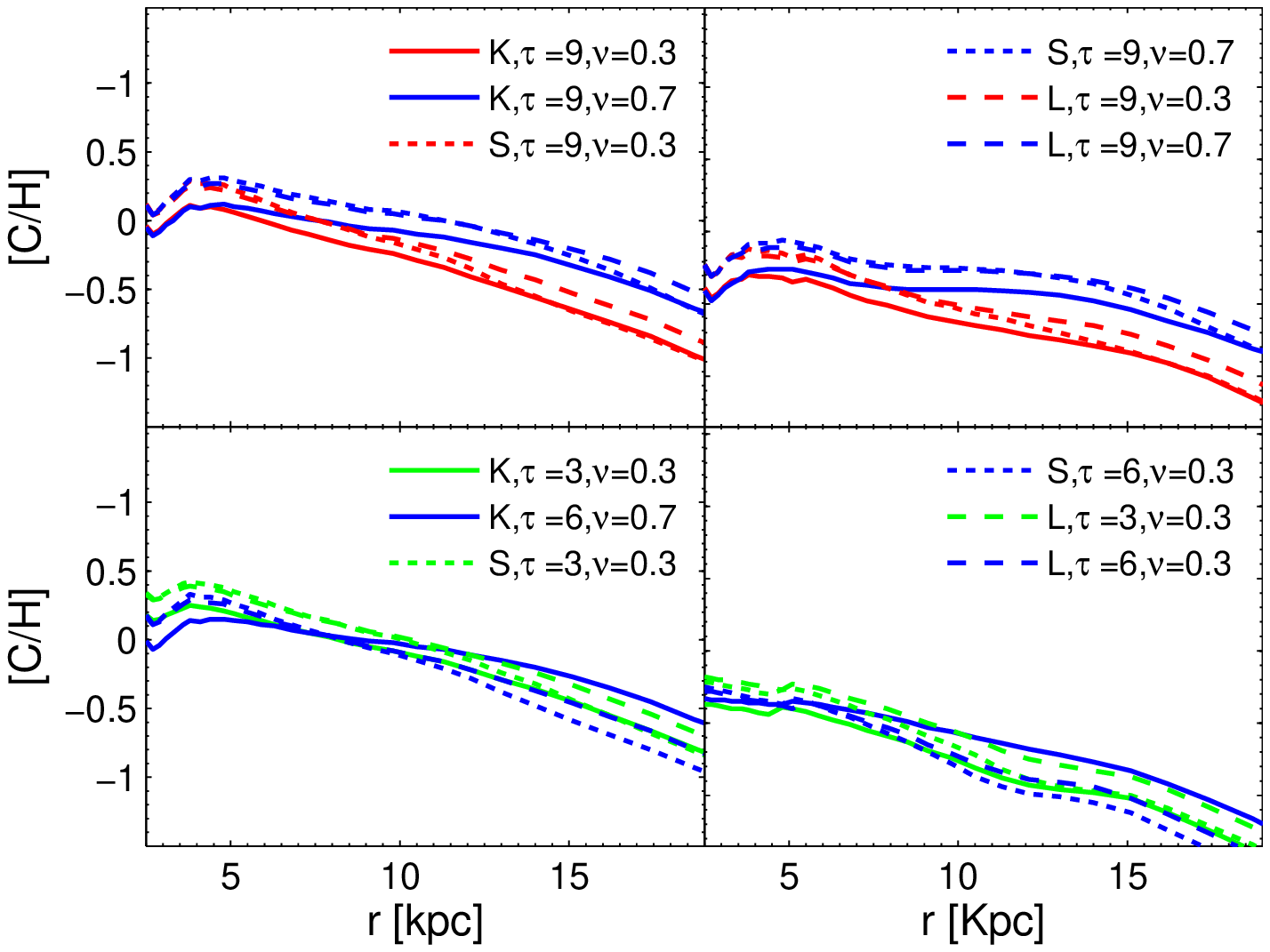}
\includegraphics[height=6.cm,width=7.5truecm]{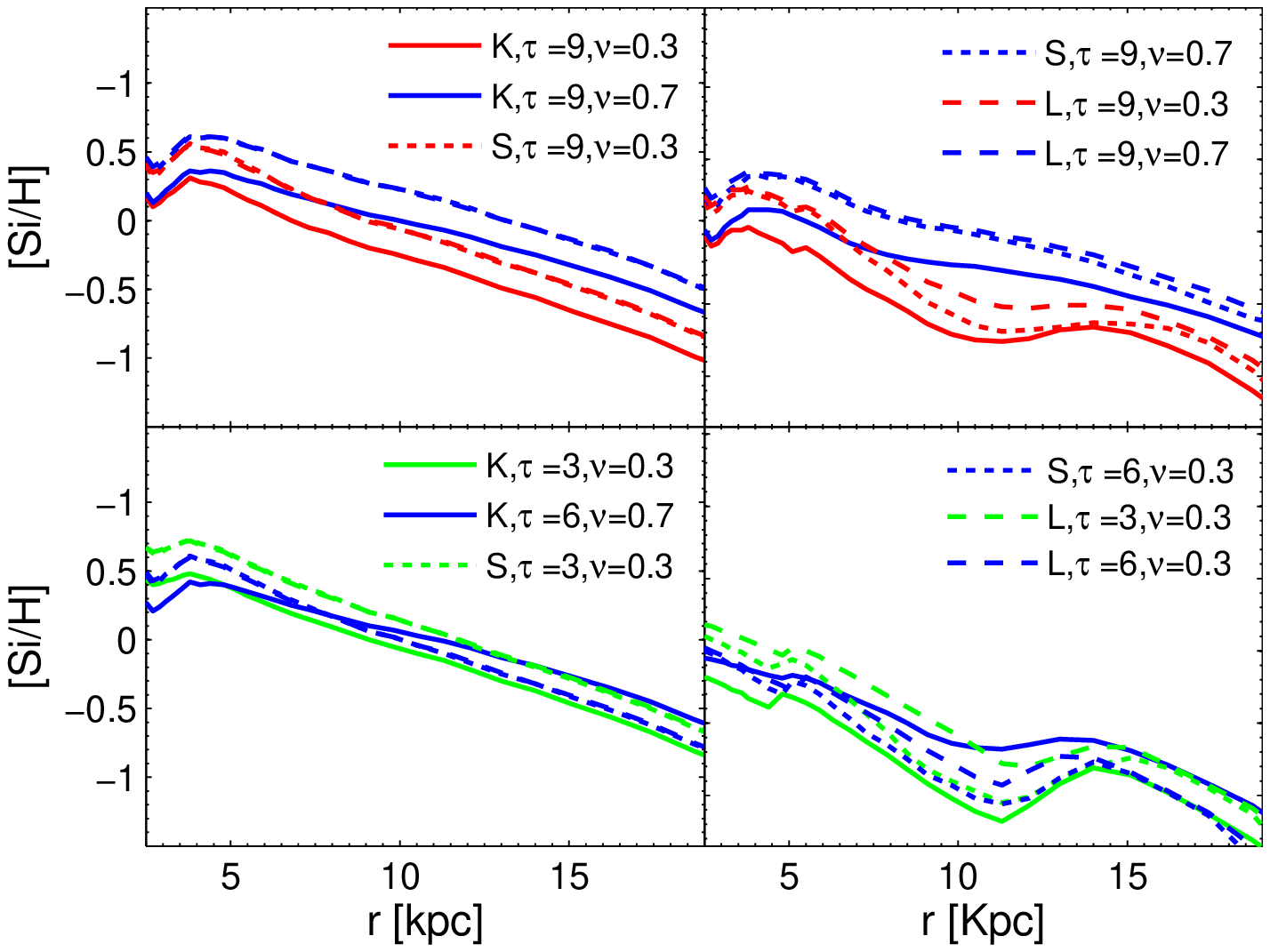}}
\vspace{-3pt}
\centerline{\hspace{-30pt}
\includegraphics[height=6.cm,width=7.5truecm]{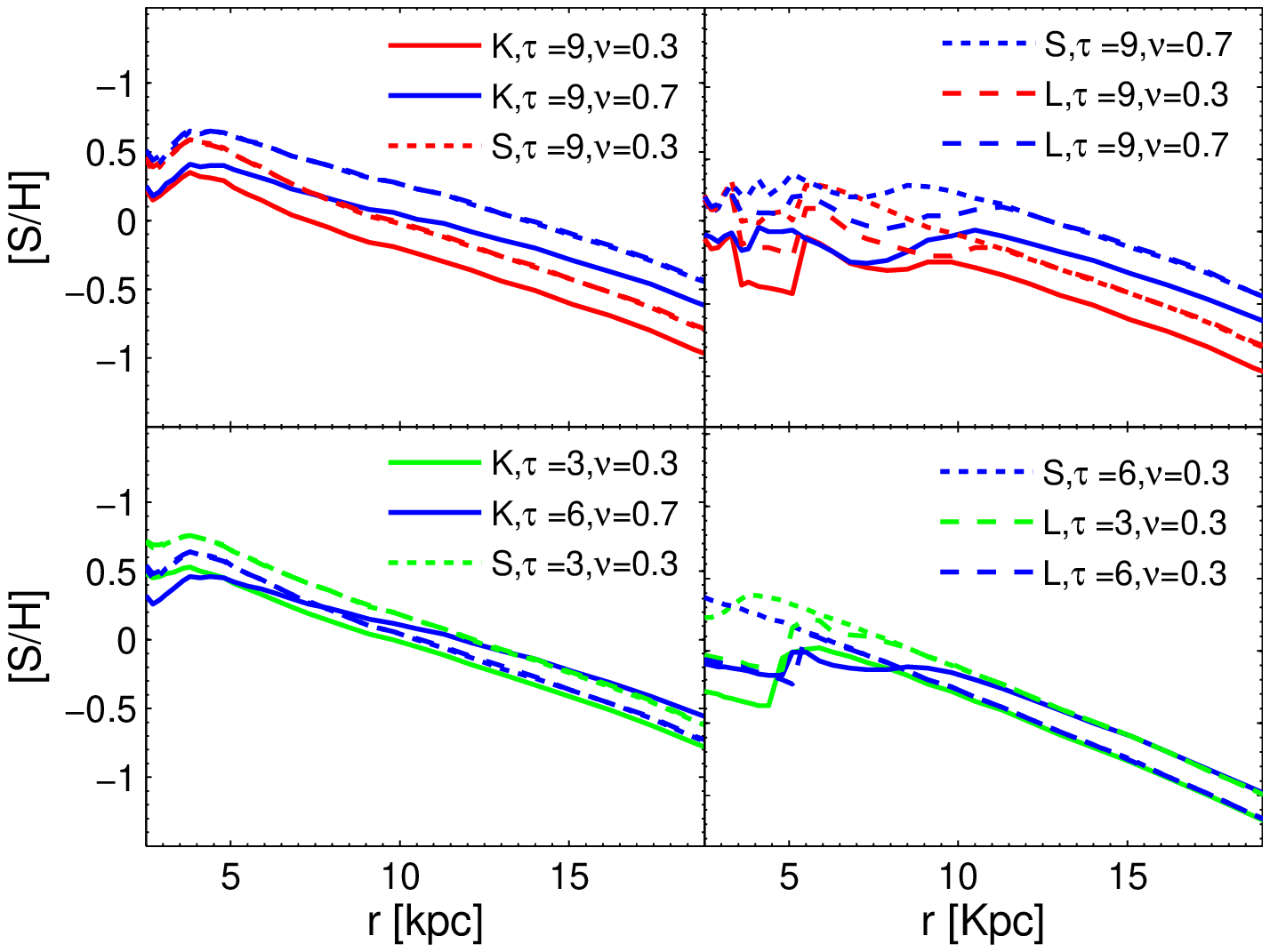}
\includegraphics[height=6.cm,width=7.5truecm]{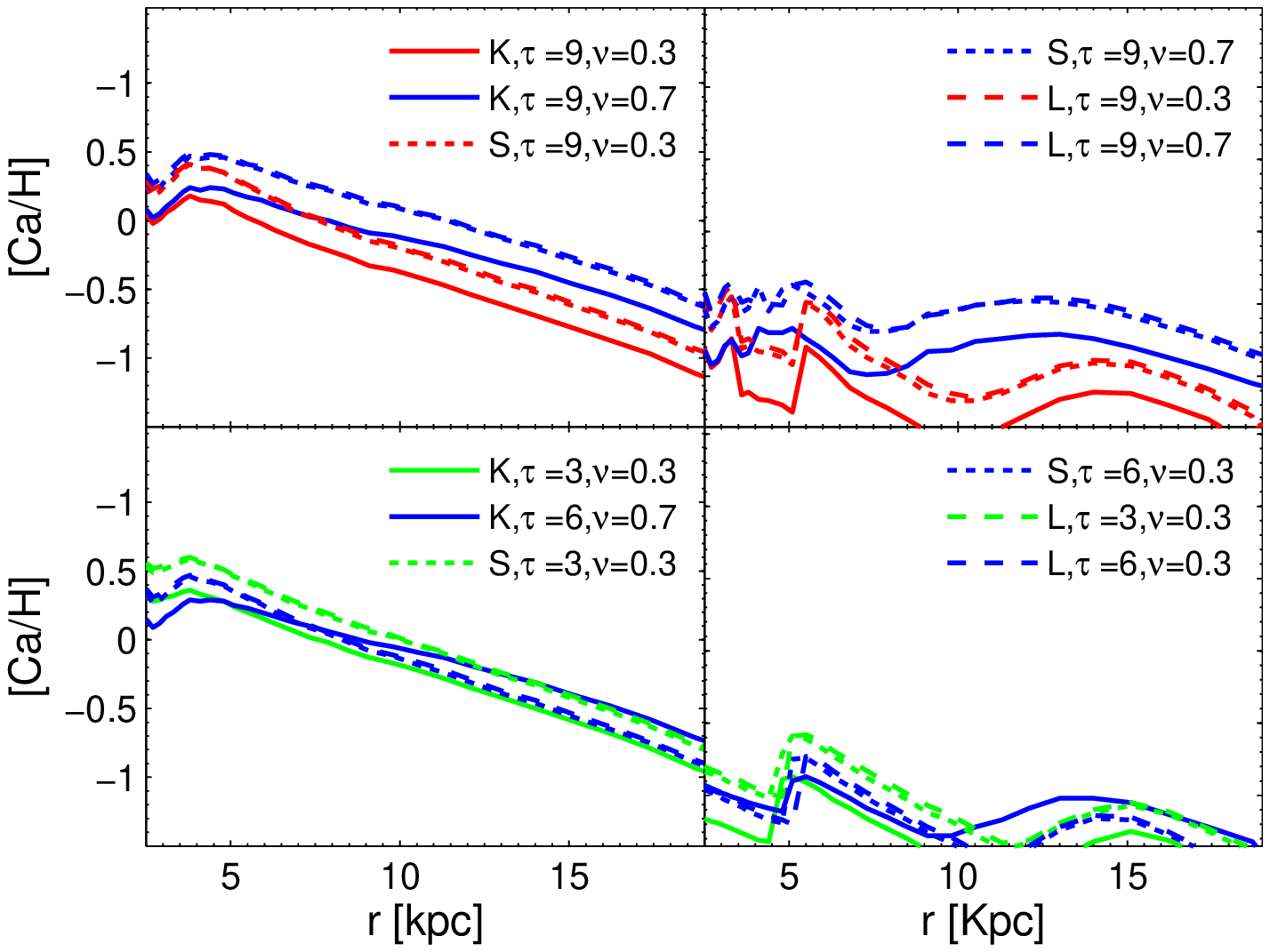}}
\caption{The Galactic abundances of $[\textrm{Mg}/\textrm{H}]$,
$[\textrm{Fe}/\textrm{H}]$, $[\textrm{C}/\textrm{H}]$,
$[\textrm{Si}/\textrm{H}]$, $[\textrm{S}/\textrm{H}]$ and
$[\textrm{Ca}/\textrm{H}]$ in the gas as a function of the
galacto-centric distance. In all diagrams we present  the radial
gradient at the current age for both the ISM component as a whole
(that is gas+dust, left panels in each plot) and  the gaseous
component \textit{alone} (right panels in each plot) taking into
account the depletion of the elements due to the presence of dust.
Twelve combinations of the parameters are considered  for both the
un-depleted (left panels) and depleted (right panels) cases.
\textit{The legend has been split between left and right panels for
the sake of clarity}. The following cases are shown. Kroupa IMF:
$(\tau,\nu) = (3,0.3)$ (continuous green line), $(\tau,\nu) =
(6,0.7)$ (continuous blue line), $(\tau,\nu) = (9,0.3)$ (continuous
red line) and $(\tau,\nu) = (9,0.7)$ (continuous blue line);
Salpeter IMF: $(\tau,\nu) = (3,0.3)$ (dotted green line),
$(\tau,\nu) = (6,0.3)$ (dotted blue line), $(\tau,\nu) = (9,0.3)$
(dotted red line), and $(\tau,\nu) = (9,0.3)$   (dotted blue line);
Larson SoNe IMF: $(\tau,\nu) = (3,0.3)$ (dashed green line),
$(\tau,\nu) = (6,0.3)$ (dashed blue line), $(\tau,\nu) = (9,0.7) $
Gyr (dashed green line) and $(\tau,\nu) = (8,0.3)$ (dashed blue
line). All theses cases are also examined in \citet{Piovan11b} for
the SoNe.} \label{EvolRadialAbun}
\end{figure*}

Finally, in the six panels of Fig. \ref{EvolRadialAbun} we show the
abundances of $[\textrm{Mg}/\textrm{H}]$,
$[\textrm{Fe}/\textrm{H}]$, $[\textrm{C}/\textrm{H}]$,
$[\textrm{Si}/\textrm{H}]$, $[\textrm{S}/\textrm{H}]$ and
$[\textrm{Ca}/\textrm{H}]$ in the gas as a function of the
galacto-centric distance. To single out the difference  between
models with and without dust depletion, in all the diagrams we show
the radial gradient at the current age for the ISM  as a whole (that
is gas+dust, left sub-panels in each plot) and for the gaseous
component  \textit{alone} (right sub-panels in each plot) including
the effect of depletion. Twelve combinations of the parameters IMF,
$\nu$ and $\tau$ are shown for both the un-depleted (left panels)
and depleted (right panels) cases.  For each group of four panels we
have therefore six un-depleted cases on the top left sub-panel and
six on the bottom left sub-panel. The same six plus six cases are
shown for the depleted gas in the top and bottom right sub-panels.
As expected, dust globally lowers the amount of elements in the gas
across the  whole   disk, but also in some cases alters
significantly what would be the slope of the gradient  in absence of
dust. In particular, in the innermost regions, where a higher
density of heavy atoms is available to accrete on the seeds, the
regularity of the gradient breaks more strikingly than in the
outermost regions where the gradient keeps roughly the original
slope. The innermost regions, that are also affected by the bar
$(\textrm{r} < \textrm{5} \, \textrm{kpc})$, present a even more
irregular behaviour, in particular for the elements most depleted
and participating to the complex process of silicates formation,
like Magnesium and Silicon. Indeed, Carbon keeps a regular gradient
as in our model no other heavy elements are supposed to participate
to the accretion process of carbonaceous grains. Iron is also very
irregular in the innermost regions, since it participates both to
the formation of silicates and iron dust grains.

\section{Discussion and conclusions}\label{Discus_Concl}

This is the last paper of a series of three
\citep{Piovan11a,Piovan11b} devoted to the study of star-dust and
dust accretion in the ISM. We have analyzed in some detail the
chemical gradients (gas and dust) across the Disk of the MW. The MW
as a whole and its Disk and SoNe have been considered as ideal
laboratories in which theories of chemical evolution and dust
formation are compared with observational data. A suitable
description of the Disk including radial flows of matter and the
presence of a inner Bar has been adopted to simultaneously explain
the bump on the inner gas density and the gradients in chemical
abundances. This model provides the frame to the formation and
evolution of the dust, one of the components of the ISM (the gas
elemental species being the others).

The main conclusions of this study can be summarized as follows:

\begin{itemize}
\item For plausible combinations of the parameters, the same that
in \citet{Piovan11b} allowed to nicely reproduce the properties of
the SoNe, both for gas and dust, we find a good general agreement
with the mass gradients observed in the MW Disk, both for the total
gas mass and the single element abundances. These latter  have been
compared with the abundances observed in different types of objects,
like Cepheids, open clusters, HII regions, O and B stars and Red
Giants. Some striking differences between models and observations
for C and N can be easily explained by means of the stellar
evolution theories.

\item The evolutionary scheme for dust formation and evolution
across the Disk, emerging from the simulations, is as follows:
SN{\ae} dominate the dust budget in the early stages in the whole
disk. As time goes on, AGB stars enter the game: they contribute to
the C-based dust budget if the metallicity keeps low,  switch to
O-rich dust as soon as the metallicity is near  solar or
super-solar. For this reason AGB stars behave in a different way in
the inner regions where the metallicity quickly increases and in the
outer regions where the metallicity keeps low for  long time. The
ISM accretion is the main contributor to the dust budget across the
whole disk. It starts very early on  in the innermost regions due to
the faster enrichment and the highest number  densities of the
elements involved in dust and much later in the outer regions where
metallicity and density are lower and where AGB stars of low-Z can
be very important for many Gyr. It is clear from this trend that a
crucial role is played by the SN{\ae}: the faster they enrich the
ISM in metals, the faster the ISM starts to play a decisive role.
Furthermore, the higher  the condensation efficiencies, the higher
is the amount of dust that can be injected during the time interval
before the ISM dominates. The uncertainties on the efficiency of
dust condensation in SN{\ae} \citep{Piovan11a} could clearly change
the picture during the early evolutionary phases
\citep{Zhukovska08,Gall11a,Gall11b}.

\item The radial distribution of the dust mass follows the behaviour
of the gas showing a bump in the inner zone and a negative slope at
the present time. It decreases moving outwards. The fractional
radial mass of silicates has also  a \textit{negative} slope,
whereas the carbonaceous grains have a \textit{positive} slope. They
also  play a more and more important role as we move toward the
outskirts. This is ultimately due to the combined effect of several
causes : (i) the yield from SN{\ae} that enrich the ISM with
elements involved in silicates and dust where star formation is more
active; (ii) the long time interval during which low-Z AGB stars
inject carbon in the outskirts of the Disk, and (iii) the delay in
the onset of the ISM accretion in the outer regions.

\item The depletion of the elements is more significant in the
inner regions, with the highest star formation and densities, while
in the outskirts we have less depletion due to lower efficiency of
dust accretion and star-dust injection.

\item As expected, the depletion significantly alters  the radial
gradients in the gas  abundances, in particular for the elements
involved in silicates, for which (at least in our simplified model)
many elements enter the process and the key element driving the
process can vary with time, making the whole thing more irregular.
\end{itemize}

We can conclude by saying that the picture drawn by  the chemical
model for the Galactic Disk in presence of dust  is  consistent with
the way in which SN{\ae}, AGB stars of different Z and ISM
contribute to the total dust budget. In particular, since the radial
properties of the elements along the disk are satisfactorily
reproduced, we are confident that the dust enrichment scenario
described is based upon a consistent general enrichment in metals of
the ISM.

\textit{Acknowledgements}. L. Piovan acknowledges A. Weiss and the
Max Planck Institut Fur AstroPhysik (Garching - Germany) for the
very warm and friendly hospitality and for providing unlimited
computational support during the visits as EARA fellow when a
significant part of this study has been carried out. The authors are
also deeply grateful to  S. Zhukovska and H. P. Gail for many
explanations and clarifications about their model of dust accretion,
T. Nozawa and H. Umeda for many fruitful discussions about SNa dust
yields. This work has been financed by the University of Padua with
the dedicated fellowship "Numerical Simulations of galaxies
(dynamical, chemical and spectrophotometric models), strategies of
parallelization in dynamical lagrangian approach, communication
cell-to-cell into hierarchical tree codes, algorithms and
optimization techniques" as part of the AACSE Strategic Research
Project.

\begin{scriptsize}
%=================== BIBLIOGRAPHY =============================%
\bibliographystyle{apj}                          % Files .bst
\bibliography{mnemonic,PiovanIII}    % Files .bib
%=================== END DOCUMENT =============================%
\end{scriptsize}

\end{document}